\documentclass[a4paper, 11pt]{extarticle}

\pdfoutput=1
\usepackage{amsfonts}
\usepackage{amssymb}
\usepackage[margin=1in]{geometry}
\usepackage{listings}
\usepackage{color}
\usepackage{graphicx}
\usepackage{float}
\usepackage{caption}
\usepackage{subcaption}
\usepackage{multicol}
\usepackage{enumitem}
\usepackage{xcolor}
\usepackage{float}
\usepackage{gensymb}
\usepackage[export]{adjustbox}

\usepackage{titlepic}
\usepackage{bm}
\usepackage{nccmath}
\usepackage{url}
\usepackage{lmodern}
\usepackage[section]{placeins}
\usepackage{braket}
\usepackage{bbm}
\usepackage{esvect}
\usepackage{setspace}
\usepackage{cancel}
\usepackage{tikz}

\usepackage{amsthm}

\usepackage[hidelinks,colorlinks=false,pdfencoding=auto,psdextra,pdfpagelabels,pdfauthor={Cameron Beetar},pdftitle={The Massive Classical Double Copy in Three Dimensions},pdftex]{hyperref}[colorlinks=false]

\begin{document}
\numberwithin{equation}{section}
        \pagenumbering{gobble} 
        \begin{titlepage}
                \centering
                \null\vspace{7em}
                {\scshape \Huge Approaches to Quantum Cosmology\\}
                \vspace{2em}
                {\scshape \Large Raymond Isichei}
                \vspace{2em}
                \begin{figure}[h]
                \centering
                \includegraphics[scale=0.25]{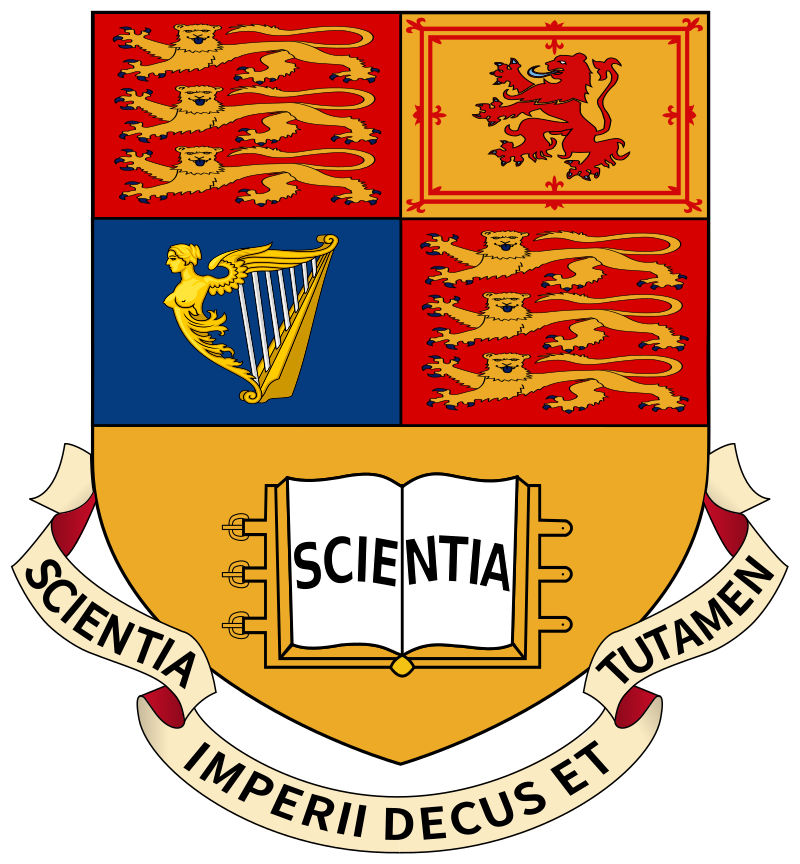}
                \end{figure}
                \vfill
                {\itshape\large Thesis Presented for partial fulfilment of the Degree of\\}
                \vspace{1.5em}
                {\scshape \Large Master of Sciences in Quantum Fields and Fundamental Forces\\}
                \vspace{1.5em}
                {\itshape\large  in the Theoretical Physics Group, Blackett Laboratory,\\
Imperial College, London SW7 2AZ, United Kingdom\\2022\\}
                \vspace{5em}
                {\scshape \Large Supervised by Prof. Jo\~ao Magueijo}
                \vspace{5em}\null
                \thispagestyle{empty}
        \end{titlepage}
        \newgeometry{margin=1in} 
        \pagenumbering{arabic} 
        \null

\tableofcontents
\newpage
\thispagestyle{plain}
\begin{center}
    \Large
    \textbf{}
        
    \vspace{0.4cm}
    \large

    \vspace{0.4cm}
    \textbf{}
       
    \vspace{0.9cm}
    \textbf{Abstract}\\
 
\end{center}
The aim of this thesis is the analysis of attempts to model the universe as a quantum mechanical system. Specifically, attempts to derive a wavefunction which describes the evolution of the universe by using the path integral approach to quantisation. This approach was pioneered by Hartle \& Hawking in the context of Euclidean path integrals. Wick rotation of the closed FRW metric is employed as the Lorentzian path integral is highly oscillatory and generally does not converge. The wavefunction of the universe is found by summation over compact complex metrics and matter fields which are regular on them. However, the Euclidean path integral suffers from many pathologies such as being unbounded from below. Recent work done by Feldbrugge, Lehners \& Turk employs the use of Picard-Lefshets Theory to deform the Lorentzian Path integral so that it becomes absolutely convergent . From this the No Boundary wavefunction of Hartle \& Hawking as well as the Tunneling wavefunction of Vilenkin can be derived. In the final section we carry out novel calculations in which the path integral of the Einstein-Cartan action for the closed FRW metric is calculated. Recent results obtained by Magueijo \& collaborators as well as previous results derived in the context of Euclidean and Lorentzian path integrals are obtained.
\newpage
\newpage
\thispagestyle{plain}
\begin{center}
    \Large
    \textbf{}
        
    \vspace{0.4cm}
    \large

    \vspace{0.4cm}
    \textbf{}
       
    \vspace{0.9cm}
    \textbf{Acknowledgments}\\
 
\end{center}
I want to extend my sincerest thanks to Professor Magueijo for the many hours he dedicated to understanding this project and for constantly challenging me at every stage of its development. Through his guidance, the study of a simple system became the best possible introduction I could have asked for. I would also like to thank Bruno for feedback and many helpful comments. Finally I would like to thank  Tim Evans for his constant support during a difficult year.

\newpage
\section{Introduction}
The singularity theorems of Penrose \& Hawking established singularities as being an inescapable aspect of General Relativity\cite{P1}\cite{H1}\cite{H2}\cite{HP}. Singularities are not simply points in spacetime in which the metric is undefined or becomes non differentiable\cite{HE}. If this were the case then one could simply use a pair of scissors to cut out these points from the manifold and declare the remaining manifold to be the entirety of spacetime . A declaration of victory over singularities would be entirely reasonable as the new manifold is singularity free according to this definition. Rather, a more complete understanding of what it means for a spacetime to have a singularity can be found in understanding the geodesics defined on that manifold. The objective of these singularity theorems was to establish that the presence of a spacetime singularity goes hand in hand with the incompleteness of null and timelike geodesics on that spacetime manifold. If a timelike or null geodesic is incomplete, the affine parameter of the geodesic cannot have arbitrary length and that observers travelling along along these incomplete geodesics in the direction of the singularity will begin to will experience growing curvature forces that become infinite.\\

The physical significance of this is that geodesics of of null or timelike observers cannot be traced arbitrarily far into the future or arbitrarily far back in the past. Thee investigation into future singularities and the geodesic incompleteness by Penrose was motivated by gravitational collapse of massive bodies and the fate of null and timelike observers within and around the Schwarzchild radius in Schwarzchild spacetime\cite{P1}. In this case, null and timelike geodesics are incomplete in the future sense. Hawking subsequently argued that as a process, expansion of the universe is quite similar to the collapse of a star and may be considered as a time reversed version of the former\cite{H1}\cite{H2}\cite{HE}. A consequence of singularity theorems in this vein is that for timelike and null observers in expanding space times there must be some notion of a beginning. The current consensus is that these singularities are artefacts of the extrapolation of classical General Relativity beyond its domain of applicability.\\

In General Relativity spacetime is physical and dynamical. Through the Einstein Field Equations we see that this dynamical spacetime interacts with matter whose interactions can be described by frameworks such as Quantum Electrodynamics and Quantum Chromodynamics. The logical conclusion of this relationship between matter and spacetime via the Einstein Field Equations is that spacetime itself must be quantised. It has been almost 90 years since the the first treatise on the quantisation of the gravitational field was written\cite{BRON} and there still does not exist a clear and unique approach to the quantisation of spacetime. There are currently several competing approaches to quantising the gravitational force including String Theory\cite{POL} and Loop Quantum Gravity\cite{ROV} amongst others. These theories differ in what is considered as the quantised degrees of freedom of which classical spacetime is the continuum limit.\\

The Planck scale is widely accepted as the scale at which Quantum gravity affects become significant. Currently there are no experimental tests which may probe effects at this scale directly or indirectly. Therefore there is no observational motivation for what the quantised degrees of freedom could be. The mathematical structures which are seen as the fundamental quantum gravitational degrees of freedom in these theories are far removed from classical spacetime. A lack of an experimental guide makes the computation of physical phenomena and predictions from these models difficult\cite{BO}.\\

Quantum cosmology is one attempt at quantising gravitational degrees of freedom. If quantum mechanics is fundamental, then the universe as a whole is a quantum mechanical system\cite{BO}. If the universe is a quantum mechanical system, then expansion of the universe should correspond in some crude way correspond to the evolution of the wavefunction. Trying to immediately quantise the universe and all the complicated interaction therein is a futile task. The task at hand is simplified by enforcing homogeneity and isotropy. This reduces infinite degrees of freedom to only a few which is a more feasible task. The description of previous and current attempts is the goal of this thesis. \\

In Chapter 2, a summary of the Borde, Guth \& Vilenkin theorem on the geodesic incompleteness of spacetimes is provided as motivation for why attempts to model the early universe quantum mechanically are important. Chapter 3 is a review of No Boundary wavefunction and how it can be derived by considering a Euclideanised closed FRW metric. Chapter 4 gives a brief overview of how, starting from the Wheeler De Witt equation, the No Boundary and Tunneling wavefunctions can be obtained as linearly independent solutions. The Euclidean path integral was chosen as the starting point because Lorentzian path integrals are highly oscillatory. Chapter 5 is a review of how Picard-Lefshets theory can be used to make Lorentzian path integrals convergent. Despite the convergence, there is significant debate concerning which of these wavefunctions is derived correctly. In Chapter 6, the path integral of the Einstein-Cartan action which enables a short derivation of previous results and new results. 

\newpage

\section{Motivation}
\subsection{Inflationary Cosmology: An end in search of a beginning?}
In the usual case of inflationary cosmological models based on the closed FRW metric and inflaton field, inflationary processes can be extrapolated eternally into the future\cite{V1}\cite{L1}. These models consist of a isolated post inflationary, or, thermalised regions which are embedded in an always inflating background. The boundaries of thermalised regions expand into this background. However, the inflationary regions which separate the thermalised regions expand at a faster rate such that post inflationary regions do not merge \cite{BV1}. In this picture the comoving volume of inflating regions vanishes as $t \rightarrow \infty$ and therefore it can be inferred that the universe in its entirety can never be considered as post inflationary. With this in mind, a reasonable question to consider is whether or not inflationary models can be extended arbitrarily far back in past making the universe past eternal. This model is attractive as there is no initial singularity since the universe would have always existed.\\

The current consensus is that singularities that appear in general relativity can only be fully understood within the context of a quantum theory of Gravity. As this theory is currently unknown, a singularity at the beginning of time therefore represents an incompleteness of inflationary cosmology. However, it is now commonly accepted that a past eternal universe model is not reconcilable with the use of the closed FRW metric in inflationary models. Using the assumption that the energy momentum tensor obeys the Weak Energy Condition, it has been proven that geodesics in inflationary space-times are past-incomplete\cite{BG1}. All forms of classical matter obey the Weak Energy Condition\cite{HE}. However, quantum fluctuations in inflationary models may violate this condition. It has been noted that increases in the Hubble parameter are associated with violations of the WEC. If spacetime is approximated as locally flat then the increase in the Hubble parameter is necessary for future eternal chaotic inflation\cite{BGV}.\\

The strongest proof of the incompleteness of null and timelike geodesics in inflationary spacetimes was given in Borde, Guth \& Vilenkin (BGV) in which the Weak Energy Condition is not assumed\cite{BGV}. Rather, BGV base their proof on the premise that in full de Sitter space, physical intuition would suggest that taking the limit as $t\rightarrow -\infty$ corresponds to exponential contraction. If in this limit, post inflationary thermalised regions exist, then it is reasonable to infer that the entire universe would be thermalised before inflation were to begin. The contracting phase can be excluded by only considering past spacetimes for which the average Hubble parameter in that spacetime is positive, or, 
\begin{equation}
    H_{av}>0.
\end{equation}
For a suitable definition of the Hubble parameter over a region of spacetime, the past incompleteness of null and timelike geodesics is a consequence of this condition. The starting point of the BGV argument is the premise that the neighbourhood of any point p in an inflating region can be approximated as homogeneous, flat and isotropic. The region around this point can be described using the flat FRW metric
\begin{equation}
    ds^{2}=-dt^{2}+a^{2}d\vec{x}^{2}.
\end{equation}
\subsection{Null Geodesic Incompleteness}
Rewriting this metric in terms of derivatives with respect to the affine parameter, the line element becomes
\begin{equation}
    \bigg(\frac{ds}{d\lambda}\bigg)^{2}=-\bigg(\frac{dt}{d\lambda}\bigg)^{2}+a^{2}\bigg(\frac{d\vec{x}}{d\lambda}\bigg)^{2}.
\end{equation}
Given that for null geodesics the left hand side of the above equation is zero[more detail], it therefore follows that
\begin{equation}
    \frac{dt}{d\lambda}=a\frac{d\vec{x}}{d\lambda}.
\end{equation}
From this relation one can deduce that $d\lambda \propto adt$. An important consideration is that of frequencies in this region of spacetime. In FRW cosmology, relative expansion of the spacetime is parameterised by the scale factor $a$. Therefore for increasing $a$ corresponding to expansion would expect the redshifting of frequencies and for there to be an inverse relationship between frequency and the scale factor. To make this more concrete, BGV consider a wave vector $k^{\mu}$ which is tangential to the geodesic. This wave vector is related to the affine parameterisation of the the geodesic by
\begin{equation}
    k^{\mu} \propto \frac{dx^{\mu}}{d \lambda}.
\end{equation}
Using the alternative but equivalent formula for the line element, we can see that if $k^{\mu}$ is a null vector then
\begin{align}
    \begin{split}
        ds^{2} = \ & g_{\mu \nu}\dot{x}^{\mu}\dot{x}^{\nu}\\
        =& g_{\mu \nu}k^{\mu}k^{\nu}\\
        =&0
    \end{split}
\end{align}
The physical frequency measured by comoving observers is given by $k^{0}$. Slightly rearranging this relevant to in the case of time it can easily be seen that,
\begin{equation}
    d\lambda \propto \frac{dt}{k^{0}}.
\end{equation}
If $k^{0} \propto a^{-1}$ then the initial relationship between the affine parameter and time differentials is recovered.\\

The question of whether or not inflationary spacetimes extend arbitrarily far back in the past can be understood in terms of the affine length of geodesics. If $a$ decreases rapidly in the past direction then the maximum affine length must be bounded. This is a result of the proportionality relationship between the differentials of the affine length and the time. Schematically the maximum affine length can be determined by integrating both sides of this relationship as
\begin{equation*}
    \underbrace{\int_{\lambda(t_{i})}^{\lambda(t_{f})}d\lambda}_{\text{Affine Length}} \propto \int_{t_{i}}^{t_{f}}a dt.
\end{equation*}
The notion the affine length can be related to the Hubble parameter averaging condition. The relation between the affine length and time can be made exact by the normalisation of the affine parameter using the scale factor at a reference time $a(t_{f})$
\begin{equation}
    d\lambda = \frac{a}{a(t_{f})}dt, 
\end{equation}
such that $d\lambda/dt=1$ when $t=t_{f}$. The Hubble parameter can be defined in terms of the scale factor and its derivatives by $H \equiv \dot{a}/a$. Multiplying by the Hubble parameter relation  and then integrating both sides of the equality
\begin{equation}
   \int_{\lambda(t_{i})}^{\lambda(t_{f})}H(\lambda)d\lambda = \int_{a(t_{i})}^{a(t_{f})}\frac{da}{a(t_{f})} \le 1.
\end{equation}
The above expression becomes an equality when $a(t_{i})=1$. The average Hubble Length over the affine parameter $H_{av}$ is given by
\begin{equation}
    \frac{1}{\lambda(t_{f})-\lambda(t_{i})}\int_{\lambda(t_{i})}^{\lambda(t_{f})}H(\lambda)d\lambda \le \frac{1}{\lambda(t_{f})-\lambda(t_{i})}.
\end{equation}
This expression indeed satisfies the averaging condition  enforced on the spacetime. Form these considerations we see that as we travel in the reverse time direction, null geodesics must have finite affine length and are therefore past incomplete.\\

\subsection{Timelike Geodesic Incompleteness}
This reasoning can be easily extended to the case of timelike geodesics parameterised by a proper time $\tau$. The standard relativistic definition of the 4 momentum of a particle mass m is give by
\begin{equation*}
    P^{\mu} \equiv m\frac{dx^{\mu}}{d\tau}.
\end{equation*}
Identifying the $x^{0}$ component as the time component and $P^{0}=E$ where $E$ is the energy measured by a comoving observer, the time component of the above expression is
\begin{equation}
    d\tau = \frac{m}{E}dt.
\end{equation}
In the context of general relativity, the standard relativistic dispersion relation can be generalised such such that the magnitude of the three momentum can be defined in terms of the metric and the three momentum vector by
\begin{equation}
    p^{2}=-g_{ij}P^{i}P^{j}\longrightarrow a^{2}(P^{i})^{2}
\end{equation}
The inverse proportionality between the magnitude of the three momentum p and the scale factor can be used as the starting point to motivate the notion of affine length  for timelike geodesics. Similar to the case of null geodesics, the three momentum can be normalised by introducing the scale factor at a reference time $t_{f}$
\begin{equation}
    p(t)= \frac{a(t_{f})}{a}p_{f}.
\end{equation}
As in the null geodesic case each side of the equation can be multiplied by the Hubble parameter to yield the relation
\begin{equation}
    \int_{t_{i}}^{t_{f}}H(\tau)d\tau=\int_{a(t_{i})}^{a(t_{f})}\frac{m}{E}\frac{da}{a}.
\end{equation}
Using the relativistic dispersion relation $E=\sqrt{p^{2}+m^{2}}$, the denominator can be rewritten such that the overall relation becomes
\begin{align}
\begin{split}
    \int_{t_{i}}^{t_{f}}H(\tau)d\tau&=\int_{a(t_{i})}^{a(t_{f})}m\ \frac{da}{\sqrt{m^{2}a^{2}+p^{2}a^{2}}}\\
    &= \int_{a(t_{i})}^{a(t_{f})}m\ \frac{da}{\sqrt{m^{2}a^{2}+p_{f}^{2}a(t_{f})^{2}}}.\\
\end{split}
\end{align}
Where the relation between 3 momentum $p(t)$ and reference time 3 momentum given was substituted into the denominator of the integral. This yields an integral with an analytic solution As in the case of null geodesics we seen that the integral of the Hubble parameter over the affine parameter is greater than zero. The solution to the integral is given by,
\begin{align}
\begin{split}
    \int_{t_{i}}^{t_{f}}H(\tau)d\tau & \le ln\Big(\frac{E_{f}+m}{p_{f}}\Big),\\
    & \le \frac{1}{2} ln\Big( \frac{\gamma +1}{\gamma - 1}\Big),
\end{split}
\end{align}
where the dispersion relation at reference time $t_{f}$ is $E_{f}\equiv \sqrt{p_{f}^{2}+m^{2}}$. The $\gamma$ term on the second line is related to $E_{f}$ and $p_{f}$ by
\begin{equation*}
    \gamma \equiv \frac{1}{\sqrt{1-v_{rel}^{2}}} \quad \text{{where}} \quad v_{rel} \equiv \frac{p_{f}}{E_{f}}.
\end{equation*}
The relative velocity or $v_{rel}$ is the relative speed of the geodesic with respect to comoving observers. The expression this equation can be averaged proper time to show that the average Hubble parameter in this region over a certain proper time interval is greater than zero which is inseparable from the fact that timelike geodesics in flat FRW spacetime cannot have arbitrary affine length. In \cite{BGV}, this argument is extended to arbitrary space-times with no assumption of homogeneity, isotropy or flatness. A general Hubble parameter which applies to arbitrary space-times is given by
\begin{equation}
    H = \frac{d}{d\tau}F(\gamma(\tau)) \quad \text{{where}} \quad     F(\gamma)= 
\begin{cases}
    \gamma^{-1},& \text{if } \kappa=0 \ \text{{(Null)}}\\
    \frac{1}{2}ln\Big( \frac{\gamma +1}{\gamma - 1}\Big),              & \text{if} \ \kappa=1 \ \text{(Timelike)}.
\end{cases}
\end{equation}
By integrating this generalised Hubble parameter over the affine parameter, a more general form of the previous inequalities
\begin{equation}
    \int_{\tau_{i}}^{\tau_{f}}H d\tau = F(\gamma_{f})-F(\gamma_{i}) \le F(\gamma_{f}),
\end{equation}
with the right hand side of the equality being true if $F(\gamma_{i})=0$. Each side of the inequality can be averaged to show that $H_{av}$ must be positive for any null or timelike geodesic and therefore these geodesics in arbitrary space-times are past incomplete.\\

The fact that null and timelike geodesics cannot have arbitrary affine length in FRW as well as general spacetimes is an indicator that classical inflationary cosmology cannot be used to avoid the nature of the singularity that is associated with the birth of the universe\cite{BV1}.
\newpage 
\section{Euclidean Path Integrals}
\subsection{Convergence via Complex Time}
Originally the question of the applicability of path integral quantisation to General Relativity was investigated entirely within the context of a Euclideanised path integrals. This is primarily due to the integrand $e^{iS}$ where $S$ is the classical action, being highly oscillatory. The most widely sued method for ensuring  the convergence of the exponential path integral is the rotation of the time coordinate into the complex plane. This convergence can be easily seen starting from the canonical action defined in terms of conjugate variables $q$, $p$ and a Hamiltonian $H(q,p)$
\begin{equation}
    S=\int_{t_{i}}^{t_{f}}dt \Bigl(p\frac{dq}{dt}-H(q,p) \Bigl).
\end{equation}
A redefinition representing a rotation of the tie coordinate into the complex plane can be defined as $t=e^{-i\epsilon}\tau$ where $\epsilon<1$. Given this rotation, the end points of the integral are suitably rotated such that $t_{i,f}=e^{-i\epsilon}\tau_{i,f}$. The above canonical integral becomes
\begin{align}
\begin{split}
    S &=\int_{\tau_{i}}^{\tau_{f}} (e^{-i\epsilon}d\tau) \  \Bigl(p\frac{dq}{d(e^{-i\epsilon} \tau)}-H(q,p) \Bigl),\\
    &=\int_{\tau_{i}}^{\tau_{f}} d\tau \Bigl(p\frac{dq}{d\tau}-e^{-i \epsilon}H(q,p) \Bigl)
\end{split}
\end{align}
Noting that the first term in the integral corresponds to a difference the product of conjugate variables at inital and final times, the second term must be responsible for the convergence properties of the integrand. Noting that the real part of the complex action $iS$ is
\begin{equation}
    Re[iS]=-\sin(\epsilon)\int_{\tau_{i}}^{\tau_{f}} d\tau \ H(q,p) <0.
\end{equation}
As the energy eigenstates of the Hamiltonian in QFTs are associated with positive energy eigenvalues, it is assumed that the above Hamiltonian is positive. Due to the overall integral being negative, as the the Hamiltonian becomes large $e^{Re[iS]}\rightarrow 0$. Therefore, by virtue of rotation of time into the complex plane, the path integral converges. The success of the complexification of time in regular QFTs was the catalyst for attempts to do the same within gravitational path integrals.
\subsection{Heuristics of Gravitational Path Integrals}
If the entire universe is to be treated as a quantum mechanical object, then the states which the universe can occupy must be specified. The dynamical behaviour of the universe can then be determined from the specified states. However, the specification of states and subsequent dynamics is not enough to ensure the uniqueness of our present universe. This uniqueness can only come from a restriction on the possible states that may be occupied. Hartle \& Hawking(HH) argued that a state is determined by specifying a wavefunction on an appropriate configuration space. Due to this, they start with the premise that that a quantum mechanical state may be constructed from a fundamental amplitude which specifies the complete history of the system being analysed\cite{HH1}.\\

The construction of gravitational path integrals depends on the extrapolation of ideas from path integrals in ordinary QM and QFTs the case of the Einstein-Hilbert action. The path integral approach to quantisation is also referred to as a \textit{sum over histories} approach. In ordinary QM the history of a single particle is a path $x(t)$. The probability amplitude $P$ for a particular path is then given by 
\begin{equation}
    P \propto e^{iS[x(t)]}.
\end{equation}
This is the fundamental amplitude which specifies the complete history of the system. In order to restrict the possible states that the system may be in, the amplitude can be restricted by the notion of superposition which in this case requires a sum over all states. For a particle which was located at position $x(t)$ and is now nowhere else at another time $t'$ or $x(t')=0$. The QM state $\psi(x,t)$ associated with this restriction is related to the amplitude by the relation
\begin{equation}
    \psi(x,t)=N\int d(x(t))\ e^{iS[x(t)]},
\end{equation}
where N is a normalisation constant and the integrand can be considered as the sum over a class C of histories which intersect position $x$ at time $t$. As mentioned previously, integrals of this type are considered to be highly oscillatory. Complexifying time ensures the convergence of this integral. The restrictions discussed can be easily generalised to the case where the particle is at $x(t)$ but was previously located at $x(t')=x'$. This yields the propagator $\braket{x,t|x',t'}$
\begin{equation}
    \braket{x,t|x',t'}=\int_{x'(t')}^{x(t)} d(x(t))\ e^{iS[x(t)]}.
\end{equation}
The ground state can be defined by the sum over the class of paths which have vanishing action in the far past. Wick rotation is employed in order to ensure the state is positive definite. The complex time parameter $\tau$ acts as a proper time coordinate with the ground state being defined as 
\begin{equation}
    \psi(x,0)=N\int d(x(\tau))^{-S[x(\tau)]}.
\end{equation}
For a well defined time coordinate and a corresponding time independent Hamiltonian, the propagator $\braket{x,t|x',t'}$ can be specialised to the case where $x'=0$ and $t=0$. Using the method of insertion of complete states, this propagator yields the path integral
\begin{align}
    \begin{split}
        \braket{x,0|0,t'} &= \sum_{n} \psi_{n}(x)\Bar{\psi_{n}}(0)e^{iE_{n}t},\\
        &= \int d(x(t)) \ e^{iS[x(t)]}.
    \end{split}
\end{align}
In direct extrapolation of path integral formalism from QM and QFT, the quantum mechanical amplitude $P_{QG}$ associated with the occurrence of a given spacetime configuration is given by\cite{HH1}
\begin{equation}
    P_{QG} \propto e^{-S_{EH}[g]},
\end{equation}
where $S_{EH}[g]$ is the Einstein-Hilbert action with positive cosmological constant $\Lambda$ and including the GHY Boundary Term
\begin{equation}
S_{EG}[g]=\frac{m_{p}^{2}}{16\pi} \Bigl(\int_{\mathcal{M}}d^{4}x\sqrt{g} \ (R-2\Lambda) + \int_{\partial \mathcal{M}}d^{3}x\sqrt{h} \ 2K\Bigl),
\end{equation}
where $R$ is the Ricci Scalar derived from the metric $g_{\mu \nu}$, $K$ is the trace of the second fundamental form and $h_{ij}$ is the induced 3-metric on the boundary $\partial \mathcal{M}$. For the first term in the action, the scalar density in the diffeomorphism invariant measure is taken as $\sqrt{g}$ rather than $\sqrt{-g}$. This is due to the Wick rotation of the time coordinate such that the manifold $\mathcal{M}$ is now a Riemannian manifold with signature $(++++)$. The above arguments can be extended to the case of the Einstein-Hilbert action with a scalar field $(\phi)$ coupled to gravity. In this case the amplitude $P_{QG}$ would be given by
\begin{equation}
\label{E1}
    P_{QG} \propto e^{iS[g,\phi]}.
\end{equation}
\newline
In order to build states from this amplitude, only spatially closed universes will be considered. For the case of the single particle, one is not interested in the entire history of the particle but rather results of observation of the particle at different times. As manifolds are compact and Riemannian, in the case of gravitational path integral one is interested in the observation of fields and geometry on different space-like hypersurfaces. Starting with the probability (\ref{E1}) and taking the sum over unobserved quantities which in this case is the full metric $g_{\mu \nu}$ and matter field $\phi$ yields
\begin{equation}
\label{E2}
    P[h_{ij}, \phi]=\int_{C} \mathcal{D}g_{\mu \nu}\mathcal{D}\phi \ e^{-S_{EH}[g,\phi]},
\end{equation}
which can be interpreted the integral over all 4-metrics and matter field $\phi$ configurations which belong to a class C that contain a 3-submanifold on which there is an induced metric $h_{ij}$ and matter field configuration. This probability can be factorised into wavefunction amplitudes
\begin{align}
\label{E3}
\begin{split}
    P[h_{ij}, \phi]&=\psi_{+}[h_{ij}, \phi] \cdot \psi_{-}[h_{ij}, \phi],\\
    &=\int_{C_{+}} \mathcal{D}g_{\mu \nu}\mathcal{D}\phi \ e^{-S_{EH}[g,\phi]} \cdot \int_{C_{-}} \mathcal{D}g_{\mu \nu}\mathcal{D}\phi \ e^{-S_{EH}[g,\phi]},
\end{split}
\end{align}
where the $C_{+}$ and $C_{-}$ terms are subclasses of the class C which specifies the quantum state and will be explained in the next section on \textit{Boundary Conditions \& The No-Boundary Proposal}. The above integral therefore represent the amplitude $\braket{h^{"}_{ij}, \phi^{"}|h^{'}_{ij}, \phi^{'}}$ of transition from an initial spatial hypersurface and matter field configuration to the final hypersurface and matter field configuration\cite{HH1}\cite{HH2}. Time cannot be specified on the hypersurfaces due to their Riemmanian nature. The dynamics of the universe are therefore dictated by the wavefunctions $\psi_{\pm}[h_{ij]}, \phi]$. If the class $ C$ is chosen to be the class of compact metrics then $\psi_{+}[h_{ij}, \phi]$ and $\psi_{-}[h_{ij}, \phi]$ are the same. This wavefunction is, rather plainly, known as \textit{The Wavefunction of  The Universe}\cite{HH2}. If the analogy with QM and QFT is to hold, then this wavefunction should obey a functional differential equation. Physically sensible boundary conditions should also be used to constrain the dynamical behavior of the universe so that at the semi-classical approximation, the wavefunction of the universe behaves as our universe. A perfectly reasonable question to ask is, what are the boundary conditions of the entire universe?

\subsection{Boundary Conditions \& The No-Boundary Proposal}
It was noted by HH that the construction of models for natural phenomena rests on two foundational principles\cite{HH2}. The first is that fields obey local laws which govern their behaviour. These local laws are differential equations which are derived from an original action $S$ by variation of this action with respect to relevant fields. In the path integral approach, the propagation of the field is derived from a path integral over all configuration weighted by $e^{iS}$. The second principle is that boundary conditions conditions specify a unique state from the set of solutions to the differential equations which are local laws.\\

In the case of the gravitational path integral, the classical state may be specified by boundary conditions as is usually the case. The quantum state is usually defined by asymptotic conditions on the class $C$ of field configurations summed over in the path integral. If boundary conditions are not important in the case of the universe as a whole and the universe were free to start arbitrarily, the universe would surely evolve arbitrarily which is against our experience of the universe as a system which evolves in a regular manner. As Wick rotation means that the path integral is a sum over Euclideanised metrics, there are two natural choices of metrics that must be considered\cite{HH1}:
\begin{enumerate}
    \item Compact metrics with matter fields that are regular on them.
    \item Non-compact metrics which are asymptotically metrics of maximal symmetry (Flat Euclidean or Flat Anti de Sitter with $\phi \rightarrow 0$ at $\infty$.
\end{enumerate}

In the relation for  $P[h_{ij}, \phi]$  given in (\ref{E2}), the class $C$ that is summed over in the probability is divided into subclasses $C_{\pm}$ when considering the probability as being composed of the inner product of amplitudes. For example, if the class $C$  is chosen to be the second class listed above, then the subclasses $C_{\pm}$ consist of two different kinds of metrics\cite{HH1}
\begin{enumerate}
    \item[a.] Asymptotically Euclidean/Anti de Sitter metrics with an inner boundary at the hypersurface $S$.
    \item[b.] Disconnected metrics which consist of a compact part at the boundary S and an asymptotically Euclidean or Anti de Sitter part without an inner boundary.
\end{enumerate}
The second type of metric cannot be excluded as disconnected metrics can be approximated by  connected metrics in which different parts are connected by different tubes with negligible action. In spite of this, the dominant contribution to the path integral to the wave function is expected to come from metrics which are near solutions to the field equations. It can be shown that metrics of the type $a$ cannot give the dominant contribution to the wavefunction\cite{HH1}. Therefore if type $2$ metrics are chosen, then the dominant contribution to the wavefunction must come from subtype $b$. However, given that metrics of this subtype are connected metrics connected by thin tubes, this type of metric is the same type of metric as type 1. HH therefore found it more natural to assume that the boundary conditions of the universe were defined by the type 1 metric. Metrics of this type were chosen to be summed over in the path integral. This proposal by HH can be summarised as;\\

    \textbf{The No-Boundary Proposal}: The Wavefunction of The Universe is defined by a sum over compact, positive definite metrics. The boundary conditions associated with this wavefunction are that the universe does not have a past spacetime boundary. The gravitational path integral is a functional integral over compact 4 metrics bounded by a final 3-metric, The Wavefunction of The Universe is therefore a probability amplitude for the universe to appear from nothing.\\

\subsection{Wheeler De Witt Equation \& Ground State Wavefunction of The Universe}
In the neighbourhood of a spacelike hypersurface $S$, a time coordinate $\tau$ can be introduced where on the surface $S$, $\tau$ is constant. The embedding of a three dimensional spacelike surface with induced metric $h_{ij}$ in a four dimensional spacetime associated with 4-metric $g_{\mu \nu}$ is described by a (3+1) decomposition of the 4-metric with time coordinate $\tau$ given by
\begin{align}
\label{E4}
    \begin{split}
        ds^{2}&=g_{\mu \nu}dx^{\mu}dx^{\nu},\\
        &=(N^{2}-N_{i}N^{i})d\tau^{2}+2N_{i}dx^{i}d\tau +h_{ij}dx^{i}dx^{j}.
    \end{split}
\end{align}
In this (3+1) decomposition, the Einstein-Hilbert action is given by
\begin{equation}
\label{E5}
    S_{EH}= \frac{m_{p}^{2}}{16\pi} \int d^{3}xd\tau \ \sqrt{h}N\Bigl(K_{ij}K^{ij}-K^{2}-^{3}R(h_{ij})+2\Lambda- \frac{16\pi}{m_{p}^{2}}L(\phi) \Bigl),
\end{equation}
where $K_{ij}$ is the second fundamental which is a measure of the extrinsic curvature associated with spatial foliations. The second fundamental form is related to the three metric by
\begin{equation}
\label{E6}
    K_{ij}=\frac{1}{2N}\Bigl(-\frac{\partial h_{ij}}{\partial t}+2D_{(i}N_{j)}\Bigl),
\end{equation}
where the second term is the covariant derivative of the Shift function $N_{j}$ with respect to a spatial coordinate. The Lagrangian given in (\ref{E5}) can be written in terms of the product of conjugate pairs and a Hamiltonian. In order to define conjugate momentum for the 3-metric $h_{ij}$ it can be seen from the above equation for the second fundamental form that $K_{ij} \approx -(2N)^{-1}\dot{h_{ij}}$. The momentum conjugate to the three momentum can be defined as
\begin{align}
\label{E7}
    \begin{split}
        \pi_{ij}=\frac{\delta S_{EH}}{\delta \dot{h_{ij}}}&=-\frac{m_{p}^{2}}{16 \pi}\sqrt{h}\Biggl(\frac{\delta(K_{ij}K^{ij})}{\delta \dot{h_{ij}}}-\frac{\delta(K^{2})}{\delta \dot{h_{ij}}}\Biggl),\\
        &=-\frac{m_{p}^{2}}{16 \pi}\sqrt{h}(K^{ij}-h^{ij}K).
    \end{split}
\end{align}
Similarly, the conjugate momentum for the scalar field $\phi$ can be defined
\begin{align}
\label{E8}
    \begin{split}
        \pi_{\phi}= \frac{\delta S_{EH}}{\delta \dot{\phi}}&=\Biggl(g^{00}\frac{\delta (\dot{\phi}^{2})}{\delta \dot{\phi}}+g^{01}\frac{\delta (\dot{\phi} \cdot \partial_{i}\phi)}{\delta \dot{\phi}} \biggl),\\
        &=\frac{1}{N}\sqrt{h}(\dot{\phi}-N^{i}\partial_{i}\phi)
    \end{split}
\end{align}
$N$ is known as the Lapse function and is characterised by the difference between elapsed coordinate time $t$ and coordinate $\tau$ defined for curves in the neighbourhood of $S$. Intuitively, $N$ can be considered as parameterising the coordinate time $t$ elapsed between spacelike hypersurface foliations of the overall 4-manifold. A variation in the Lapse function $N$ given by $\delta N$ corresponds to pushing a spatial hypersurface forwards or backwards in time. However, given that the Wavefunction of The Universe is a functional of the spatial 3-metric with boundary conditions being no past boundary and final spatial hypersurface with a 3-metric $h_{ij}$, this wavefunction is not dependent on time as parameterised by the Lapse function on the final spatial hypersurface. From this we can state that,
\begin{equation}
\label{E9}
    \frac{\delta \psi[h_{ij}, \phi]}{\delta N}=\int \mathcal{D}g_{\mu \nu}\mathcal{D}\phi \ \Biggl(\frac{\delta S_{EH}}{\delta N}\Biggl)e^{-S_{EH}[g, \phi]}=0.
\end{equation}
Classically, the term in the brackets corresponds to the Hamiltonian Constraint $H\equiv0$ in General Relativity. The variation in (\ref{E9}) can be re-expressed as
\begin{equation}
\label{E10}
    H\psi[h_{ij}, \phi]=\frac{-m_{p}^{2}}{8\pi}\sqrt{h}\Bigl(K_{ij}K^{ij}-K^{2}-^{3}R(h_{ij})+2\Lambda- \frac{16\pi}{m_{p}^{2}}L(\phi) \Bigl)\psi[h_{ij}, \phi]=0.
\end{equation}
This equation is known as the Wheeler- DeWitt (WdW) equation and is the analogue of the time-independent Schrodinger equation in Quantum Gravity. Using the operator form of $K_{ij}$, the WdW equation can be written as
\begin{equation}
\label{E11}
\Biggl(-G_{ijkl}\frac{\delta^{2}}{\delta h_{ij}\delta h_{kl}}-\sqrt{h}\Bigl[^{3}R(h_{ij})-2\Lambda + \frac{8\pi}{m_{p}^{2}}T_{nn}\Bigl(\frac{\delta}{\delta \phi}, \phi \Bigl)\Bigl] \Biggl)\psi[h_{ij}, \phi ]=0.
\end{equation}
In the above equation, the term $G_{ijkl}$ is the metric on superpsace or the space of all 3-metrics $h_{ij}$ which is the product of two 3-metric tensor with their indices permuted
\begin{equation*}
    G_{ijkl}= \frac{\sqrt{h}}{2}(h_{ik}h_{jl}+h_{il}h_{jk}-h_{ij}h_{kl}).
\end{equation*}
The WdW equation is the same for both Euclidean and Lorentzian starting points. It is worth noting that if one takes the variations of the Wavefunction of the universe with respect to the shift vector $N_{i}$ we see that
\begin{equation}
\label{E12}
    \frac{\delta \psi[h_{ij}, \phi]}{\delta N_{i}}=\int \mathcal{D}g_{\mu \nu}\mathcal{D}\phi \ \Biggl(\frac{\delta S_{EH}}{\delta N_{i}}\Biggl)e^{-S_{EH}[g, \phi]}=0.
\end{equation}
This is obvious since there is no shift vector $N_{i}$ present in the the Einstein-Hilbert action in (\ref{E5}). As a variation in the Shift vector $\delta N_{i}$ is associated with diffeomorphisms on the spatial hyper surface, the variation of the wavefunction with respect to this vector vanishing is indicative of the wavefunction being invariant under spatial diffeomorphisms. Thus, this functional is a functional of three dimensional geometry in general rather than any specific 3-metric $h_{ij}$\cite{HH1}\cite{HH2}.\\

The signature of the superspace metric $-G_{ijkl}$ is $(+-----)$. The Wheeler-de Witt equation can therefore be thought of as hyperbolic equation with time coordinate $\sqrt{h}$. Given that the boundaries of the path are purely spatial surfaces. One can impose boundary conditions on the range of $\sqrt{h}$, in this case $(0\le \sqrt{h} < \infty)$ as this ensures that the hypersurfaces are everywhere spacelike. Initially, the lower bound seems to present some conceptual problems. For a Lorentzian manifold, $\sqrt{h}=0$ is a singularity. However, for a Euclidean manifold this is not necessarily the case. For example, consider a 4-sphere of radius R given by 
\begin{equation}
\label{E13}
    x_{1}^{2}+x_{2}^{2}+x_{3}^{2}+x_{4}^{2}+x_{5}^{2}=R^{2}
\end{equation}
embedded in flat 5-dimensional space. For a surface that obeys $|x_{5}|=R'$ where $R'<R$. Then the intersection of these two surfaces is a 3-sphere with non zero radius $\sqrt{R^{2}-R'^{2}}$ such that
\begin{equation}
\label{E14}
    x_{1}^{2}+x_{2}^{2}+x_{3}^{2}+x_{4}^{2}=R^{2}-R'^{2},
\end{equation}
However, $|x_{5}|=R$ the radius of the 3-sphere sinks to zero which does not correspond to a singularity  from the perspective of the overall 5-geometry in this example.\\

As $\psi[h_{ij}, \phi]$ is defined by a path integral over compact Euclidean metrics, this automatically vanishes for 3-metrics which are not positive definite. The 3-metric $h_{ij}$ is not a quantum observable with unrestricted range. This 3-metric can be replaced with another three metric $\bar{h_{ij}}$ related to the original metric by
\begin{equation}
\label{E15}
    \bar{h_{ij}}\equiv \frac{h_{ij}}{h^{\frac{1}{3}}}.
\end{equation}
General solutions of the initial value equation can be obtained when the 3-metric of the inital spatial hypersurface is specified up to an initially unknown conformal factor. The conformal transformation above is freely specified as long as its invariant under the conformal transformation $h_{ij} \to \omega(x)^{4}h_{ij}$ where $\omega(x)$ is initially an arbitrary, unknown conformal factor which is found as part of a complete solution to the initial boundary problem\cite{Y1}. Using this conformal definition for the 3-metric, the important boundary condition on the wavefunction can be written as
\begin{equation}
\label{E16}
    \psi[\bar{h_{ij}}, \sqrt{h}, \phi]=0 \ \text{for} \ \sqrt{h}<0.
\end{equation}
Noting that $\sqrt{h}$ only has semi infinite range, it is convenient to define
\begin{equation*}
\label{E17}
   K \frac{m_{p}^{2}}{12\pi}=\text{momentum conjugate to} \ \sqrt{h}.
\end{equation*}
The trace of the second fundamental form $K$ can be either positive or negative with arbitrary value. Therefore, the range of the conjugate momentum representation can be considered as infinite. The two conjugate representations are related to each other via a Laplace transform
\begin{align}
\label{E18}
    \begin{split}
        \Phi[\bar{h_{ij}}K, \phi]&=\int_{0}^{\infty}d\sqrt{h}\ e^{-\frac{m_{p}^{2}}{12\pi}\int d^{3}x \ \sqrt{h}K}\psi[h_{ij}, \phi]\\
        \psi[h_{ij}, \phi]&= \int_{\Gamma}d\Bigl[\frac{m_{p}^{2}}{24\pi i}K\Bigl]\ e^{\frac{m_{p}^{2}}{12\pi}\int d^{3}x\ \sqrt{h}K}\Phi[\bar{h_{ij}}K, \phi].
    \end{split}
\end{align}
For each point of $S$, the contour $\Gamma$ runs from $-i\infty \to i\infty$. Assuming that $\Phi[\bar{h_{ij}}K, \phi]$ does not diverge for Large $K$, this choice of contour ensures that the wavefunction $\psi[h_{ij}, \phi]=0$ for $\sqrt{h}<0$. Due to this, the contour $\Gamma$ can be closed off in the right half K plane (EXPLAIN THIS MORE).\\

Using the previously considered fact that terms in the $(3+1)$ Einstein-Hilbert action can be considered as functional differential operators acting on the original wavefunction, an operator form of $K^{2}$ can be determined. Starting with
\begin{equation*}
\label{E19}
    \frac{\delta}{\delta \dot{h}_{kl}}=-\frac{m_{p}^{2}}{16\pi}\sqrt{h}(K^{kl}-h^{kl}K),
\end{equation*}
each side of the equation can be multiplied by $h^{-\frac{1}{2}}$ and contracted with the 3-metric to give an expression for $K$ as
\begin{align}
\label{E20}
    h_{kl}\ h^{-\frac{1}{2}} \ \frac{\delta}{\delta \dot{h}_{kl}}&= -\frac{m_{p}^{2}}{16\pi}(K-3K),\\
    &=\frac{m_{p}^{2}}{8\pi}K.
\end{align}
From the differential operator form of $K$, it is straightforward to see that the functional differential operator identity for $K^{2}$ is given by 
\begin{equation}
    \Biggl(\frac{8\pi}{m_{p}^{2}}\Biggl)^{2} \ h_{ij}\ h^{-\frac{1}{2}} \ \frac{\delta}{\delta \dot{h}_{ij}}\Biggl[h_{kl}\ h^{-\frac{1}{2}} \ \frac{\delta}{\delta \dot{h}_{kl}} \Biggl]=K^{2}.
\end{equation}
The above differential operator can be thought of as acting on the wavefunction yielding some definite value and the original wavefunction. This definite value of $K^{2}$ remains upon subsequent division by the wavefunction. An important rule of thumb which may be derived from this is
\begin{align*}
    \begin{split}
        \frac{K^{2}\psi}{\psi}&>0: \psi \ \text{corresponds to Euclidean 4-geometry in the classical limit}.\\
        &<0: \psi \ \text{corresponds to Lorentzian 4-geometry in the classical limit}.
    \end{split}
\end{align*}
These two outcomes are determined by whether the wavefunction depends on the scale $\sqrt{h}$ in an exponential or oscillatory manner. These two scenarios will become important for interpreting the behaviour of the Wavefunction for simple systems.

\subsection{Minisuperspace}
As boundary conditions such as the no boundary proposal must be postulated independently, the wavefunction $\psi[h_{ij}, \phi]$ defined by certain boundary conditions must be calculated to see if the evolution of the wavefunction is in agreement with observations of the universe we live in. However, the WdW equation is a 2nd order functional differential equation defined on the superspace $(g, \phi)$ which is an infinite dimensional manifold. There is currently no method for generally solving such differential equations. In order to find a wavefunction which obeys the WdW equation, one must first reduce the infinite degrees of freedom associated with the WdW equation to a finite dimensional manifold. \\

Generally, the boundary conditions for the WdW equation can be obtained by  using a semi-classical approximation to the path integral where
\begin{equation}
    \psi[h_{ij}, \phi]=N_{0} \ \sum_{i}A_{i}e^{-B_{i}},
\end{equation}
where $N_{0}$ is a normalisation constant, $A_{i}$ are small fluctuations about semi-classical solutions and $B_{i}$ are the actions of the classical solutions of Euclideanised field equations are compact and have given 3-metric $h_{ij}$ and matter field configuration $\phi$ on the boundary. The universe on large scales is spatially homogeneous and isotropic.  Thsi kind of universe can be excellently approximated by the closed FRW metric
\begin{equation*}
    ds^{2}=\sigma^{2}[N(\tau)^{2}\ d\tau^{2}+a(\tau)^{2} \ d\Omega_{3}^{2}]
\end{equation*}
where $\sigma^{2}=\frac{2}{3\pi m_{p}^{2}}$ and $d\Omega_{3}^{2}$ is the round metric on a 3-sphere of unit radius. The gravitational action associated with this metric and a conformal scalar field which is the simplest form of matter that can be considered is given by (alter this later to be consistent with subsequent chapters)
\begin{equation}
    S= \frac{1}{2}\int d\tau \ \Biggl(\frac{N}{a}\Biggl)
    \Biggl[- \Biggl(\frac{a}{N} \frac{da}{d\tau}  \Biggl)^{2} -a^{2} + \Biggl(\frac{a}{N} \frac{d\chi}{d\tau}\Biggl)^{2}+\chi^{2}\Biggl] \ \text{where} \ \chi=\frac{\sqrt{2}}{\pi a \sigma \phi}
    \end{equation}
As in the more general cases shown above, conjugate momenta $\pi_{a}$ $\pi_{\chi}$ can be defined such that the WdW equation associated with the variation of this action with respect to the Lapse function given by
\begin{equation}
    \frac{1}{2}\Biggl(\frac{1}{a^{p}} \frac{\partial}{\partial a}\Biggl[a^{p} \ \frac{\partial}{\partial a}\Biggl]-a^{2}-\frac{\partial^{2}}{\partial \chi^{2}}+ \chi^{2} \Biggl)\phi[a,\chi]=0
\end{equation}
can be written in terms of the squared conjugate momenta,
\begin{align}
\begin{split}
    \pi_{a}^{2}&= \frac{1}{a^{p}} \frac{\partial}{\partial a}\Biggl[a^{p} \ \frac{\partial}{\partial a}\Biggl]\\
    \pi_{\chi}^{2}&= \frac{\partial^{2}}{\partial \chi^{2}}
\end{split}.
\end{align}
(Be consistent with notation, either $\delta$ or $\partial$ not both).
For conformally invariant scalar fields, the WdW equation is separable. The wavefunction $\psi$ can be expanded in harmonic oscillator eigenstates
\begin{equation}
    \psi(a,\chi)=\sum_{n}c_{n}(a)f_{n}(\chi).
\end{equation}
Where $f_{n}(\chi)$ obey the Harmonic oscillator equation (with $E=(n+\frac{1}{2})$
\begin{equation}
    \frac{1}{2}\Biggl(-\frac{\partial^{2}}{\partial \chi^{2}} +\chi^{2}\Biggl)=\Bigl(n+\frac{1}{2}\Bigl)f_{n}(\chi)
\end{equation}
Upon separation of variables the differential equation for the gravitational part acting on function $c_{n}(a)$ is
\begin{equation}
    \frac{1}{2}\Biggl(\frac{1}{a^{p}} \frac{\partial}{\partial a}\Biggl[a^{p} \ \frac{\partial}{\partial a}\Biggl]-a^{2}\Biggl)c_{n}(a)=\Bigl(n+\frac{1}{2}\Bigl)c_{n}(a).
\end{equation}
However, there are disadvantages associated with the use of conformally invariant scalar fields. For the wavefunction defined by the gravitational path integral the ground state of the wavefunction should be defined by geometries of maximal of maximal symmetry. From previous equations we see that the ground state corresponds to the ground state of a Harmonic oscillator which is defined as $n=0$. For the gravitational equation, the ground state corresponds to the usual WdW equation shifted by $\frac{1}{2}$. From this it can be inferred that for this type of scalar field, the gravitational wavefunction is the same in both the presence and absence of matter. Given that this is not the case for the universe we live in, another type of matter field must be chosen.\\

A homogeneous and isotropic universe with a spatially constant massive scalar field which is not conformally invariant is the next simplest model to consider. The WdW equation for this universe with scalar field $\bar{\phi}$ with mass $\bar{m}$ is
\begin{align}
\begin{split}
    \frac{1}{2}\Biggl(\frac{1}{a^{p}} \frac{\partial}{\partial a}\Biggl[a^{p} \ \frac{\partial}{\partial a}\Biggl]-a^{2}-&\frac{1}{a^{2}}\frac{\partial^{2}}{\partial \phi^{2}} +a^{4}m^{2}\phi^{2} \Biggl)\psi(a,\phi)=0,
\end{split}
\end{align}
where $\phi=\sigma \bar{\phi}$ and $m= \sigma \bar{m}$. The corresponding Euclidean action in the case of a massive scalar field rather than  a conformal scalar field is given by
\begin{equation}
    S=\frac{1}{2}\int d\tau N \Biggl[-\frac{a}{N^{2}}\Biggl(\frac{da}{d\tau}\Biggl)^{2}+\frac{a^{3}}{N^{2}}\Biggl(\frac{d\phi}{d\tau}\biggl)^{2}-a+a^{3}V(\phi) \Biggl]
\end{equation}
Assuming that, semi-classically, the path integral will be dominated by classical solutions corresponding to Euclidean 4-sphere, one can take this action and vary the action with respect to $a$ and $\phi$ to find the classical field equations
\begin{align}
\begin{split}
    \frac{\delta S}{\delta \phi}&= \frac{d^{\phi}}{d\tau^{2}}+\frac{3}{Na}\frac{da}{d\tau}\frac{d\phi}{d\tau}-\frac{1}{2}\frac{dV(\phi)}{d\phi}=0,\\
    \frac{\delta S}{\delta a}&= \frac{1}{N^{2}a}\frac{d^{2}a}{d\tau^{2}}+\frac{2}{N^{2}}\Biggl(\frac{d\phi}{d\tau}\Biggl)^{2}+V(\phi)=0.\\
\end{split}
\end{align}
Setting the variation of the action with respect to the Lapse function $N$ to zero corresponds to enforcing the Hamiltonian constraint. This variation is given by
\begin{equation}
    \frac{1}{N^{2}}\Biggl(\frac{da}{d\tau} \Biggl)^{2}-\frac{a^{2}}{N^{2}}\Biggl(\frac{d\phi}{d\tau} \Biggl)^{2}-1+a^{2}V(\phi)=0.
\end{equation}
The definition of the Euclidean path integral is the sum over compact complex 4-metrics. The corresponding boundaries are 3 dimensional spatial hypersurfaces. In general the No boundary proposal can be formulated as thew universe not having a past and only a future boundary. In minisuperspace the No-Boundary proposal is the boundary condition
\begin{equation}
    a(0)=0
\end{equation}
However, if this boundary condition by itself is proposed, then any term in the field equations containing $a$ in the denominator will blow up, specifically, the second term in $\phi$ field equation. To prevent this it is specified that
\begin{equation}
    \frac{d\phi}{d\tau}\Biggl|_{\tau=0}=0.
\end{equation}
The solution to the field equations for $a(\tau)$ which satisfies this condition is given by \cite{HAL3}
\begin{equation}
    a(\tau) \approx \frac{\sin(\sqrt{V}N\tau)}{\sin(\sqrt{V}N)}
\end{equation}
Substituting this back into the equation for the variation of $N$ corresponding to its saddle point we see that
\begin{equation}
    \frac{V\bar{a}\cos^{2}(\sqrt{V}N\tau)}{\sin^{2}(\sqrt{V}N)}-1+a^{2}V=0
\end{equation}
Taking $\tau=1$ the above expression can be simplified as
\begin{equation}
    \bar{a}^{2}V\Bigl(1+\cot(\sqrt{V}N)\Bigl)=1
\end{equation}
Which can ultimately be simplified to
\begin{equation}
    \sin^{2}(\sqrt{V}N)=\bar{a}^{2}V
\end{equation}
In \cite{HH2}, HH argue that only real values of the scale factor and potential should be considered. Therefore the $\bar{a}^{2}V<1$ case is considered in which solutions are parameterised by $n\in \mathbb{Z}$. This means there are countably infinite solutions. The saddle point. The solutions for $N$ are given by
\begin{equation}
    N_{n,\pm}=\frac{1}{\sqrt{V}}\Biggl[\Biggl(n+\frac{1}{2} \Biggl)\pi \pm \cos^{-1}(\bar{a}\sqrt{V}) \Biggl].
\end{equation}
By setting $n=0$ and demanding that $\cos^{-1}(\bar{a}\sqrt{V})$ lie in the range $(0, \frac{\pi}{2})$ the solution for $a(\tau)$ is given by\cite{HAL3}
\begin{equation}
    a(\tau)\approx \frac{1}{\sqrt{V}}\sin \Biggl[\Biggl(\frac{\pi}{2} \pm \cos^{-1}(\bar{a}\sqrt{V}) \Biggl)\Biggl].
\end{equation}
Substituting this back into the action, we see that there are two possible solutions
\begin{equation}
    S_{\pm}=-\frac{1}{3V(\phi)}\Bigl[1+\Bigl(1-\bar{a}^{2}V(\phi)  \Bigl)^{\frac{3}{2}} \bigl].
\end{equation}
The solutions $S_{+}$ corresponds to the 3-sphere being closed off by more than half of a 4-sphere while the $S_{-}$ solution corresponds to the 3-sphere being closed off by less than half of a 4-sphere. However, all of these steps are not sufficient to specify a wave function as integration along different contours will lead to different saddle points being chosen which will yield different solutions. Therefore the No Boundary proposal does not yield a unique wavefunction but rather one of many possible wavefunctions. This is not surprising as, if boundary conditions have to be postulated independently and one is working in a semi-classical regime in which classical equations of motion are the dominant contribution, different boundary terms will yield different specific solutions. One usually chose a contour for which these specific saddle points will contribute dominantly. The wavefunction of the universe in minisuperspace was determined to be\cite{HH1}\cite{HH2}
\begin{equation*}
    \psi_{HH} \propto e^{\frac{\Biggl[1-1(a^{2}V(\phi)^{\frac{3}{2}}}{3V(\phi)\Biggl]}} \quad a^{2}V<1
\end{equation*}
in the classically forbidden region. In the classically allowed region it is
\begin{equation*}
    \psi_{HH} \propto e^{\frac{1}{3V(\phi)}}\cos\Biggl(\frac{(a^{2}V-1)^{\frac{3}{2}}}{3V}-\frac{\pi}{4}\Biggl)
\end{equation*}
In Chapter 4 we will see the exact definition of the No Boundary wavefunction in temrs of Airy functions.
\subsection{Gauge Fixing, Ghosts \& The Wheeler-De Witt Equation}
It was previously shown very schematically that, starting from the Euclidean Path Integral, the No-Boundary wavefunction obeys the WdW equation, this concept was put on rigorous footing by Halliwell\cite{HAL1}. Path integral calculations are carried out using the Hamiltonian form of General Relativity in which the focus is the dynamics of 3-surfaces. There are four constraints which are due to invariance under four-dimensional diffeomorphisms. Three of the constraints are momentum constraints which generate diffeomorphisms within the three surface. however, it is the fourth constraint, the Hamiltonian constraint $H=0$ which differentiates General Relativity from other gauge theories. The Hamiltonian constraint is a consequence of the the theory being invariant under time reparameterisation. However, given that the Hamiltonian is also responsible for dynamics, time symmetry and dynamics become entangled\cite{HAL1}. \\

Given these constraints, Halliwell argues that the relevant canconical quantisation procedure that should be related to path integral quantisation is Dirac's method of canonical quantisation in which constraints become operators which annihilate physical states. This was seen previously in HH's schematic derivation of the WdW equation within a path integral context. The class of reparameterisation invariant theories which Halliwell considered are those with an action
\begin{equation}
\label{H1}
    S= \int_{t'}^{t''}dt\ p_{\alpha}\dot{q}^{\alpha}-NH(p_{\alpha},q^{\alpha})
\end{equation}
where $N$ is the Lagrange multiplier which enforces the Hamiltonian constraint when the action is varied with respect to $N$. The momentum and Hamiltonian constraints are given by

\begin{align}
\begin{split}
\label{H2}
    H^{i}&=-2\pi|^{ij}_{j}=0,\\
    H&=G_{ijkl}\pi^{ij}\pi^{kl}-\sqrt{h}(^{3}R-2\Lambda)=0.
\end{split}
\end{align}
After the three surface has been integrated over, the Hamiltonian constraint reduces to expressions of the form
\begin{equation}
\label{H3}
    H=\frac{1}{2}f^{\alpha \beta}(q)p_{\alpha}p_{\beta}+V(q)=0.
\end{equation}
The metric $f^{\alpha \beta}$ can be considered as the inverse of the metric on superspace $f_{\alpha \beta}$ and like the full superspace metric $G_{ijkl}$ is of hyperbolic signature $(-,+,+,+,+,+)$. The form of the ordering problem in this case is that the superspace metric $f$ is a function of $q$ and there is therefore no clear sense of how to distribute it between the momentum operators\cite{HAL1}. In Dirac quantisation, a wavefunction $\psi$ is considered which is a function defined in superspace. The wavefunction is annihilated by the Hamiltonian to yield the WdW equation $\hat{H}q=0$. For path integral quantisation the wavefunction can be represented as
\begin{equation}
\label{H4}
    \int \mathcal{D}p\mathcal{D}q\mathcal{D}N\ e^{\int_{t'}^{t''}dt\ p_{\alpha}\dot{q}^{\alpha}-NH(p_{\alpha},q^{\alpha})}
\end{equation}
Noting that the action is reparameterisation invariant, without gauge fixing of the measure the path integral would be divergent as there would be an infinite number of contributions corresponding to an equivalent history. Given a Hamiltonian of the form (\ref{H3}), in order to gauge fix the path integral, Halliwell takes advantage of the fact that the Hamiltonian is quadratic in momenta, $p_{\alpha}$ and $N$ are free at the endpoints while $q^{\alpha}$ satisfy the boundary conditions
\begin{equation}
\label{H5}
    q^{\alpha}(t')=q^{\alpha'}, \quad q^{\alpha}(t'')=q^{\alpha''}
\end{equation}
The variation of the action $S$ with respect to each of the conjugate variables yields
\begin{align}
\label{H6}
    \begin{split}
        \dot{q}^{\alpha}=N\frac{\partial H}{\partial p_{\alpha}}=N\{q^\alpha, H\},\\
        \dot{p}_{\alpha}=-N\frac{\partial H}{\partial q^{\alpha}}=N\{p_{\alpha}, H\}.
    \end{split}
\end{align}

Variation of the action with respect to $N$ yields the Hamiltonian constraint $H(p_{\alpha},q^{\alpha})=0$. The above relations can be considered more generally as the action being invariant under reparameterisations of the form
\begin{align}
\label{H7}
    \begin{split}
        \delta q^{\alpha}& = \epsilon(t)\{q^\alpha, H\},\\
        \delta p_{\alpha}& = \epsilon(t)\{p_{\alpha}, H\},\\
        \delta N& = \dot{\epsilon}(t).
    \end{split}
\end{align}
The variation of the action as a whole $\delta S$ is then given by
\begin{equation}
\label{H8}
    \delta S= \Biggl[\epsilon(t)(p^{\alpha}\dot{q}_{\alpha}- H(p_{\alpha},q^{\alpha}))\Biggl]^{t''}_{t'}
\end{equation}
The action is invariant under this reparameterisation only for the boundary conditions
\begin{equation}
\label{H9}
    \epsilon(t')=\epsilon(t'')=0.
\end{equation}
In order to break the reparamterisation invariance given by (\ref{H7}) gauge fixing is necessary. Teitelboim has argued that a gauge fixing condition must\cite{HAL1}\cite{T1};
\begin{enumerate}
    \item [(i)] Fix the gauge completely
    \item [(ii)] Remove all residual gauge degrees of freedom
    \item [(iii)] Using the transformations given in (\ref{H7}) it must be possible to bring any configuration specified by $p$, $q$ and $N$ into one satisfying the gauge condition
\end{enumerate}
All of the above requirements are satisfied for a gauge fixing condition of the form\cite{T1}
\begin{equation}
\label{H10}
    \dot{N}= \chi (p.q.N)
\end{equation}
where $\chi$ is an arbitrary function of these variables. This condition can be imposed on the path integral measure by utilising the functional Dirac delta relation
\begin{equation}
\label{H11}
    \int dN = \int \mathcal{D}N \delta[(N-N_{0})] \int \mathcal{D}N \mathcal{D}\Pi \ e^{i \int dt \ \Pi(\dot{N}-N_{0})}
\end{equation}
where $N_{0}$ is a constant and $\Pi$ is an auxiliary field. The original action can then be written as
\begin{equation}
\label{H12}
    S+S_{gf}=\int_{t'}^{t''}dt \ (p_{\alpha}\dot{q^{\alpha}}-NH+\Pi(\dot{N}-\chi))
\end{equation}
The auxiliary field $\Pi$ will be specified to vanish at end points so that
\begin{equation}
\label{H13}
    \Pi(t')=\Pi(t'')=0.
\end{equation}
Taking the variation of the the total action with respect to the conjugate variables yields
\begin{align}
\label{H14}
    \begin{split}
        \dot{q}^{\alpha}&=N\frac{\partial H}{\partial p_{\alpha}}+ \Pi \frac{\partial \chi}{\partial p_{\alpha}} =N\{q^\alpha, H\} + \Pi\{q^{\alpha}, \chi\},\\
        \dot{p}_{\alpha}&=-N\frac{\partial H}{\partial q^{\alpha}}- \Pi\frac{\partial \chi}{\partial q^{\alpha}}=N\{p_{\alpha}, H\}+\Pi\{p_{\alpha}, \chi\},\\
    \end{split}
\end{align}
The variation of the action with respect to $\Pi$ yields the gauge condition (\ref{H10}). Variation with respect to $N$ gives
\begin{equation}
\label{H15}
    \dot{\Pi}+H=0
\end{equation}
Taking the derivative of this equation with respect to $t$ and substituting (\ref{H14}) yields
\begin{equation}
\label{H16}
    \ddot{\Pi}+\dot{H}=\ddot{\Pi}+\{\chi, H\}\Pi=0.
\end{equation}
Since $\Pi$ is required to vanish at the end points, the solution to the above equation is $\Pi(t)=0$ so that the action equipped with the boundary conditions for the auxiliary field yield the original field equations.\\

In order to ensure that the total action (\ref{H12}) is independent of the choice of gauge fixing function $\chi$, one must move to an extended phase space including non commuting ghost fields so that the action is invariant under global BRS symmetry. This is achieved via the BFV ghost method\cite{FRAD}. The path integral measure is then over the original bosonic degrees of freedom and the ghost fields. To enforce BRS symmetry, the parameter $\epsilon(t)$ is replaced with $\Lambda c(t)$ where $\Lambda$ is a constant, anti-commuting parameter and $c(t)$ is a ghost field. In order to ensure that no time derivatives are present in the action, one can make the definition $\dot{c}=\rho$. Defining $\rho$ and $\bar{\rho}$ as being anti-commuting ghost momenta enables $\dot{c}=\rho$ to be true on shell by adding the term $\bar{\rho}(\dot{c}-\rho)$ to the action. Halliwell adds $\bar{c}\dot{\rho}$ to ensure that the variation of the action with respect to $\rho$ yields $\bar{\dot{c}}=\bar{\rho}$. The initial ghost action is given by
\begin{equation}
\label{H17}
    S_{gh}=\int_{t'}^{t''} dt \ (\bar{\rho}\dot{c}+\bar{c}\dot{\rho}-\bar{\rho}\rho)
\end{equation}
The reparameterisations given in (\ref{H7}) can be rewritten as BRS transformations given by
\begin{align}
\label{H18}
\begin{split}
        \delta p^{\alpha} = -\Lambda c\frac{\partial H}{\partial q^{\alpha}}, \quad  \delta q_{\alpha} = \Lambda c\frac{\partial H}{\partial p_{\alpha} }, \quad  \delta N = \Lambda \rho.
 \end{split}
\end{align}
The variation of the original action given the BRS transformations is given by
\begin{equation}
\label{H19}
    \delta S= \int_{t'}^{t''}dt \ \Lambda(\dot{c}-\rho)H+\Biggl[\Lambda c(t)(p^{\alpha}\dot{q}_{\alpha}- H(p_{\alpha},q^{\alpha}))\Biggl]^{t''}_{t'}
\end{equation}
Given that the transformation parameter $\epsilon(t)$ vanished at the endpoints, it can be specified that
\begin{equation}
\label{H20}
    c(t')=c(t'')=0
\end{equation}
This causes the second term to vanish while the first term does not vanish because $\dot{c}=\rho$ on shell only. In order to cancel this term and make the variation of the entire action zero, the specific transformation of the other variables must be determined. Returning to the gauge fixing action and taking the variation we see
\begin{align}
\label{H21}
    \begin{split}
        \delta S_{gf}= \int_{t'}^{t''} dt \Biggl[\delta \Pi(\dot{N}-\chi)+\Pi \delta \dot{N}-\Pi \delta \chi \Biggl].\\
    \end{split}
\end{align}
By virtue of the BRS transformation for $N$ and $\dot{c}=\rho$, for the second variation we see that
\begin{equation}
\label{H22}
    \Pi \delta \dot{N}= \Pi \Lambda \dot{\rho}
\end{equation}
The third variation is given by
\begin{align}
\label{H23}
    \begin{split}
        \Pi\delta \chi&=\Pi \Biggl(\frac{\partial \chi}{\partial q^{\alpha}}\delta q_{\alpha}+\frac{\partial \chi}{\partial p_{\alpha}}\delta p_{\alpha}+\frac{\partial \chi}{\partial N}\delta N \Biggl)\\
        &=\Pi\Biggl(-\Lambda c \frac{\partial \chi}{\partial q^{\alpha}}\frac{\partial H}{\partial p_{\alpha}}+\Lambda c \frac{\partial \chi}{\partial p_{\alpha}}\frac{\partial H}{\partial q^{\alpha}}+\frac{\partial \chi}{\partial N}\Lambda \rho \Biggl)\\
        &=\Biggl(\Lambda c \{\chi, H\}+\frac{\partial \chi}{\partial N}\Lambda \rho \biggl)
    \end{split}
\end{align}
The overall variation in the gauge fixing action is given by
\begin{equation}
\label{H24}
    \delta S_{gf}= \int_{t'}^{t''} dt \ \Biggl[\delta \Pi(\dot{N}-\chi) + \Pi \Lambda \dot{\rho}- \Lambda c \{\chi, H\}+\frac{\partial \chi}{\partial N}\Lambda \rho \Biggl]
\end{equation}
The variation in the ghost action $S_{gh}$ is more straightforward with
\begin{equation}
\label{H25}
    \delta S_{gh}=\int_{t'}^{t''}[\delta \bar{\rho}(\dot{c}-\rho)+\delta \bar{c}\dot{\rho}+\bar{\rho}(\delta \dot{c}-\delta \rho)+\bar{c}\delta \dot{\rho}]
\end{equation}
The objective is to find a set of transformations such that $\delta S+ \delta S_{gf}+\delta S_{gh}=0$. If $\chi=0$, these transformations are given by
\begin{align}
\label{H26}
    \begin{split}
        \delta \Pi=0, \quad \delta c=0, \quad \delta \rho=0\\
         \delta \bar{c}=-\Lambda \Pi, \quad \delta \bar{\rho}=-\Lambda H.
    \end{split}
\end{align}
More generally, if $\chi \neq 0$, then $S{gh}$ is slightly modified
\begin{equation}
\label{H26}
    S_{gh}=\int_{t'}^{t''}dt \ \bar{\rho}\dot{c}+\bar{c}\dot{\rho}-\bar{\rho}\rho+c\{\chi, H\} \bar{c}+\rho \frac{\partial \chi }{\partial N}\bar{c},
\end{equation}
with this action being subject to the boundary conditions
\begin{equation}
\label{H27}
    \bar{c}(t')=\bar{c}(t'')=0
\end{equation}
The total action is then invariant under BRS transformations (\ref{H7}) and (\ref{H26}). Tshi variation will yield the field equations if the boundary conditions given in (\ref{H5}), (\ref{H13}), (\ref{H20}) and (\ref{H27}) hold. As the action has been gauge fixed and is BRS invariant, the path integral may be evaluated. The gauge fixed path integral. The path integral can now be defined as
\begin{align}
\label{H28}
\begin{split}
    \int G_{\chi}(q^{\alpha''}|q^{\alpha'})&=\mathcal{D}\mu e^{iS_{T}},\\
    \mathcal{D}\mu&=\mathcal{D}p_{\alpha}\mathcal{D}q^{\alpha}\mathcal{D}\Pi\mathcal{D}N\mathcal{D}\rho\mathcal{D}\bar{c}\mathcal{D}\bar{\rho}\mathcal{D}c,\\
    S_{T}&=S+S_{gf}+S_{gh}.
\end{split}
\end{align}
This path integral is independent of the choice of gauge fixing function $\chi$. Assuming that one changes variables in the path integral from the extended phase space variables to another set of variables such that the path integral measure becomes $\mathcal{D}\Tilde{\mu}$. If these new variables are related to the old variables by the BRS transformations. Then for $\Lambda = -i\int dt \ \bar{c}(\bar{\chi}-\chi)$ the action is invariant under these transformations and the new and old path integral measures are related by
\begin{equation}
\label{H29}
    \mathcal{D}\mu=\mathcal{D}\Tilde{\mu} e^{i \int dt \ \{\bar{c}(\chi-\bar{\chi}),\Omega\}},
\end{equation}
where $\Omega=cH+\rho \Pi$ is the BRS charge. The exponential factor changes $\chi$ to $\bar{\chi}$. From this it follows that $G_{\chi}=G_{\bar{\chi}}$ and is therefore independent of the function $\chi$. The ghost and Lapse integration carried out by separating the measure into time slices for $\chi=0$. Using this method $N,p_{\alpha}, \rho$ and $\bar{\rho}$ are integrated over every slice including the end point slices while $c, \bar{c}$ and $\Pi$ are integrated over time slices and are zero for the initial and final time slices while $q^{\alpha}$ on the initial and final time slices are $q^{\alpha'}$ and $q^{\alpha''}$. The skeletonisation procedure that Halliwell uses for the Lapse function and auxiliary field is as follows. Dividing both of the integrals into $n+1$ equal time slices we have\cite{HAL1}
\begin{align}
\label{H30}
\begin{split}
    \int \mathcal{D}N\mathcal{D}\Pi \  e^{i \int dt \ \Pi N}&=\int dN_{1/2}...dN_{n+1/2}\frac{1}{(2\pi)^{n}}\int d\Pi_{1}...d\Pi_{n} \ e^{i \sum_{k=1}^{n}\Pi_{k}(N_{k+1/2}-N_{k-1/2})},\\
    &=\int dN_{1/2}...dN_{n+1/2} \prod_{k=1}^{n} \delta (N_{k+1/2}-N_{k-1/2}),\\
    &=\int dN(t''-t').
\end{split}
\end{align}
In the above calculation $\Pi$ is skeletonised as a coordinate due to its boundary conditions $(\Pi_{0}=\Pi_{n+1}=0$ while $N$ is treated as conjugate momentum. The final result is due to the fact that there are n $\delta$ functions and $n+1$ integrations, the functional integration collapses to a single ordinary integration with a single $dN$ being selected. The overall path integral is given by
\begin{equation}
\label{H31}
    G(q^{\alpha'}|q^{\alpha''})= \int dN (t''-t')\int \mathcal{D}p_{\alpha}\mathcal{D}q^{\alpha} \ e^{i \int_{t'}^{t''}dt \ (p_{\alpha}\dot{q}^{\alpha}-NH)    }.
\end{equation}
Halliwell then states that this path integral has the form of a quantum mechanical propagator with time Nt given by
\begin{equation}
\label{H32}
    \int \mathcal{D}p_{\alpha}\mathcal{D}q^{\alpha} \ e^{i \int_{t'}^{t''}dt \ (p_{\alpha}\dot{q}^{\alpha}-NH)    }= \braket{q^{\alpha''}, Nt''|q^{\alpha '},Nt'}.
\end{equation}
Redefining the Lapse function such that $T=N(t''-t')$, the path integral can be thought of as the propagator
\begin{equation}
\label{H33}
    G(q^{\alpha'}|q^{\alpha''})=\int dT \ \braket{q^{\alpha''}, T|q^{\alpha '},0}.
\end{equation}
If the range of the propagator is taken as half infinite then it obeys a time independent Schrodinger equation. The action of the Hamiltonian $\hat{H}''$ defined for time $t''$ on the propagator is
\begin{align}
\label{H34}
    \begin{split}
    \hat{H''}G(q^{\alpha'}|q^{\alpha''})= &\int_{0}^{\infty}dT \ i\frac{\partial}{\partial T}\braket{q^{\alpha''}, T|q^{\alpha '},0},\\
    &i\Bigl|\braket{q^{\alpha''}, T|q^{\alpha '},0} \Bigl|^{\infty}_{0},\\
    &-i\braket{q^{\alpha''}, T|q^{\alpha '},0},\\
    &-i\delta(q^{\alpha''}-q^{\alpha'}).
    \end{split}
\end{align}
Form this we can see that for the half infinite range $0<T<\infty$,  $G(q^{\alpha'}|q^{\alpha''})$ is a Green's function of the WdW equation whereas for the integration range $-\infty < T< \infty$, the path integral yields a full solution of the WdW equation.

\subsection{The Steepest Descent Method}
To overcome various difficulties associated with the Euclideanisation of the Einstein-Hilbert action such as the conformal factor problem and ambiguities in proposed resolutions, Hartle suggested that the path integral of the Euclideanised Einstein-Hilbert action over 4-metrics should be calculated by taking the steepest descent contours in the space of complex 4-metrics. This approach is independent of whether the starting path integral is Euclidean or Lorentzian, as we will see in \textit{Lorentzian Path Integrals} section. In the case of closed FRW minisuperspace, Halliwell showed that taking different steepest descent contours in the complex $T$ plane will yield different results of the path integral\cite{HAL2}. \\

The Hamiltonian form of the Euclidean action is of the form given in (\ref{H1}) with the primary difference being the wick rotation of the time variable defined as $t\to i\tau$ where $\tau$ is the Euclidean time variable. There is no need for the label $\alpha$ as only one pair of conjugate variables is considered. The Hamiltonian of the closed FRW metric action in terms of these conjugate variables is
\begin{equation}
\label{HB1}
    H=\frac{1}{2}(-4p^{2}-\Lambda q +1).
\end{equation}
Taking $N=1$, the variation of the action with respect to the conjugate variables yields the field equations
\begin{equation}
\label{HB2}
    \dot{p}=\frac{\Lambda}{2}, \quad \dot{q}=4p.
\end{equation}
The variation of the action with respect to $N$ yields the Hamiltonian constraint $H=0$. For these calculations, Halliwell works proper time gauge $\dot{N}=0$ which corresponds to setting the gauge fixing function $\chi(q,p,N)=0$\cite{HAL1}\cite{T1}. As discussed previously, after ensuring the path integral is BRS invariant and evaluating the ghost, lapse function and auxiliary field path integrals by time slicing, the propagator between a fixed initial and final q for a Euclideanised action is given by
\begin{equation}
\label{HB3}
    G(q''|q')= \int dT \int \mathcal{D}p\mathcal{D}q \ e^{-S(p(\tau),q(\tau))}.
\end{equation}
The path integral between a fixed initial $p$ and final $q$ is given by
\begin{equation}
\label{HB4}
    G(q''|p')=\int dT \int \mathcal{D}p\mathcal{D}q \  e^{-\Tilde{S}(p(\tau),q(\tau))}.
\end{equation}
The action $\Tilde{S}$ in the second case differs from the action $S$ in the first case by virtue of a boundary term. Noting that an action $S$ of the form (\ref{H1}) can be integrated by parts to give
\begin{equation*}
\label{HB5}
    S= -\int d\tau \ (\dot{p}q-NH)+pq \Bigl|^{\tau''}_{\tau'}
\end{equation*}
If one of the conjugate variables at Euclidean inital and final time is chosen to be fixed, then by virtue of the uncertainty principle, the other must be left completely undetermined. The action for the fixed inital $p$ and fixed final $q$ scenario is then given by
\begin{equation}
    \Tilde{S}=\int_{0}^{T}d\tau (p\dot{q}-H)+p'q'
\end{equation}
where $p'$ and $q'$ are the initial values of the conjugate variables. This product can be defined as being part of the action for convenience due to the ambiguity associated with integration by parts. As previously discussed, When skeletonised Lapse integration is carried out, a convenient redefinition that can be used for the Lapse integration measure is $T=N(t''-t')$. In the case of a Euclidean time variable this becomes $T=N(\tau''-\tau')$. The contour of this integral runs through an infinite-dimensional space $(T, p(\tau),q(\tau)))$ with boundary conditions
\begin{align}
\begin{split}
    p(0)&=p', \quad p(T)=\text{{free}}\\
    q(T)&=q'', \quad q(0)=\text{{free}}
\end{split}
\end{align}
The variation of the action with respect to the conjugate variables yields the field equations given in (\ref{HB2}). The solutions to the equations of motion subject to the above boundary conditions are
\begin{align}
\begin{split}
    \bar{q}(\tau)&=-\Lambda(\tau^{2}-T^{2})-4p'(\tau-T)+q''\\
    \bar{p}(\tau)&=\frac{\Lambda \tau}{2}+p'
\end{split}
\end{align}
It must be noted that these solutions are obtained by varying the action with respect to $p$ and $q$ and enforcing the relevant boundary conditions on the saddle points, therefore they do not obey the Hamiltonian constraint which results form the variation of the action with respect to $N$. As the saddle point approximation will yield $S_{0}$ and a second order Gaussian action $S_{2}$, the full solution which does obey the Hamiltonian constraint can be written as
\begin{align}
\begin{split}
\label{E0}
    q(\tau)&=\bar{q}(\tau)+Q(\tau),\\
    p(\tau)&=\bar{p}(\tau)+P(\tau).
\end{split}
\end{align}
The action can then be written in terms of the saddle point action and the second order gaussian action as
\begin{align}
\begin{split}
    \Tilde{S}&=\Tilde{S_{0}}\braket{q'', T|q',0}+\Tilde{S}_{2}(Q(\tau)+P(\tau)),\\
    \Tilde{S_{0}}&= \frac{\Lambda^{2}T^{3}}{6}+\Lambda p'T^{2}+\Biggl[\frac{\Lambda q''-1}{2}+2p'^{2} \Biggl]T+q''p',\\
    \Tilde{S_{2}}&=\int_{0}^{T} d\tau (P\dot{Q}+2P^{2}).
\end{split}
\end{align}
The action $S_{0}$ is the result of $\bar{q}$ and $\bar{p}$ back into the original action. The overall path integral may now be written as
\begin{equation}
    G(q''|p')=\int dT \ e^{-\Tilde{S_{0}}}\int \mathcal{D}P\ \mathcal{D}Q \ e^{-\Tilde{S_{2}}}
\end{equation}
The second functional integral may be evaluated using a time slice procedure similar to the procedure for the Lapse integral shown previously. With the condition that $P$ is zero on the first slice and integrated over on the last slice while $Q$ is integrated over on the last slice and zero on the first slice results in an equal number of integrations which can be done exactly for a given contour\cite{HAL2}. For the first integral which is now just an ordinary integration in $T$, the method of steepest-descent can be used. \\

Setting the variation of $\Tilde{S_{0}}$ with respect to $T$ equal to zero yields four saddle points for $T$ corresponding to two scenarios
\begin{align}
    \begin{split}
        \Lambda q''< 1&:\quad T_{\pm}=\frac{1}{\Lambda}\Biggl(-2p'\pm (1-\Lambda q'')^{\frac{1}{2}}  \Biggl),\\
        \Lambda q''>1&:\quad T_{\pm}=\frac{1}{\Lambda}\Biggl(-2p' \pm i(\Lambda q''-1)^{\frac{1}{2}} \Biggl).
    \end{split}
\end{align}
These saddle points are both solutions of the field equations and the Hamiltonian constraint since they are derived by taking the variation of the action with respect to $T$ which is proportional to $N$ and setting the variation to be zero. The corresponding actions for these saddle points are
\begin{align}
    \begin{split}
     \Lambda q''< 1&:\quad \Tilde{S_{0}}=\mp \frac{1}{3\Lambda}(1-\Lambda q'')^{\frac{3}{2}}+\frac{1}{\Lambda}\Biggl(p'-\frac{4p'^{3}}{3} \Biggl),\\
     \Lambda q''>1&:\quad \Tilde{S_{0}}= \pm \frac{i}{3\Lambda}(\Lambda q''-1)^{\frac{3}{2}}+\frac{1}{\Lambda}\Biggl(p'-\frac{4p'^{3}}{3} \Biggl).
    \end{split}
\end{align}
Full solutions or Green's functions of the WdW equation can usually be represented as Airy functions. In this case the suitable redefinition is $T=\Tilde{T}-2p'/\Lambda$. The propagator then becomes
\begin{align}
\begin{split}
\label{HAL1}
    G(q''|p')&=e^{\Bigl(p'-\frac{4p'^{3}}{3} \Bigl)} \times \int d\Tilde{T} \ e^{\Bigl(\frac{-\Lambda^{2}\Tilde{T}^{3}}{6}+\frac{(1-\Lambda q'')\Tilde{T}}{2}  \Bigl)}\\
    &= e^{\Bigl(p'-\frac{4p'^{3}}{3} \Bigl)}\times \text{{Ai}}\Biggl[\frac{1-\Lambda q''}{(2\Lambda)^{2/3}} \Biggl]
\end{split}
\end{align}
In order to arrive at this result, the contour chosen for $\Lambda q''<1$ is the contour $ABD$ which corresponds to the contour $EFG$ for $\Lambda q''>1$. This contour may be distorted into the contour $Re(\Tilde{T})=0$ corresponding to the imaginary $\Tilde{T}$ axis. Given this, the second line of the above equation follows directly from the first. The second contour that one may take is the contour $ABC$ for $\Lambda q''<1$ which is $EF$ for $\Lambda q''>1$. The result of the path integral taken along these contours in the complex $\Tilde{T}$ plane is 
\begin{equation}
\label{HAL2}
    \Tilde{G}(q''|p')= e^{\Bigl(p'-\frac{4p'^{3}}{3} \Bigl)} \times \Biggl[ \text{{Ai}}\Biggl[\frac{1-\Lambda q''}{(2\Lambda)^{2/3}}\Biggl]+i\text{{Bi}}\Biggl[\frac{1-\Lambda q''}{(2\Lambda)^{2/3}}\Biggl]\Biggl]
\end{equation}
This contour can ultimately be deformed into $(+i \infty,0) \ \bigcup \ (0, \infty)$. Other contours taken will be linear combinations of these contours. The remaining case that hasn't been specified is the propagator between fixed initial $q'$ and final $q''$. In \cite{HAL2} Halliwell also derived this propagator subject to these boundary conditions by the same process. Taking the variation of the action with respect to conjugate variables and solving the field equations subject tot these boundary conditions yields particular solutions $\bar{q}(\tau)$ and $\bar{p}(\tau)$ which are not solutions to the Hamiltonian constraint.  Using the saddle point approximation, and substituting these solutions into the action yields
\begin{align}
\begin{split}
    G(q''|q')&= \int \frac{dT}{\sqrt{T}} \ e^{-S_{0}\braket{a'', T|q',0}},\\
    S_{0}&=\frac{\Lambda^{2}T^{3}}{24}+ \Biggl[\frac{\Lambda(q''+q')}{4}-\frac{1}{2} \Biggl]T-\frac{(q''-q')^{2}}{8T}.
\end{split}
\end{align}
In using the saddle point approximation for this action it can be found that there are four saddle points whose position will depend on whether $\Lambda q''$ and $\Lambda q'$ are greater or less than one. Halliwell was able to show that by taking specific contours  in the same manner as shown above, the propagator between fixed final $q''$ and initial $q'$ can be written in terms of the product of Airy functions given by
\begin{equation}
\label{HAL3}
    G(q''|q')=\frac{1}{\Lambda^{1/3}}\Biggl(\text{{Ai}}\Biggl[\frac{1-\Lambda q''}{(2\Lambda)^{2/3}} \Biggl] \times \text{{Ai}}\Biggl[\frac{1-\Lambda q'}{(2\Lambda)^{2/3}} \Biggl] \Biggl)
\end{equation}
\begin{figure}[h!]
    \centering
    \begin{minipage}{0.48\textwidth}
    \includegraphics[width=1.00\columnwidth]{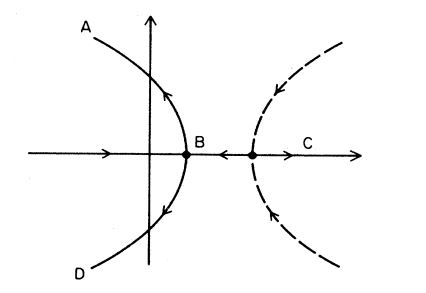}
    \caption{The steepest descent paths in the complex $T$ plane for $\Lambda q''<1$. The paths of steepest descent are represented by arrows. The propagator \ref{HAL1} is the result of taking the $ABD$ contour. The propagator \ref{HAL2} corresponds to $ABC$. Adapted from \cite{HAL2}}
    \label{fig:my_label}
    \end{minipage}\hfill
    \begin{minipage}{0.48\textwidth}
    \centering
    \includegraphics[width=1.00\columnwidth]{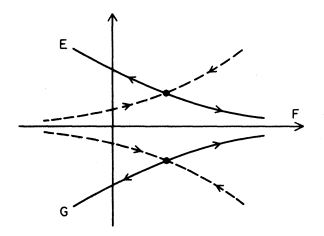}
    \caption{The steepest descent paths in the complex $T$ plane for $\Lambda q''>1$. The paths of steepest descent are represented by arrows. The propagator \ref{HAL1} is the result of taking the $EFG$ contour. The propagator \ref{HAL2} corresponds to $EF$. Adapted from \cite{HAL2}}
    \label{fig:my_label}
    \end{minipage}
\end{figure}\\
\newpage
\section{Canonical Quantisation}
\subsection{Boundary Conditions \& The Tunneling Proposal}
If the wavefunction of the universe is dependent on the boundary conditions imposed on the classical equations of motion, then a worthwhile investigation is whether or not different boundary conditions can yield universes that, at the classical level, may an approximation to the universe we live in. The solution to the Friedmann equations for $k=1$ is given by
\begin{equation}
    a=H^{-1}\cosh(Ht) \ \text{{where}} \ H= \sqrt{\frac{8\pi G \rho_{v}}{3}}.
\end{equation}
This solution represents a contracting universe for $t < 0$, an expanding universe for $t>0$ and the universe at minimum size with $a_{min}=H^{-1}$. Vilenkin considered this behaviour to be analogous to a particle bouncing off a potential barrier at $a_{min}$. However, given that quantum mechanically, particles may tunnel through a potential barrier, the initial state of the universe may be the result of a quantum tunneling event. The result of the tunneling event is the emergence of an initial universe that is not a singularity but instead with finite size Vilenkin's proposal for the Wavefunction of The Universe and its inital conditions in minisuperspace can be formulated as \cite{V3}\cite{V4}\cite{V5}
\\

    \textbf{The Tunneling Proposal \& Boundary Conditions}: The universe emerged from a bounce point described by solutions to the Euclideanised Friedmann equations. The properties of this universe are
    \begin{enumerate}
    \item The emergence of this universe is due to quantum mechanical tunneling from $a=0$ to $a_{min}=H^{-1}$.
    \item The universe had zero velocity ($\dot{a}=0$) upon emergence.
    \item The evolution of this universe is only defined by $t>0$.
    \end{enumerate}
The tunneling proposal can therefore be understood to define a de Sitter expanding universe which emerged from nothing in the sense of no classical spacetime. Vilenkin's first approach to finding the tunneling amplitude was to solve the WdW equation directly. Like the Hartle \& Hawking approach to finding the wavefunction of the universe, the infinite degrees of freedom associated with the entirety of superspace are truncated by the assumption of a closed universe that is homogeneous, isotropic with a scalar field $\phi$ which is restricted to a constant value at one of the extreme of the potential. The WdW equation takes the form (assuming youre at one of the extreme of the potential)
\begin{align}
\begin{split}
    &\frac{1}{2}\Biggl(\frac{1}{a^{p}} \frac{\partial}{\partial a}\Biggl[a^{p} \ \frac{\partial}{\partial a}\Biggl]-U(a)\Biggl)\psi(a)\\
    &U(a)=\Biggl[ \frac{3\pi}{2G}\Biggl]^{2}a^{2}(1-H^{2}a^{2})
\end{split}
\end{align}
The variation of the factor ordering parameter $p$ is assumed to only affect $\psi(a)$ for $a<\sqrt{G}$ which is a physical scenario that is not considered in this case. Choosing $p=0$, the WdW equation becomes a 1D Schrodinger equation for a particle described by coordinate $a$. This particle has zero energy and is moving in the potential $U(a)$. The region $U(a)<0$ corresponds to the Euclidean or Quantum region of the potential for which the wavefunction is exponential whilst the region $U(a)>0$ corresponds to the Lorentzian region for which the wavefunction is classical.
\begin{align}
\begin{split}
    \psi_{\pm}^{(1)}(a)&=e^{\pm i\int_{a_{min}}^{a}\sqrt{-2U(a')}da'\mp \frac{i\pi}{4}} \quad \text{(Classically Allowed Regions)}\\
    \psi_{\pm}^{(2)}(a)&=e^{\pm\int_{0}^{H^{-1}}|\sqrt{-2U(a')}|da'} \quad \text{(Under barrier/Quantum Region)}
\end{split}
\end{align}
From the solutions of the WdW, it can be seen that choosing either of the solution corresponds to choosing an outgoing wave for $a>a_{min}$. Choosing the positive solution corresponds to choosing he outgoing wave to describe a contracting universe while choosing the negative wavefunction corresponds to the universe expanding\cite{V5}. The underbarrier wavefunction can be determined using the WKB connection formula. The specific choice of solutions corresponding to the Tunneling amplitude is given by
\begin{align}
\begin{split}
\psi_{T}(a>H^{-1})&=\psi_{-}^{(1)}(a) \quad \text{(Classical Region)}\\
\psi_{T}(a<H^{-1})&=\psi_{+}^{(2)}(a)+\frac{i}{2}\psi_{-}^{(2)}(a) \quad \text{(Under barrier Region)}
\end{split}
\end{align}
For the Tunneling amplitude we can see that $a=a_{min}$  the underbarrier part of the wavefunction can be approximated by  $\psi_{T}(a<H^{-1})\approx \psi_{+}^{(2)}(a)$ which is in line with the inital intuition about tunneling. Despite the fact that the evolution of the universe is only defined for $t>0$, we see that this approximation is in with the initial discussion of bounces. The emergence of the universe around $a=a_{min}$ can be considered as approximately the end of a contracting phase of a bounce solution. As mentioned in the chapter on the Euclidean Path Integrals, probabilities associated with a given amplitude can be understood to be proportional to the exponential of the euclideanised Einstein-Hilbert action. Initially Vilenkin's result for the probability of tunneling was
\begin{equation*}
    P \propto e^{\Bigl(\frac{3}{8}G^{2}\rho_{v}\Bigl)}
\end{equation*}
However this is contrary to the general rule that the argument of the exponential should be minus the euclideanised Einstein-Hilbert action. As the Euclidenaised action is positive definite, then there cannot be any sign change and thus the probability for the de Sitter instanton would be large and positive. If the exponential is negative and the argument is large then this is in agreement with the intuitive understanding that the probability for tunneling should be extremely small. Vilenkin's argument for the correction of the sign\cite{V4} is that the underbarrier wavefunction contains growing and decreasing exponentials of equal magnitude. In this regime, gravitational contributions dominate $\psi(g,\phi)$. Taking the absolute value of the gravitational part leads to the correct result. This can also be seen by taking the tunneling amplitude ratio\cite{V5}
\begin{align}
\begin{split}
    \frac{\psi_{T}(H^{-1})}{\psi_{T}(0)}&=e^{-\int_{0}^{H^{-1}}|\sqrt{-2U(a')}|da'}\\
    &=e^{\Bigl(-\frac{3}{16G^{2}p_{v}}\Bigl)}
\end{split}
\end{align}
The Tunneling amplitude is, of course, is not the unique linear combination of under barrier and classical wavefunctions which is a solution to the WdW equation. The solution can be specified to be real so that for both the classically allowed and underbarrier region none of the wavefunctions or linear combinations have complex coefficients. This real wavefunction is none other than the Hartle-Hawking Wavefunction of The Universe discussed in the previous chapter. The Hartle-hawking wavefunction is given by
\begin{align}
\begin{split}
\psi_{HH}(a>H^{-1})&=\psi_{-}^{(2)}(a) \quad \text{(Classical Region)}\\
\psi_{HH}(a<H^{-1})&=\psi_{+}^{(1)}(a)+\psi_{-}^{(1)}(a) \quad \text{(Under barrier Region)}
\end{split}
\end{align}
Physically, this solution can be understood to represent a particle bouncing off the potential barrier at $a_{min}=H^{-1}$. Therefore the Hartle-Hawking wavefunction can be understood to be a contracting and re-expanding universe. As mentioned previously the original interpretation of this is a superposition of two universes, one contracting and one expanding. As the wavefunction of the universe describes describes the behaviour of the universe and any observers within that universe, these universes evolved independently of each other\cite{HH1}. As the Hartle-Hawking wavefunction is real, it is also time symmetric which is not the case for the Tunneling wavefunction.\\

These considerations can be extended to the case in which there is a scalar field minimally coupled to gravity. In this case the WdW equation takes a new form with the primary difference being that the potential becomes
\begin{equation}
    U(a, \phi)= \Biggl[\frac{3 \pi}{2G} \Biggl]^{2}a^{2}\Biggl[k-\frac{8\pi}{3}Ga^{2}V(\phi) \Biggl]
\end{equation}
The same rules concerning the Euclidean and Lorentzian regions apply provided that the kinetic energy of the field is small. The width of the potential barrier $U(a, \phi)$ has a minimum at the maximum of the scalar field potential. As a consequence of this, tunneling is most likely to occur at the maximum value of $V(\phi)$. \\

For the purposes of solving the WdW equation, minisuperspace can be considered as a 2D manifold defined by the ranges $0<a<\infty$ and $-\infty<\phi<\infty $ with a non singular boundary at $a=0, |\phi|<\infty$. For ease of analysis the change of variables is considered $\alpha= ln(a)$ is considered. Employing the WKB approximation with a matter field corresponds to solving the Hamilton-Jacobi equation
\begin{equation}
    \Biggl[\frac{\partial S_{n}}{\partial \alpha}\Bigg]^{2}+\Biggl[\frac{\partial S_{n}}{\partial \phi}\Biggl]^{2}+U(a, \phi)=0.
\end{equation}
For small values of $a$ corresponding to the limit $\alpha \rightarrow - \infty$, the potential $U(a, \phi) \rightarrow 0$. The solutions of the Hamilton-Jacobi equations are given by
\begin{equation}
    \psi_{k}=e^{ik(\alpha \mp \phi)}
\end{equation}
The parameter $k$ signifies the state of the universe relative to a singularity with $k>0$ corresponding to a universe collapsing to a singularity and $k<0$ corresponding to a universe expanding out of a singularity. In this case the tunneling boundary condition corresponds to enforcing only $k<0$ modes. There is a non singular boundary in the chronological past as $k\rightarrow 0$ for which the overall wavefunction is a constant. This corresponds to the early stages of the evolution of the universe. As the Tunneling amplitude of the universe was first considered within the context of inflation, the scalar field potential $V(\phi)$ is assumed to be a slowly varying function of $\phi$ that obeys
\begin{equation}
    \Bigg|\frac{1}{V}\frac{dV}{d\phi}<<1 \Bigg|
\end{equation}
Assuming that $\bar{V}(\phi)$(whats the relation to original v) is small compared to the Planck scale so that $|V|<<1$ (they seem to be related by EH constant prefactors). Using the slow rolling assumption and the fact that $\psi(a=0, \phi)=const$, the second derivatives wavefunction with respect to the scalar field vanish. For a specific choice of $p$ and using the substitution $z=U(a, \phi)$ the WdW equation becomes (show proof in your own words) an Airy equation
\begin{equation}
   \Biggl( \frac{d^{2}}{dz^{2}}+z \Biggl) \psi=0
\end{equation}
The airy equation has solutions
\begin{align}
    \begin{split}
        {\rm Ai}(z) \equiv \frac{1}{2\sqrt{\pi}}z^{-\frac{1}{4}} \ e^{-\xi} \quad & { \rm Ai}(-z)\equiv\frac{1}{\sqrt{\pi}}z^{-\frac{1}{4}} \ \sin \Bigl(\xi + \frac{\pi}{4}\Bigl)\\
        {\rm Bi}(z) \equiv \frac{1}{\sqrt{\pi}}z^{-\frac{1}{4}} \ e^{\xi} \quad &  { \rm Bi}(-z)\equiv\frac{1}{\sqrt{\pi}}z^{-\frac{1}{4}} \ \cos \Bigl(\xi + \frac{\pi}{4}\Bigl)
    \end{split}
\end{align}
where $\xi \equiv \frac{2}{3}z^{3/2}$. These solutions are defined asymptotically in the limit as $z \to \infty$. The Tunneling amplitude associated with tunneling boundary conditions is defined in terms of linear combinations of exponential wavefunctions whose argument contains the integral
\begin{align}
\begin{split}
    \int da'\sqrt{-2U(a')}&= \sqrt{2}\frac{3 \pi}{2G}\int da' \ a' \sqrt{H^{2}a'^{2}-1}\\
    &=\sqrt{2}\frac{3 \pi}{2G} \frac{1}{3H^{2}}(H^{2}a'^{2}-1)^{\frac{3}{2}}
\end{split}
\end{align}
With some algebra we can see that this is the $\xi$ parameter in the exponential of the Airy functions. This is not surprising as the minisuperspace Tunneling and Hartle-Hawking wavefunctions are both semi-classical solutions to the WdW equation. For second order, non linear differential equations like the WdW equation that are not tractable analytically, enforcing the WKB approximation by conditions such as setting the first derivative to zero transforms the original differential equation into the Airy equation. The discussion or finding the Tunneling amplitude can then be reformulated as a question of finding the correct Airy function or linear combination of Airy functions that display the expected behavior previously discussed.\\

For the Tunneling amplitude, only an outgoing wave is present in the classically allowed region. This corresponds to the condition that (show this heuristically)
\begin{equation*}
    i \frac{1}{\psi} \frac{\partial \psi}{\partial a} >0
\end{equation*}
The tunneling amplitude which displays this behaviour is defined in terms of a linear combination of Airy functions defined in terms of $z$ that is normalised by the same combination of Airy functions at $z_{0}=z(a=0)=-(2V)^{-\frac{2}{3}}$
\begin{equation}
    \psi_{T}(a>H^{-1})=\frac{{\rm Ai}(-z)+i{\rm Bi}(-z)}{{\rm Ai}(-z_{0})+i{\rm Bi}(-z_{0})}.
\end{equation}
This tunneling amplitude is defined for scalar field potential $V(\phi)=0$. For negative values of $V(\phi)$, $-z$ and $-z_{0}$. Negative values of these parameters can be thought of as the analytic continuation of the absolute values of these parameters where $V(\phi)=e^{-i\pi}|V(\phi)|$, $-z=e^{\frac{2\pi i}{3}}|z|$ and $-z_{0}=e^{\frac{2\pi i}{3}}|z_{0}|$. Using  this analytic continuation and the identity
\begin{equation}
    {\rm Ai}(e^{\frac{2\pi i}{3}}z)+i{\rm Bi}(e^{\frac{2\pi i}{3}}z)=2e^{\frac{pi i}{3}}{\rm Ai}(z),
\end{equation}
we see that for $V(\phi)<0$, the tunneling amplitude is 
\begin{equation}
    \psi_{T}=\frac{{\rm Ai}(|z|)}{{\rm Ai}(|z_{0}|)},
\end{equation}
which is a real wavefunction. By making the relevant substitutions( which are?) so that the wavefunction can be written in terms of the relevant cosmological parameters (pretty sure $V$ is either $\Lambda$ or $H$...or both if theyre related?). The classically allowed region corresponds to $z$ being large and positive while $z_{0}$ large and negative.  The Tunneling and Hartle-hawking wavefunctions in the classically allowed regions are given by 
\begin{align}
\begin{split}
    \psi_{T}(a^{2}V>1)&=e^{\frac{i\pi}{4}}(a^{2}V-1)^{-\frac{1}{4}} \ {\rm exp}\Biggl(  -\frac{1+i(a^{2}V-1)^{\frac{3}{2}}}{3V}\Biggl)\\
    \psi_{HH}(a^{2}V>1)&=2(a^{2}V-1)^{-\frac{1}{4}} e^{\frac{1}{3V}}\cos\Biggl(\frac{(a^{2}V-1)^{\frac{3}{2}}}{3V}-\frac{\pi}{4}\Biggl).
\end{split}
\end{align}
In the underbarrier regions the wavefunctions are given by 
\begin{align}
    \begin{split}
         \psi_{T}(a^{2}V<1)&=(1-a^{2}V)^{-\frac{1}{4}}\ {\rm exp}\Biggl(  \frac{(1-a^{2}V)^{\frac{3}{2}}-1}{3V}\Biggl)\\
         \psi_{HH}(a^{2}V<1)&=(1-a^{2}V)^{-\frac{1}{4}} \ {\rm exp}\Biggl(  \frac{1-(1-a^{2}V)^{\frac{3}{2}}}{3V}\Biggl)
    \end{split}
\end{align}
Having derived these wavefunctions, the next pressing question is that of interpretation. Vilenkin postulated that very soon after the nucleation of the universe, the evolution of the universe becomes classical. The wavefunctions in the classically allowed regions are interpreted as representing an ensemble of classically allowed universes. One important aspect of interpretation is that of probabilities. For any given solution $\psi$ of the WdW equation, a conserved current can be defined
\begin{equation}
    j^{l}=\frac{i}{2}(\psi^{*}\nabla^{l}\psi-\psi\nabla^{l}\psi^{*})
\end{equation}
where $\nabla^{i}$ is a covariant derivative on superspace such that $\nabla^{i}\nabla_{i}=\nabla^{2}$\cite{V5} and the indices can be raised and lowered with an appropriate metric on superspace. For minisuperspace the currents for the scale factor and scalar field can be considered. These currents are 
\begin{align}
\begin{split}
    j^{a}&=\frac{i}{2}a^{p}(\psi^{*}\partial_{a}\psi-\psi \partial_{a}\psi^{*})\\
    j^{\psi}&=-\frac{i}{2}a^{p-2}(\psi^{*}\partial_{\phi}\psi-\psi \partial_{\phi} \psi^{*}).
\end{split}
\end{align}
In the case of minisuperspace usually $p=-1$Associated with these currents is the continuity equation
\begin{equation}
    \partial_{a}J^{a}+\partial_{\phi}J^{\phi}=0
\end{equation}
The current component $j^{a}$ can be thought of as the probability density for $\phi$ at a given value of $a$. From the continuity equation coupled with this interpretation, it can immediately be inferred that the conservation of probabilities in minisuperspace is given by
\begin{equation}
    \partial_{a}\int d\phi \ j^{a} = 0.
\end{equation}
For minisuperspace defined by the solution for the Friedmann equation $a=H^{-1}\cosh(Ht)$ (where $V^{2}=H$) and constant $\phi$, probabilities are never negative so $a$ can be considered as a good time variable. The difference in probability densities $\rho(a, \phi)$ and $\rho(a, \phi)d\phi$ can be considered as the probability for a scalar field to be between $\phi$ and $\phi+d\phi$ which is independent of $a$ and with $\phi$ approximately constant on classical trajectories which is in line with the slow roll variation of $V(\phi)$ . This is the same probability density associated with the tunneling probability discussed earlier. In light of the discussion of currents one crucial addition is that the the tunneling probability is only defined for $V(\phi)>0$. For $V(\phi)<0$, $\psi_{T}$ is real and the associated probability density is then zero since the currents would be complex. Currents and probabilities are easier to discuss if the wavefunction is complex which is not the case for the Hartle-Hawking wavefunction.\\

\newpage
\section{Lorentzian Path Integrals}
In the section on Euclidean Path Integrals, we see how the Euclideanisation of the path integrals is justified by the convergence of the initially rapidly varying phase $e^{iS}$. For a path itne The Euclideanisation of gravitational path integrals is in essence, trying to imitate the success of Wick rotation in QFTs. Despite this, the Euclidean Einstein-Hilbert action is also unbounded from below. The conformal factor problem can be summarised as gradients in the scale factor of the metric contributing negatively to the Euclidean action and also rendering it unbounded from below. On the other hand, starting off with the homogeneous WdW equation and trying to find the Wavefunction of The Universe also presents presents its own challenges. There are, in theory, an infinite number of choices of boundary conditions on superspace that could be chosen to find some wavefunction of the universe. In addition to this there seems to be additional choice as to whether the solutions are real or complex. The HH wavefunction corresponding to the No-Boundary proposal is a real wavefunction which is considered as the superposition of an expanding and contracting universe. However, it is then difficult to discuss probabilities since the components of the probability currents are then complex. On the other hand, the Vilenkin wavefunction compounding to the Tunneling approach is a complex wavefunction describing an expanding universe so one can discuss probability currents.\\
 
Feldbrugge, Lehners \& Turok (FLT) instead chose to deal with the Lorentzian path integral rather than enact any sort of Euclideanisation technique. In doing so all calculations will be dealing with real, Lorentzian metrics rather than complex metrics that result from Wick rotation. FLT argue that a Lorentzian path integral naturally incorporates notions of causality and unitarity. The ambiguities in the selection of boundary terms associated with starting off with the WdW equation defined superspace is also claimed to be dealt with by Lorentzian path integrals as the boundary conditions are specified by initial and final 3-geometry.\\

\subsection{Picard-Lefshetz Theory \& The Steepest Descent Method}
The $e^{\frac{iS}{\hbar}}$ term in the path integral is not absolutely convergent but rather, conditionally convergent. Picard-Lefshetz theory can be used to turn conditionally convergent integrals into absolutely convergent integrals\cite{FLT}. This is achieved by deforming the contour of integration from the real axis to the the complex plane in such a way that the integral in such a way that the integral is made absolutely convergent. The limits of integration, which in the case of a gravitational path integral, correspond to the 3-boundary geometries remain fixed while the integral over the 4-geometries are deformed. FLT argue that for a theory in which spacetime is dynamical a malleable contour over physical quantities is more natural than Wick rotation.\\

Lorentzian path integrals are defined as integrals over the phase $e^{\frac{iS}{\hbar}}$ where S is a real function of the dynamical variables of the theory under consideration. The contour of integration for these variables is deformed into the complex plane to to run through a saddle point. Within the framework of Picard-Lefshetz Theory, this deformation is possible of one can slide down a contour of steepest ascent from the saddle point to the real axis which is usually the original contour of integration. It can be shown that the real part of the exponent in the path integral $Re[\frac{iS_{cl}}{\hbar}]$ always decreases when going up point of steepest ascent as it starts off at zero on the real axis and must be negative at any saddle point. The real part of the exponent completely determines the semi-classical factor in the quantum mechanical amplitude. In Picard-Lefshetz Theory, since the real part of the exponent is always decreasing, the semi-classical component can only suppress and not enhance the quantum mechanical amplitude. FLT argue that the appropriate contours for the path integral are completely specified by\cite{FLT};
\begin{enumerate}
    \item They are continuously deformable to contours running over real Lorentzian spacetime metrics
    \item They follow steepest descent contours along which the path integral is absolutely convergent.
\end{enumerate}
This process corresponds to considering semi-classical, Lorentzian amplitudes in General Relativity. A consequence of these criterion is that path integrals for minisuperspace cosmologies cannot be deformed to a Euclidean contour. FLT take the stance that there does exist such a Euclidean contour for which, if the steepest descent method is applied to the path integral over this contour, yields the HH wavefunction. However this contour is unrelated to the contour for Lorentzian path integrals that they consider. The primary point of contention between FLT and proponents of the No Boundary wavefunction such as Halliwell and Hartle is this choice of contour as the choice of contour will determine what saddle points contribute semi-classically to the path integral in Picard-Lefshets Theory. Different saddle point contributions will then yield different wavefunctions which will either be full solutions to the WdW equation or solutions to the inhomogeneous WdW equation.\\

Picard-Lefshets Theory is applicable to oscillatory integrals of the form
\begin{equation}
\label{L1}
    I=\int_{D}dx \ e^{\frac{iS[x]}{\hbar}}
\end{equation}
where $S[x]$ is a real valued function, $\hbar$ is a real parameter and $D$ is the domain of the integral. This domain is usually defined by singularities of the integrand. As discussed within the context of the WKB approximation, in the context of quantum mechanics, one is usually concerned with small values of $\hbar$. The limit as $\hbar \to 0$ corresponds to the classical limit. The premise of Picard-Lefshets Theory is to consider $S[x]$ as a holomorphic function of $x \in \mathbb{C}$ which is the complex plane. By Cauchy's theorem, the integration contour can be deformed from the real domain $D$, on the real axis into a contour $\mathcal{C}$ in the complex $x$-plane. The aim of this deformation is to mould $\mathcal{C}$ into a steepest descent contour which passes through one or more critical points of $S[x]$ where $\partial_{x}S=0$. $Re[iS[x]]$ which is the real part of the exponent and controls the magnitude of the integrand has a saddle point in the real two dimensional plane $(Re[x], Im[x])$ plane. The steepest descent contour  through the saddle point is then defined as the path along which $Re[iS[x]]$ decreases most rapidly\cite{FLT}.\\

FLT introduce this concept through the simple example of the action $S[x]=x^{2}$. This action has a critical point at $x=0$. Defining $x=Re[x]+Im[x]$, the action $iS[x]$ can be written as

\begin{align}
\label{L2}
    \begin{split}
        iS[x]=i(Re[x]+iIm[x])^{2}\\
        Re[iS[x]]=-2Re[x]Im[x]
        \end{split}
\end{align}
As the magnitude of the integrand is determined by the real part of the exponent, we see that the integrand decreases most rapidly along the contour $Re[x]=Im[x]$ which is therefore the steepest descent contour. The magnitude increases most rapidly along the contour $-Re[x]=Im[x]$ which is the steepest ascent contour.\\

The action can be re-expressed as the sum of its real and complex parts with the argument $x$ also having a real and complex part
\begin{equation*}
   S[x]=h[u^{1}(\lambda)+iu^{2}(\lambda)]+iH[u^{1}(\lambda)+iu^{2}(\lambda)].
\end{equation*}

Downward flow is the defined by
\begin{equation}
\label{L3}
    \frac{du^{1}}{d\lambda}=-g^{ij}\frac{\partial h}{\partial u^{j}},
\end{equation}
where $\lambda$ is a parameter along the flow and $g_{ij}$ is a Riemannian metric induced on the complex plane. The real part of the exponent $h$ decreases along this flow away from critical points. This is due to the fact that
\begin{equation}
\label{L4}
    \frac{dh}{d\lambda}=\sum_{i}\frac{\partial h}{\partial u^{i}}\frac{du^{i}}{d\lambda}=\sum_{i}\Bigl(\frac{\partial h}{\partial u^{i}}\Bigl)^{2}<0.
\end{equation}
The fastest rate of decrease occurs in the direction of steepest descent. This is the direction which maximises the magnitude of the gradient. The reason why a metric was introduced is to properly define the notion of gradient. To see how this metric can be specified, one can take the definitions
\begin{align}
\label{L5}
\begin{split}
    u&=Re[x]+iIm[x],\\
    \bar{u}&=Re[x]-iIm[x].\\
\end{split}
\end{align}
In these coordinates, it can be inferred that
\begin{align}
\label{L6}
    \begin{split}
        u\bar{u}&=Re[x]^{2}+Im[x]^{2}=|x|^{2},\\
        d(u\bar{u})&=d(Re[x]^{2}+Im[x]^{2})=|dx|^{2}.
    \end{split}
\end{align}
The terms in the metric are $g_{uu}=0$, $g_{\bar{u}\bar{u}}=0$ and $g_{u\bar{u}}=g_{\bar{u}u}=\frac{1}{2}$. Adapting the previously defined downward flow relation to this metric yields
\begin{align}
\label{L7}
    \begin{split}
        \frac{du}{d\lambda}&=-g^{u\bar{u}}\frac{\partial h}{\partial \bar{u}}=-2\frac{\partial h}{\partial \bar{u}},\\
        \frac{d\bar{u}}{d\lambda}&=-g^{\bar{u}u}\frac{\partial h}{\partial u}=-2\frac{\partial h}{\partial u}.
    \end{split}
\end{align}
Noting that the real and complex parts of the exponent can be re-expressed as
\begin{equation}
\label{L8}
    h=\frac{\mathcal{I}+\bar{\mathcal{I}}}{2}, \quad \quad  H= \frac{\mathcal{I}-\bar{\mathcal{I}}}{2i}.
\end{equation}
Substituting these forms of the real and complex parts of the exponents into the downward flow formulae in this metric yield the relation,
\begin{align}
\label{L9}
\begin{split}
     \frac{du}{d\lambda}&=-\frac{\partial \bar{\mathcal{I}}}{\partial \bar{u}},\\
     \frac{d\bar{u}}{d\lambda}&=-\frac{\partial \mathcal{I}}{\partial u}.
\end{split}
\end{align}
We may also see that the complex part of the wavefunction is conserved along these flows. Using the definition of the complex part of the exponents in terms of the exponent and its conjugate given in (\ref{L8})
\begin{align}
\label{L10}
\begin{split}
    \frac{dH}{d\lambda}&=\frac{1}{2i}\frac{d(\mathcal{I}-\bar{\mathcal{I}})}{d\lambda},\\
    &=\frac{1}{2i}\Bigl(\frac{\partial \mathcal{I}}{\partial u}\frac{du}{d\lambda}-\frac{\partial \bar{\mathcal{I}}}{\partial \bar{u}}\frac{d\bar{u}}{d\lambda}\Bigl),\\
    &=0.
    \end{split}
\end{align}
The last line of (\ref{L10}) follows from the substitution of (\ref{L9}) into the second line of the above equation. From these considerations we see that the generally highly oscillatory integrand $e^{\frac{iS[x]}{\hbar}}$ does not oscillate when considering downward flows. This is due to the real part monotonically decreasing along flows while the complex part is conserved. These two properties ensure that the integral converges as rapidly as possible\cite{FLT}. At the saddle point the $\lambda =-\infty$ which then runs to positive values as $h$ decreases going down the flow. The Lefshets Thimble associated with a saddle is the set of all flows which leave the saddle in this way. Similar to the above considerations, upward flows can be defined by the relation
\begin{equation}
\label{L11}
    \frac{d u^{i}}{d\lambda}=+ g^{ij}\frac{\partial h}{\partial u^{j}}.
\end{equation}
The complex part of the exponent is also conserved along these flows. A general rule of thumb is that any steepest descent contour $\mathcal{J}_{\sigma}$ originating from a saddle point, ends on a singularity where $h \to -\infty$. A steepest scent contour $\mathcal{K}_{\sigma}$ beginning on a saddle point will end on a singularity where $h \to +\infty$. Steepest descent and steepest ascent contours will intersect each other at a single saddle point where both are defined. This intersection can be defined by an intersection number
\begin{equation}
\label{L10}
    \text{{Int}}(\mathcal{J}_{\sigma}, \mathcal{K}_{\sigma'})=\delta_{\sigma \sigma'}.
\end{equation}
As mentioned previously the goal of Picard-Lefshets Theory is to deform the original contour to be a contour composed of a number of steepest descent curves on which the integrand is absolutely convergent. The goal is therefore,to deform the original contour into
\begin{equation}
\label{L11}
    \mathcal{C}=\sum_{\sigma}n_{\sigma}\mathcal{J}_{\sigma}
\end{equation}
 where $n_{\sigma}$ may take the values $0$ or $\pm1$ depending on orientation of the contour of the thimble. Merging the previous two equations (\ref{L10}) and (\ref{L11}), we see that
 \begin{equation*}
     n_{\sigma}=\text{{Int}}(\mathcal{C}, \mathcal{K}_{\sigma})= \text{{Int}}(D, \mathcal{K}_{\sigma}).
 \end{equation*}
The intersection number $n_{\sigma}$ is topological and wont change if the contour is deformed back to the original domain. From this it can be inferred that a prerequisite for the a given thimble $\mathcal{J}_{\sigma}$ to be relevant is that a contour of steepest ascent must intersect the original domain D. FLT describe this visually by noting that there should be no obstacle to smoothly moving the intersection point from the real axis to along $\mathcal{K}_{\sigma}$ to the saddle point $p_{\sigma}$. At this saddle point $p_{\sigma}$, there will be Lefshets thimbles $\mathcal{J}_{\sigma}$ leaving the saddle point.  Therefore the contour has been deformed onto the Lefshets  thimble on which the integrand is convergent.\\
\begin{figure}
    \centering
    \includegraphics{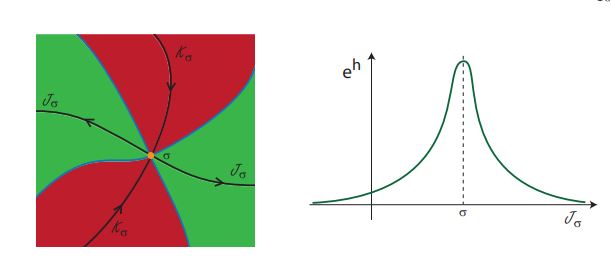}
    \caption{The point $\sigma$ is a saddle point from which Lefshets thimbles $\mathcal{J}_{\sigma}$ (steepest descent contours) flow downwards and steepest ascent contours $\mathcal{K}_{\sigma}$. The green wedges $J_{\sigma}$ containing the steepest descent contours are regions in which the magnitude of the real part of the action is lower than the saddle point. This is illustrated by the figure on the right The red wedges $K_{\sigma}$ are regions in which the magnitude of the real part of the wave function is greater than its value at the saddle point. Adapted from \cite{FLT}}
    \label{fig:my_label}
\end{figure}
Usually, more than one Lefshets thimble contributes to an integral. In this case the contour is deformed to a sum of thimbles. Once the contour has been deformed from the domain $D$ corresponding to the real axis to run through Lefshets thimbles associated with the saddle points which contribute to the path path integral the original integral (\ref{L1}), can be rewritten as
\begin{equation}
\label{L12}
    I=\int_{D}dx \ e^{\frac{iS[x]}{\hbar}}=\int_{C}dx \ e^{\frac{iS[x]}{\hbar}}=\sum_{\sigma}n_{\sigma} \int_{\mathcal{J}_{\sigma}}dx \ e^{\frac{iS[x]}{\hbar}}
\end{equation}
As the magnitude of the integral is determined by the real part of the exponential. The integral over a thimble is deemed to be absolutely convergent if 
\begin{equation}
\label{L13}
    \Biggl|\int_{\mathcal{J}_{\sigma}}dx \ e^{\frac{iS[x]}{\hbar}} \Biggl| \ \le \ \int_{\mathcal{J}_{\sigma}} |dx| \ \biggl|e^{\frac{iS[x]}{\hbar}} \biggl|  \ = \ \int_{\mathcal{J}_{\sigma}} |dx|e^{h(x)} \ < \ \infty.
\end{equation}
The length of the curve can be defined as $l= \int |dx|$. The integral then converges if $(x(l))<-ln(l)+A$ for a constant $A$ as $l \to \infty$. Thus, the original integral can be expressed as the sum of absolutely convergent integrals along steepest descent contours. Using the saddle point approximation with an expansion in $\hbar$\cite{FLT}
\begin{align}
\label{L14}
    I&=\int_{D}dx \ e^{\frac{iS[x]}{\hbar}}=\sum_{\sigma}n_{\sigma}e^{iH(p_{\sigma})}\int_{\mathcal{J}_{\sigma}}dx \ e^{h}.\\
    &\approx \sum_{\sigma}n_{\sigma}e^{\frac{iS(p_{\sigma})}{\hbar}}[A_{\sigma}+ \mathcal{O}(\hbar)]
\end{align}
where $A_{\sigma}$ is the result of the leading order Gaussian integral about the saddle point $p_{\sigma}$. The method of steepest descent can be thought of as a sophisticated modification of the regular WKB approximation used to evaluate path integrals in the semi-classical regime.

\subsection{Lorentzian Path Integrals in Quantum Cosmology: Minsuperspace}
The steepest descent method can be applied to Lorentzian Path integrals within the simplest cosmological models\cite{FLT}. In this case, a homogeneous, isotropic and closed universe. The starting point is the regular Einstein-Hilbert action wit the metric being the closed FRW metric. The use of BFV ghosts, breaks time reparameterisation invariance and fixes the gauge to be the constant or proper time gauge $\dot{N}=0$\cite{HAL1}. For the case of minisuperspace, the reduced path integral is given by
\begin{equation}
    G[a_{1}|a_{0}]= \int_{0^{+}}^{\infty}dN\int_{a_{0}}^{a_{1}}\mathcal{D}a \ e^{\frac{iS(N,a)}{\hbar}}
\end{equation}
FLT interpret the first path integral as representing the quantum mechanical amplitude for the universe to evolve from $a_{0}$ to $a_{1}$ in proper time $N$. As $N$ represents proper time, the integration limits indicate that paths of every proper duration in the range $0 < N < \infty$ should be considered\cite{T1}. This choice of integration leads to the notion of $a_{0}$ preceding $a_{1}$ or vice versa. In doing so, there is in an inherent notion of causality for the Lorentzian path integral. The ordering $a_{1}>a_{0}$ represents an expanding universe while $a_{1}<a_{0}$ represents a contracting universe. The direction of the arrow of time is therefore controlled by propagator rather than boundary conditions that are imposed later. The minisuperspace action is given by
\begin{equation}
    S=6\pi^{2}\int_{0}^{1} dt N\Biggl(-a\frac{\dot{a}^{2}}{N^{2}}+a\Bigl(k-\frac{\Lambda}{3}a^{2}\Bigl)\Biggl).
\end{equation}
This integral can be simplified by the redefinition
\begin{equation*}
    N(t) \to \frac{N(t)}{a}
\end{equation*}
which converts the action to an action that is quadratic in $a$ and its derivative. For convenience, one can define $q(t)=a^{2}$ with the path integral over $q(t)$ being performed exactly. It is expected that the redefinition of the Lapse function in the exponent should also correspond to the changing of the path integral measure. It has been shown that the Hamiltonians of theories expressed in terms of $q=a^{2}$ and $a$ are the same\cite{HAL1}. The first functional integral may be evaluated by using the semicalssical approximation. The classical equations of motion for $a$ can then be solved to yield the classical solution $a=a_{c}(Nt)$ which holds for arbitrary initial $a_{0}$ and final $a_{1}$\cite{FLT}. With these redefinitions the action becomes
\begin{equation}
\label{LB-1}
    S=6\pi^{2}\int_{0}^{1} dt N\Biggl(-\frac{\dot{q}^{2}}{N^{2}}+\Bigl(k-\frac{\Lambda}{3}q\Bigl)\Biggl)
\end{equation}
The equations of motion derived by the variation of the action with respect to $q(t)$ and the constraint which is derived from the variation of the action with respect to $N$ are given by
\begin{equation}
\label{LB2}
    \ddot{q}=\frac{2\Lambda}{3}N^{2}; \quad \frac{3}{4N^{2}}\dot{q}^{2}+3k=\Lambda q
\end{equation}
As in the case of steepest descent analysis of Euclidean path integrals, solutions to the field equations given boundary conditions do not satisfy the Hamiltonian constraint as the constraint corresponds to the variation of the action with respect to $N$.
With the arbitrary boundary conditions $q(0)=q_{0}$ and $q(1)=q_{1}$, the general solution tot he first equation is given by
\begin{equation}
\label{LB-2}
    \bar{q}=\frac{\Lambda}{3}N^{2}t^{2}+\Biggl(-\frac{\Lambda}{3}N^{2}+q_{1}-q_{0}\Biggl)t+q_{0}.
\end{equation}
The full solution which does satisfy the Hamiltonian constraint can be written in a similar fashion as (\ref{E0}) with the primary difference being that in this case Lorentzian time $t$ is taken as fundamental rather than Euclidean time $\tau$
\begin{equation}
    q(t)=\bar{q}(t)+\mathcal{Q}(t)
\end{equation}
In doing this the saddle point approximation is employed such that
\begin{equation}
    G[q_{1}|q_{0}]=\int_{0}^{\infty}dN \ e^{\frac{2\pi^{2}iS_{0}}{\hbar}}\int_{Q[0]=0}^{Q[1]=0} d\mathcal{Q} \ e^{\frac{2\pi^{2}iS_{2}}{\hbar}}.
\end{equation}
with the actions $S_{0}$ and $s_{2}$ being
\begin{align}
\begin{split}
    S_{0}=\int_{0}^{1} dt \ \Biggl(-\frac{\dot{\bar{q}}^{2}}{N^{2}}+\Bigl(k-\frac{\Lambda}{3}\bar{q}\Bigl)\Biggl).
\end{split}
\end{align}
The path integral over $Q$ is a Gaussian integral and can therefore can be evaluated
\begin{equation}
    \int_{Q[0]=0}^{Q[1]=0} d\mathcal{Q} \ e^{\frac{2\pi^{2}iS_{2}}{\hbar}}=\sqrt{\frac{3\pi i}{2N\hbar}}
\end{equation}
The resulting integral is an ordinary integral in the Lapse function of the form
\begin{equation}
    G[q_{1}|q_{0}]=\sqrt{\frac{3\pi i}{2 \hbar}}\int_{0}^{\infty} \frac{dN}{\sqrt{N}}\ e^{\frac{2\pi^{2}iS_{0}}{\hbar}}
\end{equation}
This integral possess singularities at its endpoints  $N=0$ and $N=\infty$.  FLT proved in general that integrals of this form can be deformed into a convergent steepest descent integral given a suitable cutoff\cite{FLT}. Substituting the solution $\bar{q}$ into the action $S_{0}$ yields
\begin{equation}
\label{LB0}
    S_{0}= N^{3}\frac{\Lambda^{2}}{36}+N\Biggl(-\frac{\Lambda}{2}(q_{0}+q_{1})+3k\Biggl)+\frac{1}{N}\Biggl(-\frac{3}{4}(q_{1}-q_{0})^{2}\Biggl)
\end{equation}
Once again employing the saddle point applying the saddle point approximation in $N$ we see that the four saddle points are given by
\begin{align}
\label{LB1}
\begin{split}
    \frac{\partial S_{0}}{\partial N}\Biggl|_{N=N_{s}}&=\Lambda^{2}N_{s}^{4}+(-6\Lambda(q_{0}+q_{1})+36)N_{s}^{2}+9(q_{1}-q_{0})^{2}=0\\
    N_{s}&=c_{1}\frac{3}{\Lambda}\Biggl[\Biggl(\frac{\Lambda}{3}q_{0}-k \Biggl)^{\frac{1}{2}} +c_{2}\Biggl(\frac{\Lambda}{3}q_{1}-k  \Biggl)^{\frac{1}{2}}  \Biggl]
\end{split}
\end{align}
with each of the saddle points being defined by the unique selection of the pair $c_{1}, c_{2} \in \{-1,1\}$. For $k=1$ corresponding to spherical geometry these saddle points will be complex for $q_{0,1}<k$. This corresponds to the underbarrier region of the potential where there is non classical behavior. This can be interpreted as non-Lorentzian geometries dominating the path integral. For $k=0$ flat and $k=1$ hyperbolic geometries, these saddle points are always real. These saddle points can be substituted back into the action $S_{0}$ to yield
\begin{equation}
    S_{0}^{saddle}-c_{1}\frac{6}{\Lambda}\Biggl[(\frac{\Lambda}{3}q_{0}-k \Biggl)^{\frac{3}{2}} +c_{2}\Biggl(\frac{\Lambda}{3}q_{1}-k  \Biggl)^{\frac{3}{2}} \Biggl]
\end{equation}
In the context of Picard-Lefshets Theory, each of these contours has an associated contour of steepest descent $\mathcal{J}_{\sigma}$ and contour of steepest ascent $\mathcal{K}_{\sigma}$. The original contour can be written as a sum over Lefshets thimbles
\begin{equation}
    (0^{+}, \infty)= \sum_{\sigma}n_{\sigma}\mathcal{J}_{\sigma}
\end{equation}
The propagator can be approximated by the saddle point approximation in the limit $\hbar \to 0$
\begin{align}
\begin{split}
    G[q_{1}|q_{0}]&=\sum_{\sigma}n_{\sigma}\sqrt{\frac{3\pi i}{2 \hbar}}\int_{\mathcal{J}_{\sigma}}\frac{dN}{\sqrt{N}} \ e^{2\pi^{2}iS_{0}/\hbar}\\
    &\approx \sum_{\sigma}n_{\sigma}\sqrt{\frac{3\pi i}{2 \hbar}}\frac{e^{2\pi^{2}iS^{saddle}_{0}/\hbar}}{\sqrt{N_{s}}}\int_{\mathcal{J}_{\sigma}}dN \ e^{\frac{i\pi^{2}}{\hbar}S_{0,NN(N-N_{s})^{2}}}[1+\mathcal{O}(\sqrt{\hbar})]
\end{split}
\end{align}
Using the definition $N-N_{s}\equiv ne^{i\theta}$ where $\theta$ is the angle of the Lefshets thimble with respect tot the positive $N$ axis. Substituting this into the above approximate propagator
\begin{align}
    \begin{split}
        G[q_{1}|q_{0}]&\approx \sum_{\sigma}n_{\sigma}\sqrt{\frac{3\pi i}{2 \hbar}}\frac{e^{2\pi^{2}iS^{saddle}_{0}/\hbar}}{\sqrt{N_{s}}} e^{i\theta_{\sigma}}\int_{\mathcal{J}_{\sigma}}dn \ e^{-\frac{\pi^{2}}{\hbar}|S_{0, NN}|n^{2}}[1+\mathcal{O}(\sqrt{\hbar})]\\
        &\approx\sum_{\sigma}n_{\sigma}\sqrt{\frac{3i}{2N_{s}|S_{0,NN}|}}e^{i\theta_{\sigma}e^{2\pi^{2}iS_{0}^{Saddle}}/\hbar}[1+\mathcal{O}(\sqrt{\hbar})]
    \end{split}
\end{align}
FLT discuss the use of Picard-Lefshets theory and the saddle point approximation for four distinct boundary conditions for $k=1$ geometry. The two boundary conditions that are relevant to this thesis are;\

\begin{enumerate}
    \item Both saddle points are real: $q_{1}\ge q_{0}>\frac{3}{\Lambda}$ corresponding to a classical universe.
    \item One root is complex: $q_{1}>\frac{3}{\Lambda}>q_{0}$ which includes the No-boundary proposal.
\end{enumerate}
\begin{center}
    \textbf{Classical Universe}
\end{center}
In the first case, the four saddle points given by (\ref{LB1}) are real. However, the domain of integration is $0<N<\infty$. Therefore the two negative saddle points do not contribute to the path integral semi-classically.  The two saddle points in the positive complex plane are specified by $c_{1}=1$ and $c_{2}=\pm 1$. These saddle points are given by
\begin{equation}
    N^{+}_{s\pm}=\sqrt{\frac{3}{\Lambda}}\Biggl[\Biggl(q_{1}-\frac{3}{\Lambda} \Biggl)^{\frac{1}{2}} \pm  \Biggl(q_{0}-\frac{3}{\Lambda} \Biggl)^{\frac{1}{2}} \Biggl].
\end{equation}
In this special case, the two saddle points lie on the positive real $N$ axis. Therefore their curves of steepest ascent automatically intersect the positive real $N$ axis which is the domain of integration. As a result, both of these saddle points contribute to the path integral semi-classically. Physically, $q_{0}$ and $q_{1}$ may either sit on the same side of the de Sitter hyperboloid or may be separated by the waist. By taking the first derivative of the solution to the equations of motion $\bar{q}(t)$ at $t=0$ it is found that $N^{+}_{s+}$ is the decreasing solution and corresponds to the waist of de Sitter space sitting between $q_{0}$ and $q_{1}$.  The saddle point $N^{+}_{s-}$ is the solution for which $q_{0}$ and $q_{1}$ are on the same side of the de Sitter hyperboloid. Given that these two saddle points will contribute to the semi-classical propagator, the propagator can be written as
\begin{align}
    \begin{split}
        G[q_{1}|q_{0}]&\approx \Biggl(\frac{3i}{4\Lambda\sqrt{\Bigl(q_{0}-\frac{3}{\Lambda}\Bigl)\Bigl(q_{1}-\frac{3}{\Lambda}\Bigl)}}  \Biggl) \cdot (e^{-i\frac{\pi}{4}}e^{iS(N_{s-})/\hbar}+(e^{i\frac{\pi}{4}}e^{iS(N_{s+})/\hbar})\\
        &\approx \Biggl(\frac{e^{i\frac{\pi}{4}}3^{\frac{1}{2}}}{[(\Lambda q_{0}-3)(\Lambda q_{1}-3)]^{\frac{1}{4}}} \Biggl) \times \cos \Biggl(\frac{4\pi^{2}\Lambda^{\frac{1}{2}}}{3^{\frac{1}{2}}\hbar} \Biggl(q_{0}-\frac{3}{\Lambda}  \Biggl)^{\frac{3}{2}}-\frac{\pi}{4} \Biggl) \times e^{-i4\pi^{2}\Lambda^{1/2}/3^{1/2}\hbar (q_{1}-3/\Lambda)^{3/2}}
    \end{split}
\end{align}
\begin{figure}[h!]
    \centering
    \includegraphics{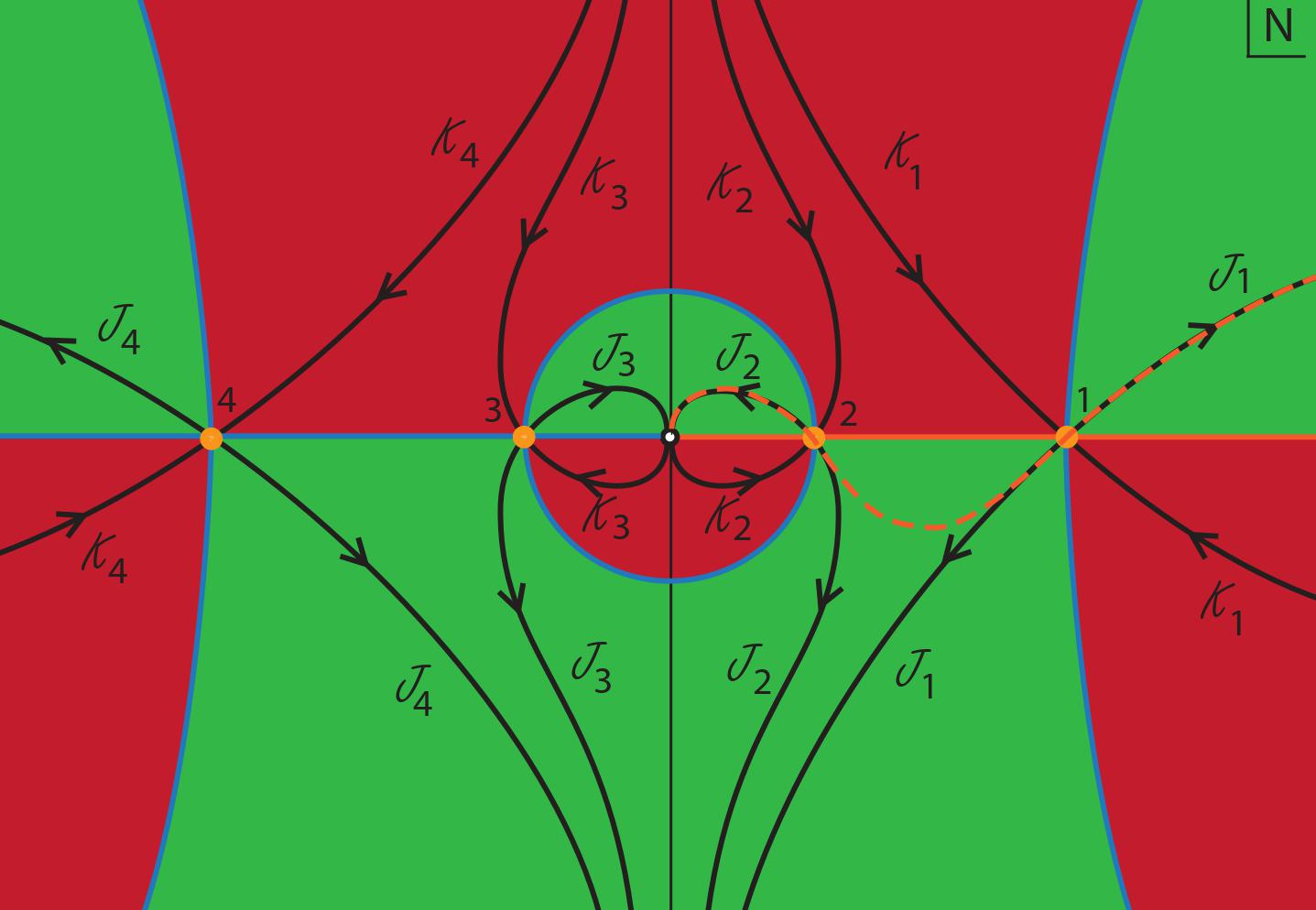}
    \caption{The steepest descent contours $\mathcal{J}_{\sigma}$ and steepest ascent contours $\mathcal{K}_{\sigma}$ which enter and leave saddle points in the complex $N$ plane for the boundary conditions $q_{1}>q_{0}\ge \frac{3}{\Lambda}$. The green wedges $J_{\sigma}$ correspond to regions in which the real part of the exponent has a value lower than the saddle point. The red wedges $K_{\sigma}$ are regions where the real part of the exponent has a value greater than the saddle point. The original contour is the positive N axis (the solid orange line). This contour is then deformed onto the steepest descent contours $\mathcal{J}_{1}$ and $\mathcal{J}_{2}$ which leave saddle points 1 and 2 (orange dashed line). The integral is absolutely convergent along this contour. Adapted from \cite{FLT}.}
    \label{fig:my_label}
\end{figure}
\begin{center}
    \textbf{No Boundary Conditions }
\end{center}
The HH No Boundary proposal can be formulated as the rule that the only metrics that should be summed over in the path integral are those whose boundary is the final spatial hypersurface. A complementary definition is the sum of metrics that do not have a past boundary. In closed FRW minisuperspace these boundary conditions correspond to $q_{0}=0$ which represents no past spacetime boundary. One must then find a 4-metric that is regular at this boundary. This is possible for spherical 3-geometry corresponding to $k=1$. The boundary condition in combination with the constraint (\ref{LB2}) yields\cite{FLT}
\begin{equation}
     {\dot{q}}^{2}=-4Nk
\end{equation}

The final spatial hypersurface can be taken to correspond to a late time configuration where the universe is large represented by $q_{1}>k$. The saddle points of this action are
\begin{equation}
\label{LB4}
    N_{s,NB(1)}=+\frac{3}{\Lambda}\Biggl[i\pm \Biggl(\frac{\Lambda}{3}q_{1}-1  
    \Biggl)^{\frac{1}{2}} \Biggl], \quad  N_{s,NB(2)}=-\frac{3}{\Lambda}\Biggl[i\pm \Biggl(\frac{\Lambda}{3}q_{1}-1  
    \Biggl)^{\frac{1}{2}} \Biggl].
\end{equation}
The actions obtained when these saddle points are substituted back into the original action are given by
\begin{equation}
    S_{0,NB(1)}=+\frac{6}{\Lambda}\Biggl[-i\pm \Biggl(\frac{\Lambda}{3}q_{1}-1  
    \Biggl)^{\frac{3}{2}} \Biggl], \quad 
    S_{0, NB(2)}=-\frac{6}{\Lambda}\Biggl[-i\pm \Biggl(\frac{\Lambda}{3}q_{1}-1  
    \Biggl)^{\frac{3}{2}} \Biggl]
\end{equation}
Two of these saddle points will be in the upper and lower complex plane where $N$ is positive while two will be in the upper and lower complex plane where $N$ is negative. Once again, the relevant original contour is defined as the positive real $N$ axis $0<N<\infty$, the two saddle points in the negative complex plane $N_{s, NB(2)}$ will not contribute to the path integral. For these boundary conditions, the steepest ascent contour associated with the saddle points in the upper complex plane intersect the real $N$ axis. Therefore the real $N$ axis can be deformed to be the sum of Lefshets thimbles $\mathcal{J}_{1}$ and $\mathcal{J}_{2}$. However as the integration contour is strictly positive, the original domain of integration can only be deformed to the thimble $\mathcal{J}_{1}$ along which the integral is absolutely convergent. As a consequence of this, the only saddle point which contributes to the path integral is the saddle point in the positive upper complex plane
\begin{equation}
     N^{+}_{s,NB(1)}=+\frac{3}{\Lambda}\Biggl[i+ \Biggl(\frac{\Lambda}{3}q_{1}-1  
    \Biggl)^{\frac{1}{2}} \Biggl]
\end{equation}
whose action is given by
\begin{equation}
    S^{+}_{0, NB(1)}=-\frac{6}{\Lambda}\Biggl[-i+ \Biggl(\frac{\Lambda}{3}q_{1}-1  
    \Biggl)^{\frac{3}{2}} \Biggl].
\end{equation}
Using the saddle point approximation in the $\hbar \to 0$ limit the propagator has the approximate form
\begin{equation}
    G[q_{1}|0]\approx e^{i\frac{\pi}{4}}\frac{3^{\frac{1}{4}}}{2(\Lambda q_{1}-3)^{\frac{1}{4}}}e^{\frac{-12\pi^{2}}{\hbar \Lambda}-i4\pi^{2}\sqrt{\frac{\Lambda}{3}}(q_{1}-\frac{3}{\Lambda})^{\frac{3}{2}}/\hbar}
\end{equation}
Saddle points in the upper half plane given in (\ref{LB4}) will have weighting $e^{i 1\pi^{2}S_{0}} \sim e^{-12\pi^{2}/(\hbar\Lambda)}$ while saddle points in the lower half plane will have weighting $e^{i 1\pi^{2}S_{0}} \sim e^{12\pi^{2}/(\hbar\Lambda)}$. The HH result has weighting $e^{12\pi^{2}/(\hbar\Lambda)}$ while this propagator has the inverse. As the HH result was derived using a Euclidean path integral which cannot be related to the Lorentzian path integral, FLT take this as conclusive proof that the Hartle-Hawking wavefunction of the universe lacks physical motivation as the contour and saddle points of the Lorentzian path integral are highly constrained and lead to the opposite result. The difference between the Euclidean and Lorentzian results was initially interpreted by FLT as being due to the nature of Wick rotation. In the proper time gauge $\dot{N}=0$, at the level of the Einstein-Hilbert action, the Wick rotation $t \to it$ may interpreted as a redefinition of $N$ as $N \to iN$\cite{FLT}. Thus, Euclidean time is associated with taking the imaginary $N$ axis as being the contour. But given that there are no steepest ascent contours which intersect the imaginary axis, the method of steepest descent can't be used as no saddle points would contribute to the integral defined over this contour.

\begin{figure}[h!]
    \centering
    \includegraphics{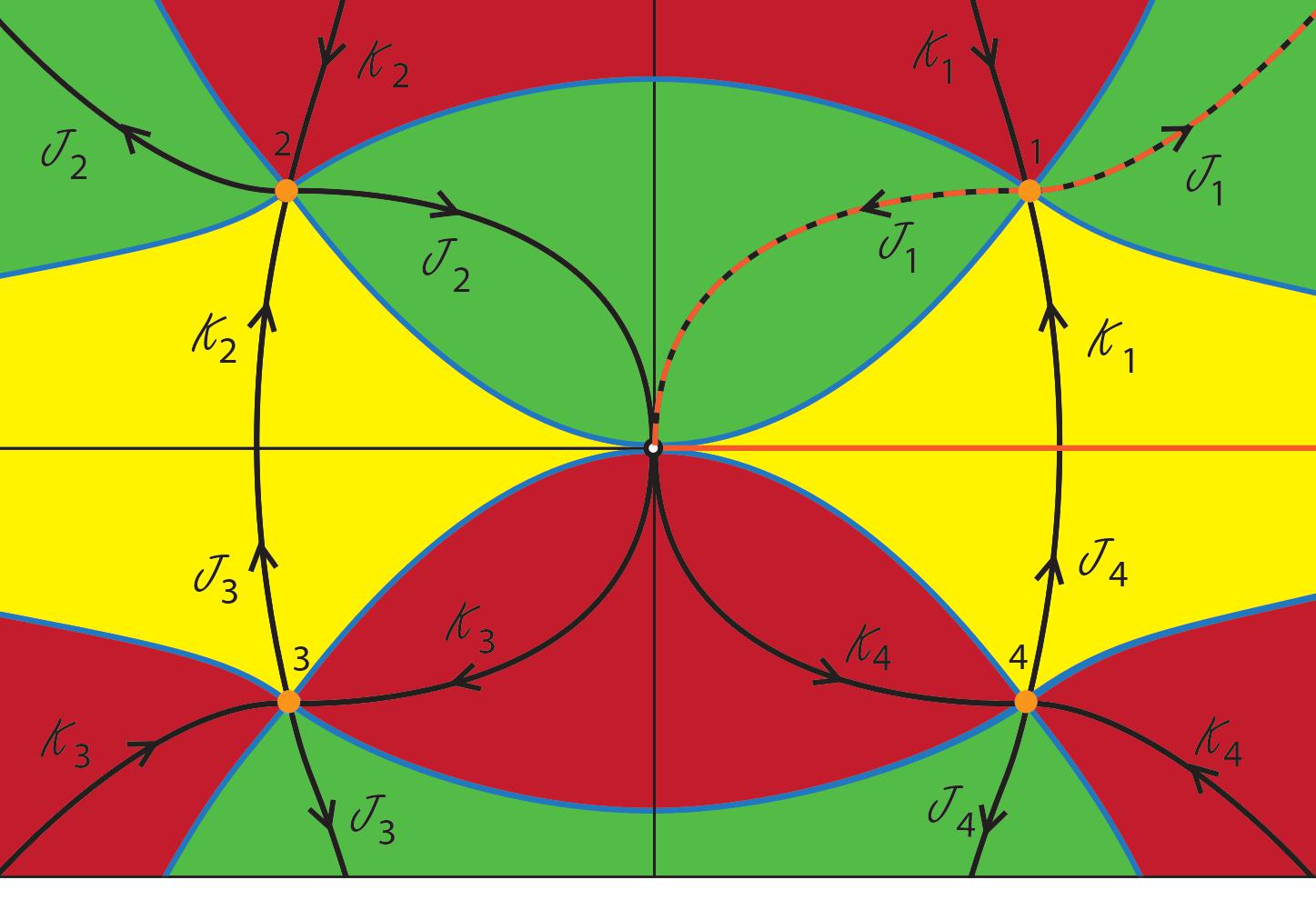} 
    \caption{A depiction of the saddle points and their associated steepest descent and ascent contours for No Boundary proposal boundary conditions $q_{1}>q_{0}=0$. The yellow regions correspond to regions in which the real part of the exponential has a value that is intermediate with respect to the two saddle points above and below it. As the steepest descent contour $\mathcal{K}_{1}$ intersects the original contour of integration (the orange dashed line), the contour can be deformed to the steepest descent contour $\mathcal{J}_{1}$.Adapted from \cite{FLT}.}
    \label{fig:my_label}
\end{figure}

\subsection{Hartle-Hawking Wavefunction from The Lorentzian Path Integral}
The arguments of FLT assumed that the only possible way to derive the HH wavefunction was by starting with a Euclidean path integral. However, Diaz Dorronsoro et al\cite{DD} showed that it is possible to derive the HH result starting from the minisuperspace Lorentzian path integral. While the No-Boundary Proposal boundary condition of $q(0)=0$ remains unchanged, the primary difference in approach lies in the choice of contour. The contour in this case is defined as the entire real $N$ axis $-\infty < N < \infty$. As argued in previous work by Halliwell on steepest descent contours within the context of Euclidean path integrals, extending the range over the entire real line yields wavefunctions which are genuine solutions to the homogeneous WdW equation whereas a half infinite range yields Green's functions which are solutions to the inhomogeneous WdW equation.\\

The starting point of the analysis of Diaz Dorronsoro et al\cite{DD} is the same as that of FLT, the Lorentzian Einstein-Hilbert action of the closed FRW metric in minisuperspace. Once again, the variation of the action with respect to $q=a^{2}$ yields the equations of motion while the variation with respect to $N$ yields the constraint that $q$ must obey. Evaluation of the integral over $q$ yields
\begin{equation}
\psi_{NB}(q_{1})=\sqrt{\frac{3\pi i}{2 \hbar}}\int_{\mathcal{C}}\frac{dN}{\sqrt{N}} \ e^{2\pi^{2}iS_{0}(N,q_{1})/\hbar}
\end{equation}
where the reduced action $S_{0}$ which uis a result of the substitution of the solution of the equations of motion for $q$ when $q_{0}=0$ is
\begin{equation}
    S=\frac{\Lambda^{2}}{36}N^{3}+\Biggl(3-\frac{\Lambda q_{1}}{2} \Biggl)N-\frac{3q_{1}^{2}}{4N}.
\end{equation}
This is simply (\ref{LB0}) with $q_{0}=0$ to suit the No Boundary proposal conditions. In \cite{DD} Diaz Dorronsoro et al interpret the result of the path integral calculation as being a wavefunction since they are considering the result of the path integral as being true solutions to the WdW equation. The integration contour is specifically taken as $\mathcal{C}=(-\infty, \infty) \downarrow$ to avoid the singularity at $N=0$. In the case where the final boundary is late time classical universe where $q_{1}>\frac{3}{\Lambda}$, the four saddle points of the action $S_{0}$ are given by
\begin{equation}
    N_{s}=\pm \frac{3}{\Lambda}\Biggl(i\pm\sqrt{\frac{\Lambda q_{1}}{3}-1}\Biggl)
\end{equation}
In this case, they find that it is the two saddle points in the lower complex plane that contribute to the path integral. Noting that the action is a real function of $N$ such that $\overline{S_{0}(N)}=S_{0}(\overline{N})$, this leads to a degeneracy between steepest ascent and descent curves. Conceptually the steepest ascent and descent contours which leave one saddle point may be the steepest descent and steepest ascent contour which leave saddle point which is its reflection about the real $N$ axis due to the overall action being real. This is dealt with by introducing small complex perturbations which break symmetry. The limit of the symmetry breaking perturbations can then be taken to zero. In this case, in the limit where the perturbations are zero, the integration contour also runs through the saddle points in the upper half plane. The contribution of the saddle points in the upper half plane is exponentially suppressed so they are ignored.\\

Taking account of the angles at which the Lefshets thimble passes through the saddle point as well as the prefactor arising from the Gaussian integral in the $\hbar \to 0$ limit they arrive at the wavefunction
\begin{equation}
\label{LB4}
    \psi_{NB}(q_{1})=\frac{e^{+12\pi^{2}/\hbar \Lambda}}{\Bigl(\frac{\Lambda q_{1}}{3}-1 \Bigl)^{\frac{1}{4}}}\cos\Biggl[\frac{12\pi^{2}}{\hbar \Lambda} \Biggl(\frac{\Lambda q_{1}}{3}-1 \Biggl)^{\frac{3}{2}}+\frac{3\pi}{4} \Biggl]
\end{equation}
Diaz-Dorronsoro et al emphasize that this is a real wavefunction with positive exponential weighting $e^{+12\pi^{2}/\hbar \Lambda}$. This wavefunction is exactly equal to the No Boundary wavefunction derived from a Euclidean path integral. They note that for large values of the scale factor $q_{1}$ the wavefunction on the spatial hypersurface takes is of the WKB form and predicts classical scale factor evolution.\\

The primary reason why the contour $\mathcal{C}=(-\infty, \infty) \downarrow$ is  chosen is so that the resultant wavefunction in the Diaz et al interpretation, is a genuine solution of the WdW equation. The WdW equation acting on a wavefunction $\psi_{NB}$ is given by
\begin{equation}
    \hbar^{2}\frac{\partial^{2} \psi_{NB}}{\partial q_{1}^{2}}+12\pi^{4}(\Lambda q_{1}-3)\psi_{NB}(q_{1})=0
\end{equation}
Substituting in the the form of $\psi_{NB}(q_{1})$ in (\ref{LB4}), this yields
\begin{equation}
    \hbar^{2}\frac{\partial^{2} \psi_{NB}}{\partial q_{1}^{2}}+12\pi^{4}(\Lambda q_{1}-3)\psi_{NB}(q_{1})=6\pi^{2}i\sqrt{\frac{3 \pi i}{2}}\Biggl[\frac{e^{2\pi^{2}iS_{0}(N,q_{1})/\hbar}}{\sqrt{N}}  \Biggl]^{\infty}_{\infty}
\end{equation}
As the right hand side completely vanishes, $\psi_{NB}(q_{1})$ is a solution of the WdW equation. If the range were half infinite. Then the result of the wavefunction would have been a Green's function with a delta function on the right hand side which is not seen by Diaz Dorronsoro et al to be a true solution of the WdW equation\cite{DD}. An important aspect of the FLT argument was that the half infinite contour should be chosen as it endows the propagator with an aspect of causality\cite{FLT}. However, Diaz Dorronsoro et al argue that the contour of Lapse integration is not physically equivalent to the arrow of time which defines other physical processes. The histories of a given spatial hypersurface are curves in the superspace of 4-geometries with no physical notion of one being before or after the other in a temporal sense. If the reversal of the sign of $N$ merely corresponds to a reparameterisation of these curves in superspace and does not signify physical time reversal, then taking the contour to be the entire real $N$ axis is allowable.\\

\begin{figure}[h!]
    \centering
    \includegraphics{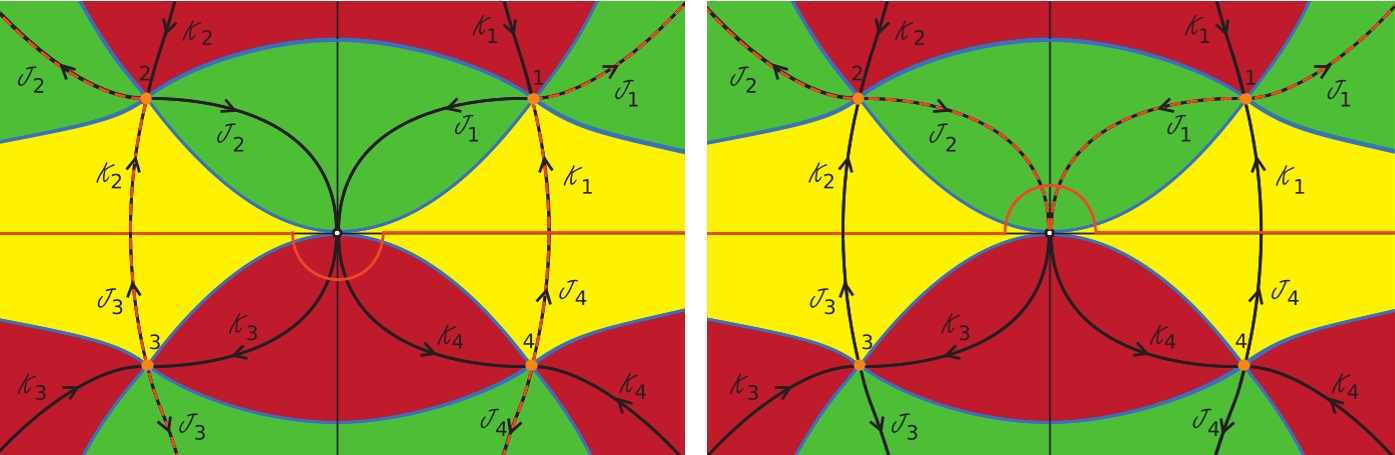}
    \caption{\textit{Left}: The no boundary contour taken by Diaz Doronsorro et al in \cite{DD} where the interval is the entire real line which goes below the contour at $N=0$. It is argued that the contour should be deformed to pass through all four saddle points with the contribution from the saddle points 1 and 2 being exponentially suppressed relative to saddle points 3 and 4. \textit{Right}: FLT argue that if the contour over the entire real $N$ line is taken, it must be the contour which passes above $N=0$\cite{FLT2}. This is due to the rule that that a steepest ascent contour must intersect the original contour of integration. For the contour on the right, these are $\mathcal{K}_{1}$ and $\mathcal{K}_{2}$. For this contour only saddle points 1 and 2 dominate so the No Boundary wavefunction is not the result.  }
    \label{fig:my_label}
\end{figure}

\subsection{Tunneling Wavefunction from The Lorentzian Path Integral}
In the context of minisuperspace, the Tunneling and No-Boundary conditions are actually the same boundary condition, namely that $q(0)=a(0)^{2}=0$. The interpretation of this boundary condition is slightly different with Vilenkin's interpretation of this boundary condition being that this represents no classical spacetime. The universe then quantum mechanically tunnels to a de Sitter universe of radius $a_{min}=H^{-1}$. In the No-Boundary proposal, the universe is considered to not have a past boundary with the only relevant boundary being the final spatial hypersurface. FLT concluded their analysis by noting that using the steepest descent method over the contour $0<N<\infty$ favours Vilenkin's tunneling approach due to the resultant propagator having the negative exponential weighting $e^{i 1\pi^{2}S_{0}} \sim e^{-12\pi^{2}/(\hbar\Lambda)}$. This was firmly shown by Vilenkin \& Yamada\cite{VY}.\\

The starting point of this analysis is the familiar Lorentzian Einstein-Hilbert action for the minisuperspace closed FRW metric. The Path integral assumes the proper time gauge $\dot{N}=0$ with the contour of $N$ taken as $0<N<\infty$. The path integral is interpreted as a Green's function of the WdW equation which obeys
\begin{equation}
    \mathcal{H}G(a_{1}|a_{0})=-i\delta(a_{1}-a_{0})
\end{equation}
Therefore in the limit as $a_{0} \to 0$, the Green's function is then a solution of the WdW equation in the entire range. As in the previous cases one may find the equations of the motion (\ref{LB2}) from the action (\ref{LB-1}), find the solution unique to the boundary conditions as in (\ref{LB-2}) and substitute this solution for $\bar{q}$ back into the action to yield $S_{0}$. At this point Picard Lefshets theory can then be employed to find which saddle points contribute to the semi-classical evaluation of the path integral by deformation of the contour to their Lefshets thimbles. Given the nature of the contour, the dominant contribution will come from the saddle point in the upper, positive quadrant of the complex plane which leads to the familiar weighting.

\newpage
\section{An Off Piste Approach to a Simple System}
The goal of this thesis has been to present a roughly chronological narrative of the development of path integral approaches to quantum cosmology. Despite the shift from the Euclidean path integral to the Lorentzian path integral as the starting point for calculations, the recipe used to derive the wavefunction or propagator has remained the same. To recapitulate, this involves the variation of the action with respect to the scale factor and/or conjugate momenta to yield the closed FRW field equations. The variation of the FRW action with respect to lapse function $N$ yields the Hamiltonian constraint. A general solution which does obey the Hamiltonian constraint can be written as the sum of the solution to the field equations subject to boundary conditions of choice which does not obey the constraint and a part which does obey the Hamiltonian constraint for which the corresponding path integral is Gaussian and can be evaluated trivially. Upon substituting the boundary condition specific solution of the field equations into the action, the action is solely a function of $N$. Taking the variation of this action with respect to $N$ will yield saddle points in the complex plane which will give the dominant contributions to the path integral. The original contour is then deformed such that these saddle points are included in the contour and the path integral along this contour is convergent.\\

However, despite all the variations considered, there seems to be very little room for variation of the recipe used to find the propagator/wavefunction. This is, in part, due to the fact that specifying that classical equations of motion and the saddle points in $N$ are the dominant contributions to the path integral semi-classically is highly constraining. Therefore if entirely new results, or new methods for calculating previous results are to be found, they can only be found by changing the starting point of the calculation. Recent work carried out by Mageuijo and collaborators has examined the canonical quantisation using the closed FRW metric in minisuperspace. This work differs from the contributions shown in previous sections as the starting point is the Einstein-Cartan in tetrad formalism rather than Einstein-Hilbert action. As a \\

It is important to point out that the notation used in the following sections is that of considering the path integral as a Green's function of the WdW equation. However, this is just notation for the sake of convenience. As we will see, the primary new result of the subsequent sections is that, from the closed FRW Einstein-Cartan action path integral, a  propagator/wavefunction which is the Fourier Transform of both the No-Boundary and Tunneling wavefunctions can be found. \\

This final section significantly overlaps with calculations that will appear in a future work by the author\cite{RI}\\

\subsection{Schematic rationale}
Initially, we will disregard the calculations and machinery developed in previous sections. As this schematic reasoning breaks down, results and techniques from previous sections will be used. The starting point for these calculation is the Einstein-Cartan action for the closed FRW metric in minisuperspace.\\

The evaluation of these Path Integrals (In the case of Approach 1) closely mirror calculations carried out by Vilenkin and Yamada\cite{VY}  in which the leading order term for a homogeneous de Sitter model Path Integral is evaluated. For these calculations, the gauge in which the lapse function $N(t)$ is constant is chosen. This leads to a simplification of the path integral over super-space form an initial amplitude ($g_{0}, \phi_{0}$) to a final amplitude ($g_{1}, \phi_{1}$)\cite{VY},
\begin{equation}
    G(g_{0}, \phi_{0};g_{1}, \phi_{1})=\int_{g_{0}, \phi_{0}}^{g_{1}, \phi_{1}}\mathcal{D}g \mathcal{D}\phi \ e^{iS},
\end{equation}
to a path integral over the scale function a and the Lapse function $N$ given by (no matter fields are under consideration),
\begin{equation}
    G^{(0)}= \int_{0}^{\infty}dN\int \mathcal{D}a \ e^{iS^{(0)}[a.N]}.
\end{equation}
Lapse integration is conducted over a semi-infinite range ($N > 0$) t ensure that conformal time employed in the calculations occurs after conformal time $-\infty$ such that geometrically identical histories linked by time reflection are not included in the path integral. V\& Y introduce introduce a new conformal time coordinate and introduce boundary conditions for $a$ to yield a Green's function of the inhomogeneous WdW equation. These calculations were carried out for an action that is only dependent on the scale factor. It can be assumed that if one uses the path integral approach for minisuperspace with an action $S=S[a,b,N]$, then there exists additional choice of what variables to integrate in the path integral. If once again, Lapse integration is taken over a semi-infinite range and path integrals can be evaluated exactly, then it follows that integration over the Lapse function and one conjugate variable, should yield a Green's function in the remaining conjugate variable. Schematically this is given by
\begin{align}
\begin{split}
    &G^{(0)}(b)=\int_{0}^{\infty} dN \int \mathcal{D}a \ e^{iS^{(0)}[a,b,N]},\\
    &G^{(0)}(a)=\int_{0}^{\infty} dN \int \mathcal{D}b \ e^{iS^{(0)}[a,b,N]}.
\end{split}
\end{align}
Assuming that integration is exact, there is no need to impose boundary conditions and as such the above Green's functions obey the following equations
\begin{align}
\begin{split}
    &\mathcal{H}G(a)=-i\delta(a) \longrightarrow \mathcal{H}G(a)=0, \quad a\neq a(0),\\
    &\mathcal{H}G(b)=-i\delta(b) \longrightarrow \mathcal{H}G(b)=0.
\end{split}
\end{align}

\subsection{Canonical Quantisation of the Einstein-Cartan Action}
Starting with the closed FRW metric, the components may be rewritten in tetrad formalism, Cartan's first and second equations in the case of no torsion may be used to derive connection 1-forms and curvature 2-forms and these may be substituted into the Einstein-Cartan action to yield the FRW Einstein-Cartan action for a mixture of perfect fluids\cite{BC1}\cite{B2}\cite{B3}
\begin{equation}
    S=\frac{3V_{c}}{8\pi G_{0}}\int dt \ \Biggl[a^{2}\dot{b}-Na\Biggl[-(b^{2}+k)+\sum_{i}\frac{m_{i}}{a^{1+3w_{i}}}\Biggl]\Biggl],
\end{equation}
where $V_{c}$ is the comoving volume resulting from the integral $\int d^{3}x$ and $G_{0}$ is Newton's constant. The term $w_{i}$ defines the equation of state of the individual fluids. 

In this approach, the Poisson bracket of the conjugate variables $a^{2}$ and $b$ is thought of as the classical limit of the quantum mechanical commutation relation between these two variables as,
\begin{equation}
    \{b,a^{2}\}=\frac{1}{6kV_{c}}\quad \longrightarrow \quad [b, a^{2}]= i\frac{l_{p}^{2}}{3V_{c}}= i\textbf{h},
\end{equation}
where $\textbf{h}$ can be thought of as an effective Planck's constant related to the original constant $\hbar$ by $l_{p}=\sqrt{8 \pi G_{N}\hbar}$. In the connection representation where $a^{2}$ is analogous to a momentum operator, this operator is given by,
\begin{equation}
    a^{2}=-i\frac{l_{p}^{2}}{3V_{c}}\frac{\partial}{\partial b}=- i \textbf{h} \frac{\partial}{\partial b}.
\end{equation}
Taking the variation of the general Einstein-Hilbert action reduced to minisuperspace with respect to the Lapse function leads to the Hamiltonian constraint (in the case of a single fluid)\cite{JMA},
\begin{equation}
    \frac{\delta S_{0}}{\delta N}= 6kV_{c}\Bigl(-(b^{2}+k)+ \frac{m}{a^{1+3w}}\Bigl)=0,
\end{equation}
with the standard Wheeler DeWitt equation given by,
\begin{equation}
    H_{0}\psi_{s}(b,m)=0.
\end{equation}
Using the fact that we are working in the connection representation, the above equation can be altered such that the equation we are trying to solve becomes\cite{JMA},
\begin{align}
\begin{split}
    &\mathcal{H}_{0}\psi_{s}(b,m)=0,\\
    &(h_{\alpha}(b)a^{2}-\alpha)\psi_{s}(b,m) =0,\\
    &\Bigl(h_{\alpha}(b)\frac{\partial}{\partial b}-\alpha \Bigl)\psi_{s}(b,m)=0,
\end{split}
\end{align}
where $h_{\alpha}(b)= (b^{2}+k)^{\frac{2}{1+3w}}$ and $\alpha = m^{\frac{2}{1+3w}}$. The above equation can be transformed into a plane wave equation in the connection be with the definition,
\begin{equation}
    h_{\alpha}(b)\frac{\partial}{\partial b} \equiv \frac{\partial}{\partial X_{\alpha}(b)},
\end{equation}
for which using the chain rule leads to the relation,
\begin{equation}
    X_{\alpha}(b)= \int \frac{db}{h_{\alpha}(b)}.
\end{equation}
This series of manipulations and definitions allows for the Wheeler DeWitt equation to be written as a plane wave equation in $X_{\alpha}$ given by,
\begin{equation}
    \Bigl(-i\mathbf{h}\frac{\partial}{\partial X_{\alpha}(b)}- \alpha \Bigl)\psi_{s}(b, \alpha)=0.
\end{equation}
Omitting a normalisation constant, the general solutions for the plane wave equation given above are given by,
\begin{equation}
    \psi_{s}(b, \alpha)= e^{i \frac{3V_{C}}{l_{p}^{2}}\alpha X_{\alpha}(b)}.
\end{equation}
This solution will differ for the three different cases as each $\alpha$ depends on the unique equation of state w. The three states corresponding to the solutions for the unique equations of state are given by\cite{JMA},
\begin{align}
\begin{split}
    &\text{{Dust}}(w=0) = e^{i3V_{c}m_{D}^{2}\int \frac{db}{(b^{2}+k)^{2}}}\\
    &\text{{Radiation}}(w=1/3) = e^{i3V_{c}m_{R}\int \frac{db}{(b^{2}+k)}}\\
    &\text{{Lambda}}(w=-1) = e^{i3V_{c}\cdot \frac{3}{\Lambda}\int db(b^{2}+k)}.
    \end{split}
\end{align}
The correct canonical context to think of the WdW equation is that of Dirac quantisation. As the relationship between Dirac quantisation and the path integral in the ADM formalism has been firmly established\cite{HAL1}, it should follow that the connection representation states which obey the Hamiltonian constraint should also be the result of a path integral calculation.

\subsection{The Fourier Transform}
As shown above, the most general solution for the Hamiltonian constraint in the connection representation for the FRW metric with cosmological constant $\frac{\Lambda}{3}$ is the Chern-Simons functional. However, this is a solution to the Hamiltonian constraint in the connection representation. Equivalently, one could choose the metric representation in which the connection operator can be written as a derivative of the $a^{2}$ operator given by
\begin{equation}
    b=\frac{il_{p}^{2}}{3V_{c}}\frac{d}{d(a^{2})}
\end{equation}
The Hamiltonian constraint in the metric representation is therfore given by
\begin{equation}
    \Biggl(\frac{il_{p}^{2}}{3V_{c}}\Biggl)^{2}\frac{d}{d(a^{2})}\Biggl(\frac{d}{d(a^{2})}\Biggl)+k-\frac{\Lambda}{3}a^{2}=0.
\end{equation}
Using the relation
\begin{equation}
    \frac{d}{d(a^{2})}=\frac{1}{a}\frac{d}{da},
\end{equation}
the Hamiltonian constraint can be simplified as
\begin{align}
    \begin{split}
        &\Biggl(\frac{il_{p}^{2}}{3V_{c}}\Biggl)^{2}\frac{1}{a}\frac{d}{da}\Biggl(\frac{1}{a}\frac{d}{da}\Biggl)+k-\frac{\Lambda}{3}a^{2}=0,\\
        &\Biggl(\frac{il_{p}^{2}}{3V_{c}}\Biggl)^{2}\Biggl(\frac{1}{a^{2}}\frac{d^{2}}{da^{2}}-\frac{1}{a^{3}}\frac{d}{da}\biggl)+k-\frac{\Lambda}{3}a^{2}=0,\\
        &\frac{d^{2}}{da^{2}}-\frac{1}{a}\frac{d}{da}+\Biggl(\frac{3V_{c}}{il_{p}^{2}}\Biggl)^{2}a^{2}\Biggl(k-\frac{\Lambda}{3}a^{2}\Biggl)=0.
    \end{split}
\end{align}
This is the usual WdW equation with specific ordering. Specifically, $k=1$ for a closed FRW universe and one can set $V_{c}=2\pi^{2}$. The ordering parameter from previous sections can also be set with $p=-1$ as in previous sections. This is expected as the Einstein-Cartan action reduces to the Einstein-Hilbert action in the case of no torsion. As shown previously, depending on the boundary conditions chosen, the solution of the WdW equation can either be the HH\cite{HH1}\cite{HH2} or Vilenkin wavefunctions. The Hamiltonian constraint in the connection representation for which the Chern-Simons is a solution, is a first order differential equation while the WdW equation is second order with two unique solutions. The question of how the solutions in the different representations are related is pertinent. The answer to this question was put forward by Magueijo who noted that as these are solutions to the WdW equation in complementary variables, they must be related by a Fourier Transform given by\cite{JM2}
\begin{equation}
    \psi_{a^{2}}(a^{2})=\frac{3V_{c}}{l_{p}^{2}}\int \frac{db}{\sqrt{2\pi}}\ e^{-i\frac{3V_{c}}{l_{p}^{2}}a^{2}b}\psi_{b}(b)
\end{equation}
The central result of \cite{JM2} is that the insertion of the Chern-Simons state into the above equation yields the integral representation of an Airy function which may be Ai, Bi or a linear combination. As shown in previous sections the Tunneling wavefunction is a complex wavefunction which is a linear combination of Airy functions as $\psi_{T} \propto \text{{Ai}(-z)}+i\text{{Bi}}(-z)$. The No-Boundary wavefunction is a real wavefunction defined as $\psi_{HH}\propto \text{{Ai}(-z)}$. This Fourier transform relation yields the integral representation of the Airy function by defining
\begin{equation}
    z=-\Biggl(\frac{9V_{c}}{\Lambda l_{p}^{2}}\Biggl)^{\frac{2}{3}}, \quad t=\ \Biggl(\frac{9V_{c}}{\Lambda l_{p}^{2}}\Biggl)^{\frac{1}{3}}b.
\end{equation}
Inferring from the general theory of integral representations of Airy functions, Mageuijo\cite{JM2} was able to show that the Chern Simmons state is dual to both the HH and Tunneling wavefunction due to the freedom of the choice of contour in the complex $b$ plane. If the contour of integration in the Fourier Transform is the real $b$ axis $(-\infty, \infty)$ or $D_{1}$ then the No-Boundary wavefunction is the result of the Fourier Transform. On the other hand if the contour is taken as $D_{2}$ $(-i\infty,0) \ \bigcup \ (0, \infty)$ then the Tunneling wavefunction is the result of Fourier Transform defined over this contour. Magueijo suggests that one should not see the Chern-Simons state (defined for the appropriate domain) and the No Boundary/Tunneling wavefunctions as being different since they are expressions of the same quantum state in two different representations as
\begin{align}
\begin{split}
    \psi_{T}(a^{2})&=\braket{a^{2}|\psi_{T}},\\
    \psi_{CS}(b;b \in D_{2})&=\braket{b|\psi_{T}}
\end{split}
\end{align}
This is the same for the HH wavefunction with
\begin{align}
\begin{split}
    \psi_{HH}(a^{2})&=\braket{a^{2}|\psi_{HH}},\\
    \psi_{CS}(b;b \in D_{1})&=\braket{b|\psi_{HH}}
\end{split}
\end{align}\\

Let us initially treat the propagator as some regular exponential function in multiple variables. If this is the case then we may try to naively use exponential identities where relevant to solve the integral with the connection representation states derived by Magueijo as the final results.
\begin{center}
    \textbf{Dust Calculation}
\end{center}
Inserting the equation of state $w=0$ into the most general minisuperspace Hamiltonian yields the action,
\begin{equation}
    S^{(0)}= 6\kappa V_{c}\int dt \ \Bigl(a^{2}\dot{b}-Na \Bigr[-(b^{2}+k) + \frac{m_{D}}{a} \Bigr]\Bigl),
\end{equation}
where $m_{D}$ is a constant associated with the specific equation of state for dust. The overall path integral can then be written as,
\begin{equation}
     G^{(0)}(b)=\int_{0}^{\infty}dN\int \mathcal{D}a \ e^{i6\kappa V_{c}\int dt \Bigl(a^{2}\dot{b}-Na \Bigr[-(b^{2}+k) + \frac{m_{D}}{a} \Bigr]\Bigl)}.
\end{equation}
The action in the path integral can be rewritten as,
\begin{equation*}
    6kV_{c}\int dt \ (a^{2}\dot{b}+Na(b^{2}+k)-Nm_{D}).
\end{equation*}
Therefore, the path integral over $a$ can be easily evaluated as the functional equivalent of the exponential integral
\begin{equation}
    \int_{-\infty}^{\infty}dx \ e^{-(ax^{2}+bx+c)}= \sqrt{\frac{\pi}{a}}e^{\frac{b^{2}}{4a}-c}.
\end{equation}
The result of the evaluation of the exponential integral is given by,
\begin{equation}
    G^{(0)}(b)=\sqrt{-\frac{\pi}{\dot{b}}}\int_{0}^{\infty}dN \ e^{i\int dt \Bigl(-\frac{N^{2}(b^{2}+k)^{2}}{4 \dot{b}}-Nm_{D}\Bigl)} 
\end{equation}
As the integral in the exponent is quadratic in N, the saddle point approximation can be used in which the solutions for $\frac{\partial S^{0}}{\partial N}=0$ can be substituted back into the original path integral to yield an exact value for the path integral over N. Employing the saddle point approximation over N for the action in the above integral yields the root,
\begin{equation}
    N= -\frac{2m_{D}\dot{b}}{(b^{2}+k)^{2}}.
\end{equation}
Substituting this back into the above action yields the solution to the overall path integral
\begin{equation}
   G^{(0)}(b)= \sqrt{-\frac{\pi}{\dot{b}}}e^{im^{2}_{D}\int dt \frac{\dot{b}}{(b^{2}+k)^{2}}}
\end{equation}
This is almost in agreement with the solution of the Hamiltonian constraint for dust. However, there is a prefactor which is the result of the $\dot{b}$ in the integral being the highest order term for the corresponding identity.
\begin{center}
    \textbf{Radiation Calculation}
\end{center}
A similar calculation can be used to evaluate the path integral for the radiation equation of state corresponding to $w=\frac{1}{3}$. The full path integral for radiation is given by,
\begin{equation}
    S^{(0)}= 6\kappa V_{c}\int dt \Bigl(a^{2}\dot{b}-Na \Bigr[-(b^{2}+k) + \frac{m_{R}}{a^{2}} \Bigr]\Bigl),
\end{equation}
Noting that the multiplication of terms in the brackets by the proceeding term leads to $a$ and $a^{-1}$ terms which lead to the overall path integral being unsuitable for evaluation using exponential integral identities. This problem can be resolved by a suitable redefinition of $N$ given by,
\begin{equation*}
    N\longrightarrow \Bar{N}, \quad \Bar{N}=\frac{N}{a}.
\end{equation*}
As mentioned in Chapter 2, Halliwell showed that the Hamiltonian operator in the WdW equation remains unchanged under this transformation and therefore it is allowed \cite{HAL2}.
The above action can then be written as,
\begin{equation*}
    S^{(0)}= 6\kappa V_{c}\int dt \Bigl(a^{2}\dot{b}-N \Bigr[-(b^{2}+k) + \frac{m_{R}}{a^{2}} \Bigr]\Bigl).
\end{equation*}
As in the dust calculation, the path integral with this action in the exponent can be considered as the functional equivalent of the regular exponential integral,
\begin{equation}
    \int_{-\infty}^{\infty}dxe^{-\alpha x^{2}}e^{- \frac{\beta}{x^{2}}}= \sqrt{\frac{\pi}{\alpha}}e^{-2\sqrt{\alpha \beta}}.
\end{equation}
With the identification of $\alpha=-\dot{b}$ , $\beta =Nm_{R}$ and the second term in the action being independent of $a$ due to the redefinition of $N$, the path integral over $a$ can be solved analytically to yield the path integral over N,
\begin{equation}
  G^{(0)}(b)=  \sqrt{-\frac{\pi}{\dot{b}}}\int_{0}^{\infty}dNe^{i\int dt (-2\sqrt{-\dot{b}Nm_{R}}+N(b^{2}+k)}.
\end{equation}
As in the case of the dust path integral the saddle point approximation for the action can be employed in which the specific value for $N$ is found to be,
\begin{equation}
    N=-\frac{\dot{b}m_{R}}{(b^{2}+k)^{2}}
\end{equation}
The substitution of this value for N back into into the action yields the final result,
\begin{equation}
    G^{(0)}(b)=\sqrt{-\frac{\pi}{\dot{b}}}e^{-3im_{R}\int dt \Bigl(\frac{\dot{b}}{(b^{2}+k)}\Bigl)}.
\end{equation}
Once again we find that this is the solution for the Hamiltonian constraint in the case of radiation in the connection representation but it is also multiplied by a prefactor that is an artefact of the use of regular exponential identities.

\subsection{Dust and Radiation: Path Integral over $N$ and $a$}
Its apparent that the naive use of exponential identities is only yields an approximation to an answer. The motivation for the use of exponential identities is that the Einstein-Cartan action contains $a$ while the Einstein-Cartan action contains $a^{2}$. However, the action is linear in $N$ and measure for the Lapse function in the constant gauge is $dN$. For Dust and Radiation the lapse function path integral can be carried out
 
 \begin{center}
    \textbf{Radiation Calculation}
\end{center}
Starting with a path integral containing the action for radiation given in (12) and utilising the redefinition of $N$ given by $N\rightarrow Na$, the overall path integral can be written as,
\begin{equation}
\int dN \int \mathcal{D}a^{2} \ e^{i \int dt (\dot{b} +N(b^{2}+k))a^{2}+Nm_{R}}= \int dN \ \delta(\dot{b} +N(b^{2}+k))\ e^{i \int dt Nm_{R}}.
\end{equation}
Enforcing the Dirac delta condition leads to the vanishing of the path integral unless,
\begin{equation*}
    N=-\frac{\dot{b}}{(b^{2}+k)}.
\end{equation*}
Inserting this condition into the previous path integral leads to,
\begin{equation}
    G^{(0)}(b)=e^{im_{R}\int dt \Bigl(\frac{\dot{b}}{(b^{2}+k)}\Bigl)}.
\end{equation}
As in the case of Lambda, the result of integrating over $\mathcal{D}a^{2}$ rather than $\mathcal{D}a$ is a much cleaner calculation with no need of the saddle point approximation in N as this is fixed by a Dirac Dirac delta function. Once again, a $-\dot{b}^{-1/2}$ prefactor is absent due to the exponential integral identities not being used.

\begin{center}
    \textbf{Dust Calculation}
\end{center}
For this calculation, the evaluation of the $a^{2}$ integral prior to integration over the lapse function is quite tedious. It is worth noting that the integration limits for the lapse function and squared scale factor are independent of each other. Given this, and the fact the degrees of freedom are bosonic (meaning there is no sign change upon reversing order of integration) and not conjugate to each other (meaning there is no Fourier Transform relationship between both variables), reversing the order of integration of the path integral should not change the final solution. Therefore the path integral for dust can also be written as,
\begin{equation}
    G^{(0)}(b)=\int \mathcal{D}a^{2}\int dN \ e^{i6\kappa V_{c}\int dt \Bigl(a^{2}\dot{b}-Na \Bigr[-(b^{2}+k)+ \frac{m_{D}}{a} \Bigr]\Bigl)}
\end{equation}
This leads to a Dirac delta function such that,
\begin{equation}
    G^{(0)}(b)=\int \mathcal{D}a^{2} \ \delta(m_{D}-(b^{2}+k)\cdot a)e^{i\int dt a^{2}\dot{b} }.
\end{equation}
The Dirac delta function fixes the value of $a$ and therefore fixes the value of $a^{2}$ as
\begin{equation*}
    a=\frac{m_{D}}{(b^{2}+k)}\quad \longrightarrow \quad a^{2}= \frac{m_{D}^{2}}{(b^{2}+k)^{2}}.
\end{equation*}
Inserting this into the exponential term leads to the identification of the Greens function as
\begin{equation}
    G^{(0)}(b)=e^{im_{D}^{2}\int dt \frac{\dot{b}}{(b^{2}+k)^{2}}}.
\end{equation}
Which is the same exponential term as shown in Approach 1 without the $-\dot{b}^{-1/2}$ factor. This method of reversing the order of integration can be Lambda and Radiation cases and also yields Dirac delta functions which fix the $a$ to be a certain value due to the presence of a single power of the lapse function $N$ in all three path integrals. These results should not be surprising since the carrying out the $N$ integral in this case leads a delta term. But the term in the square brackets is the Hamiltonian. This method can therefore be seen as a complementary way deriving solutions to the Hamiltonian constraints in the connection representation.

\subsection{ Lambda : Path Integral over $N$ and $a^{2}$}
 In Chapter 2 we discussed how, in the context of Euclidean Path Integrals, Halliwell was able to show that the path integral over conjugate variables and $N$ can be made finite by making the action BRS invariant which in turn, allows us to fix in to be in the proper time gauge\cite{HAL1}. Assuming that the ghost and auxiliary field integrals are carried out, the remaining path integral is simply a path integral over a constant $N$ and conjugate variables. Therefore the full phase space measure in the proper time gauge for the Einstein-Cartan action is $\mathcal{D}\mu=dN\mathcal{D}b\mathcal{D}a^{2}$. In the following calculations, we still use the assumption that one only integrates over $a^{2}$ and $N$ to get a propagator or wavefunction int he connection representation. The base variable of Quantum Cosmology conjugate to $b$ is $a^{2}$. As $\Lambda$ is the only one of the three actions that contains $a^{2}$ in its Hamiltonian, only $\Lambda$ can be considered for a full phase space path integral.\\

 For the Einstein-Cartan action in the case of positive cosmological constant $\Lambda$ can be carried out by either evaluating the $N$ or $a^{2}$ integral first. In the case of evaluating the $N$ integral corresponding to solving the Hamiltonian constraint. The path integral is given by
\begin{equation}
    G^{(0)}(b)=\int \mathcal{D}a^{2}\int dN e^{i\int dt \ a^{2}\dot{b}-N\Bigl[-(b^{2}+k)+\frac{\Lambda}{3} a^{2} \Bigl]}.
\end{equation}
Immediately taking the Hamiltonian constraint leads to
\begin{align}
    G^{(0)}(b)&=\int \mathcal{D}a^{2}\delta \Biggl(-(b^{2}+k)+\frac{\Lambda}{3}a^{2}\Biggl)e^{i\int dt \ a^{2}\dot{b}},\\
    &=e^{i\frac{3}{\Lambda}\int db \ (b^{2}+k)}
\end{align}
Given that we are interested in the propagator between final spatial hypersurfaces, we specify the limits of integration to be $t_{f}$ and $t_{i}$ such that $b(t_{f})=b_{f}$ and $b(t_{i})=b_{i}$. Then the above propagator yields
\begin{equation}
    G^{(0)}(b_{f}|b_{i})=e^{i\frac{3}{\Lambda}\Bigl(\frac{b^{3}_{f}}{3}+kb_{f} \Bigl)}\times e^{-i\frac{3}{\Lambda}\Bigl(\frac{b^{i}_{i}}{3}+kb_{i} \Bigl)}.
\end{equation}
If $b_{i}$ is set to zero which would correspond to the instant of Tunneling in Vilenkin approach, the then the above becomes a single Chern-Simons state and is in line with Mageuijo's result. Given that $a^{2}$ and $b$ are conjugate variables, the Fourier Transform of the end points can be taken to yield the initial or final point in the conjugate representation. Taking the Fourier Transform of the Chern-Simons function at $t_{f}$ we see that,
\begin{align}
\begin{split}
   G^{(0)}(a_{f}|b_{i})&= \frac{3V_{c}}{\sqrt{2\pi}l_{p}^{2}}\int db_{f} \ e^{i\frac{3}{\Lambda}\Bigl(\frac{b^{3}_{f}}{3}+kb_{f} \Bigl)-i\frac{3V_{c}}{l_{p}^{2}}a^{2}_{f}b_{f}}\times e^{-i\frac{3}{\Lambda}\Bigl(\frac{b^{i}_{i}}{3}+kb_{i} \Bigl)}\\
   &=\text{{Ai}}(z_{f}) \times e^{-i\frac{3}{\Lambda}\Bigl(\frac{b^{i}_{i}}{3}+kb_{i} \Bigl)},\\
   &\text{{or}}\\
   &=(\text{{Ai}}(z_{f})+i\text{{Bi}}(z_{f}) \times e^{-i\frac{3}{\Lambda}\Bigl(\frac{b^{i}_{i}}{3}+kb_{i} \Bigl)}
 \end{split}
\end{align}
This is the exact result result derived by Halliwell\cite{HAL2} for the propagator between initial $q$ and final $p$ ass shown in (\ref{HAL1}) and (\ref{HAL2}). The Airy function associated with final $a^{2}$ depends on the contour in the complex $b$ plane that is taken. Taking the Fourier Transform of the inital Chern-Simmons function, we arrive at the result
\begin{align}
    \begin{split}
        G(a_{f}|a_{i})_{HH}&=\text{{Ai}}(z_{f})\times \text{{Ai}}(z_{i}),\\
        G(a_{f}|a_{i})_{T}&=(\text{{Ai}}(z_{f})+i\text{{Bi}}(z_{f}) \times 
        (\text{{Ai}}(z_{i})+i\text{{Bi}}(z_{i}).
    \end{split}
\end{align}
This result is also identical to the result derived by Halliwell using the steepest descent method within the context of Euclidean Path integrals shown in (\ref{HAL3}). From this we see that if we start from the Einstein-Cartan action, deriving the propagators for the various combinations of initial and final conjugate variables is quicker than the calculations shown in previous chapters. The simplicity of the calculation is due to the Einstein-Cartan action being Linear in $N$ and $a^{2}$ (or quadratic in $a$). As these are the same powers of the path integral measures, the path integral is fixed by Dirac delta relations.

\begin{center}
    \textbf{Physical Significance of Lambda Calculation}
\end{center}
Evaluating the $a^{2}$ integral for Lambda first we see that
\begin{equation}
   G^{(0)}(b_{f}|b_{i}) \int dN \ \delta \Biggl(\dot{b}-N\frac{\Lambda}{3}  \Biggl) \ e^{i\int dt \ N(b^{2}+k)}
\end{equation}
This will yield the same Chern Simons, or, product of Chern Simons propagators depending on the initial $b_{i}$. As previously discussed, there is much debate concerning the concerning which contour should be taken in the complex $N$\cite{FLT}\cite{FLT2}\cite{HAL3} when the starting point for calculations is the Einstein-Hilbert action. Here we see that the integral for the Chern-Simons propagator will vanish unless this condition is satisfied which is the field equation for $b$. Given that $b=\dot{a}$ on shell, $\dot{b}=\ddot{a}$ on shell. Therefore the above equation tells us that the acceleration of the universe is constant. In addition to this, if we take $\Lambda$ to be positive which is the case if we want to consider expanding de Sitter spacetimes, this means that $N$ must also be positive. This, at first glance, seems to disagree with Halliwell and Diaz Doronsorro's statement that the $N$ is merely a mathematical parameter which parameterises histories in superspace\cite{HAL3}. However, the extent to which one can follow through with this reasoning is unclear and it may turn out that upon greater investigation N cannot be thought of as a causal parameter.

\subsection{Extension of Path Integral Approach to Unimodular Gravity (in Minisuperspace)}
In the case of unimodular gravity, constants of nature which may appear as fixed parameters in the Einstein-Hilbert action may be promoted to integration constants by the transformation of the base action as[JM], 
\begin{equation}
    S_{0}\longrightarrow S = S_{0}-\int d^{4}x \ (\partial_{\mu}T^{\mu}_{U})\alpha \quad or \quad S = S_{0}+\int d^{4}x \ (\partial_{\mu}\alpha)T^{\mu}_{U}.
\end{equation}
These two are equivalent up to a boundary term. Generically, $\alpha$ is a constant while $T_{U}^{\mu}$ is a vector density. Enforcing that the symmetry of the action should be the group of volume preserving diffeomorphisms leads to only the time component $T^{0}_{U}$ being the only relevant physical component of this vector density. The above equation becomes
\begin{equation*}
    S_{0}\longrightarrow S = S_{0}-\int dt \ \dot{T^{0}_{U}}\alpha \quad or \quad S = S_{0}+\int dt \ \dot{\alpha}T^{0}_{U}.
\end{equation*}
The time component of the vector density can be thought of as being conjugate to the constant of nature and as such obey the classical Poisson bracket and quantum mechanical commutator,
\begin{equation}
    \{T,a^{2}\}=\frac{1}{6kV_{c}}\quad \longrightarrow \quad [b, a^{2}]= i\frac{l_{p}^{2}}{3V_{c}}= i\textbf{h}
\end{equation}
As in the case of the previous conjugate variables, one can work in the $T$ representation in which the constant of nature $\alpha$ is  conjugate momentum like differential operator. This converts the WdW equation given  into an effective Schrodinger equation given by\cite{JM1},
\begin{equation}
    \Bigl(-i\mathbf{h}\frac{\partial}{\partial X_{\alpha}(b)}- i\textbf{h}\frac{\partial}{\partial T_{\alpha}} \Bigl)\psi(b, T_{\alpha})=0.
\end{equation}
In this case the s subscript denoting a spatial wave function has been removed since $T_{\alpha}$ plays the role of a time variable. The monochromatic solutions of this effective Schrodinger equation are given by,
\begin{align}
\begin{split}
    &\psi(b, T_{\alpha})=\psi_{s}(b,\alpha)e^{-\frac{i}{\textbf{h}}\alpha T_{\alpha}},\\
    &\psi(b, T_{\alpha})=e^{i \frac{3V_{C}}{l_{p}^{2}}\alpha (X_{\alpha}(b)-T_{\alpha})}.
\end{split}
\end{align}
\begin{center}
    \textbf{Minisuperspace Path Integral Version}
\end{center}
Once the Green's functions associated with the base theory have been derived, extension to the unimodular case appears to be trivial. Assuming that the path integration in the proper time gauge and $a$, then the additional integral  added in the case of unimodular gravity is a constant with respect to both of these variables and as such is not manipulated by any calculation. As an example, in the case of Lambda, the trivial unimodular extension of the Green's function is given by (going to change format for the sake of clarity),
\begin{align}
\begin{split}
   G^{(0)}(b,T_{3/\Lambda})&= \int d\phi \int \mathcal{D}a^{2}\int dN e^{i\int dt \ a^{2}\dot{b}-N\Bigl[-(b^{2}+k)+\frac{\Lambda}{3} a^{2} \Bigl]-i\phi T_{3/\Lambda}},\\
   &=\int d\phi \  \text{exp}(X_{3/\Lambda}-T_{3/\Lambda}),\\
   &= \delta(X_{3/\Lambda}-T_{3/\Lambda}),
\end{split}
\end{align}
where $X_{3/\Lambda}$ is the Chern-Simons function or product of Chern-Simons functions at defined times.
The rationale applies to both the dust and radiation cases which yield the unimodular Greens functions $G(b, T_{m_{D}})$ and $G(b, T_{m_{R}})$ respectively.

\begin{center}
    \textbf{Setting Up The Unimodular Calculation}
\end{center}
 Usually when propagators are considered, both conjugate variables are present in the path integral measure much like the Phase space propagator. In the unimodular case, this would mean an additional path integral over $\phi$ should be added to the unimodular set up calculation. In this case, this is due to the fact that the one can immediately evaluate the path integral for conjugate time. In this case
\begin{align}
\begin{split}
    G^{'(0)}(T_{3/\Lambda})&=\int\mathcal{D}T_{3/\Lambda}\int \mathcal{D}\phi \text{{exp}}(i \int dt \dot{T}_{3/\Lambda}\phi-\phi T'_{3/\Lambda})\\
    &= \int \mathcal{D}\phi \ \delta(\dot{\phi})\text{exp}(-\phi \bar{T'}_{3/ \Lambda})\\
    &=\Pi_{i}\int d\phi_{i} \ \delta(\dot{\phi_{i}})\text{exp}(-\phi_{i} \bar{T'}_{3/ \Lambda(i)}).
\end{split}
\end{align}
Where $T'_{3/\Lambda}$ are the inital values of these variables with $\phi$ free and $T'$ fixed.The delta condition therefore selects the $\phi_{i}$ that are constant between start and end paths. As in the case of the Lapse function $N$ in the constant gauge, this reduces the path integral to a normal integral over a range of values and the unimodular phase space propagator can be written as
\begin{equation}
    G^{'(0)}(T'_{3/\Lambda})= \int d\phi\ \text{exp}(-\phi \bar{T'}_{3/ \Lambda}).
\end{equation}However, as in the case of $a^{2}$ and $b$, the equivalence between canonical quantisation and path integral approaches only holds when one of the conjugate variables are integrated over rather than both. If both of the conjugate variables are integrated over then the overall unimodular propagator is a delta function
\begin{align}
\begin{split}
     G^{'(0)}(b, T_{3/\Lambda}) &= \int d\phi\ \text{exp}\Bigl(i \phi\Bigl(X_{3/\Lambda}-T_{3/\Lambda} \Bigl) \Bigl),\\
     & = \delta\Bigl(X_{3/\Lambda}-T_{3/\Lambda} \Bigl),
\end{split}
\end{align}

{\subsection{Removing Divergences}
\subsubsection{Adaptation of The Gell-Mann Low Theorem}
To find the Green's functions in the metric or connection representation, we consider a path integral over one of the conjugate variables and the lapse function rather than the entire path integral over all of phase space. If the path integral over the entirety of phase space is considered, then integration over the Lapse function and the metric variable yields (assuming that the initial connection $b(t_{i})$ vanishes) a final path integral over $b$ which can be evaluated as
\begin{align}
\begin{split}
    G(b_{f})&=\int_{0}^{b_{f}}\mathcal{D}b \ e^{\frac{3}{\Lambda}\Bigl(\frac{b_{f}^{3}}{3}+kb_{f} \Bigl)},\\
    &= \lim_{N \rightarrow \infty}\prod_{i=1}^{N} \int_{0}^{b_{f}}db_{i} \ e^{\frac{3}{\Lambda}\Bigl(\frac{b_{f}^{3}}{3}+kb_{f} \Bigl)},\\
    &= (b_{f})^{\infty} \times e^{\frac{3}{\Lambda}\Bigl(\frac{b_{f}^{3}}{3}+kb_{f} \Bigl).}
\end{split}
\end{align}
This is clearly horribly divergent with the divergence coming from the result of the additional path integral over the connection that was initially ignored. Given that initially ignoring the path integral over phase space yields the Chern-Simons functional which is in agreement with canonical quantisation methods, there must therefore be a consistent way to remove the divergence and obtain the desired result. This can be done by the adaptation of the Gell-Mann-Low Theorem to this case. Usually the path integral can be thought of as the ratio of the path integral with a current source and the free path integral. The free path integral of regular QFTs corresponds to the path integral with the classical action without the Hamiltonian constraint given by
\begin{equation}
    G_{cl}(b)= \int_{0}^{b_{f}}\mathcal{D}b \int dN \int \mathcal{D}a^{2} \ e^{i\int dt \ a^{2}\dot{b}}.
\end{equation}
Using the fact that for the FRW metric in the Einstein-Cartan formalism, the variables are related by $\dot{b}=\frac{\Lambda}{3}Na$, this can be substituted into the above action to yield
\begin{equation}
    G_{cl}(b)= \int_{0}^{b_{f}}\mathcal{D}b \int dN \int \mathcal{D}a^{2} \ e^{i\frac{\Lambda}{3}\int dt \ Na^{3}}
\end{equation}
By performing the redefinition $N \rightarrow \frac{N}{a}$, then either the Lapse function integral can be evaluated to give $\delta(a^{2})$ which causes the second integral to vanish or the metric integral can be evaluated to yield $\delta(N)$ which causes the Lapse integral to vanish. The final result is that the classical path integral is given by
\begin{align}
\begin{split}
    G_{cl}(b)&= \int_{0}^{b_{f}}\mathcal{D}b \ = \ (b_{f})^{\infty}
\end{split}
\end{align}In the case of regular QFT's, a source term coupled to the scalar field can be added to the action such that ratio of the sourced path integral and the free path integral defines the generating functional. In this case we see that, when taking the ratio of the full path integral and the classical path integral the distinguishing factor is the product of the Hamiltonian and Lapse function which is completely analogous to the field $\times$ source term in QFTs. The ratio between the two path integrals is 
\begin{align}
\begin{split}
    \frac{G(b_{f})}{G_{cl}(b_{f})}&=\frac{\int_{0}^{b_{f}}\mathcal{D}b \int \mathcal{D}N \int \mathcal{D}a^{2} \ e^{i\int dt \ (a^{2}\dot{b}-NH) }}{\int_{0}^{b_{f}}\mathcal{D}b \int \mathcal{D}N \int \mathcal{D}a^{2} \ e^{i\frac{\Lambda}{3}\int dt \ a^{2}\dot{b} }},\\
    &= \frac{\int_{0}^{b_{f}}\mathcal{D}b\ e^{\frac{3}{\Lambda}\Bigl(\frac{b_{f}^{3}}{3}+kb_{f} \Bigl)}}{\int_{0}^{b_{f}}\mathcal{D}b},\\
    &=e^{\frac{3}{\Lambda}\Bigl(\frac{b_{f}^{3}}{3}+kb_{f} \Bigl)}.
\end{split}
\end{align}Therefore the remaining divergent path integral can be removed with the result being in complete agreement with the canonical quantisation method. Given this justification, integral over one half of the conjugate variables and the lapse function can be considered by virtue of this normalisation. Despite this, it may be argued that the integral is not divergent, but rather can be made finite by a skeletonisation procedure similar to that shown in (\ref{H30}) where only $N$ finite time steps $b_{i+1}<...b_{i+N}$ rather than an infinite number of time steps are considered. As there is no dependence on time at the boundaries $b_{i+N} \neq b_{f}$. This seems unlikely as the skeletonisation integrals are defined by skeletonising conjugate coordinate and momenta. This yields a Dirac delta function which converts a path integral in one of the conjugate variables to a normal integral and in this case $a^{2}$ has already been integrated over. Ignoring that point assuming it is possible, then the result of this would be $(b_{f})^{N}$ which is finite. The classical action  would also yield $(b_{f})^{N}$. Whether the above prefactor is finite or divergent, this ratio will cause these terms to cancel yielding just the Chern-Simons function.

\subsection{Delayed Gauge Fixing}
In the above calculation, it was implicitly assumed that the Lapse function is in the constant gauge which is the gauge that calculations of this sort usually employ\cite{HAL1}. However, given that this is a different action, it is worth attempting to fix this action to be constant in different ways. The Lapse function being in the constant gauge can be expressed in terms of a full path integral with an additional path integral in an auxiliary field $b$ and an additional Fourier Transform relation
\begin{align}
\begin{split}
    G(b_{f})&=\int \mathcal{D}\bar{N} \int \mathcal{D}a^{2} \int^{b_{f}}_{0} \mathcal{D}b \int \mathcal{D}B \ \delta([\bar{N}-N_{0}])e^{i \int dt \ (a^{2}\dot{b}-\bar{N}H)}\\
    &= \int \mathcal{D}\bar{N} \int \mathcal{D}a^{2} \int^{b_{f}}_{0} \mathcal{D}b \int \mathcal{D}B \ e^{i \int dt \ (a^{2}\dot{b}-\bar{N}H +B(\bar{N}-N_{0}))}
\end{split}
\end{align}Upon evaluating the full Lapse function path integral a Dirac delta term arises. This can be used to fix either $a^{2}$ or $b$ 
\begin{equation}
    G(b_{f})=\int \mathcal{D}a^{2} \int^{b_{f}}_{0} \mathcal{D}b \int \mathcal{D}B \delta[(H-B)]e^{i \int dt \ (a^{2}\dot{b})-BN_{0}}
\end{equation}The term within the Dirac delta corresponds to the Wheeler De-Witt equation using metric ordering with the additional auxiliary field from delayed gauge fixing given by
\begin{equation}
    -(b^{2}+k) + \frac{\Lambda}{3}a^{2}-B=0.
\end{equation}Fixing $a^{2}$ then substituting this into the first term in the exponential yields
\begin{equation}
    \int \mathcal{D}b \int \mathcal{D}Be^{i \int dt \ \frac{3}{\Lambda}(\dot{b}(b^{2}+k+B))-BN_{0}}
\end{equation}This integral can be rewritten in terms of the Chern-Simons functional and remaining terms containing the auxiliary field $B$
\begin{align}
\begin{split}
     G(b_{f})&=e^{i\frac{3}{\Lambda}\Bigl(\frac{b_{f}^{3}}{3}+kb_{f} \Bigl)} \ \int^{b_{f}}_{0} \mathcal{D}b \int \mathcal{D}B \ e^{i\int dt \ B\Bigl(\frac{3}{\Lambda}\dot{b}-N_{0}\Bigl)},\\
     &=e^{i\frac{3}{\Lambda}\Bigl(\frac{b_{f}^{3}}{3}+kb_{f} \Bigl)} \ \int^{b_{f}}_{0} \mathcal{D}b \ \delta\Bigl(\frac{3}{\Lambda}\dot{b}-N_{0}\Bigl).
\end{split}
\end{align} As in the previous cases, the presence of a derivative within the Dirac delta reduces the full path integral in $b$ to a normal integral which gives a finite value for the propagator with
\begin{equation}
    G(b_{f})=b_{f} \times e^{i\frac{3}{\Lambda}\Bigl(\frac{b_{f}^{3}}{3}+kb_{f} \Bigl)}.
\end{equation}It is worth noting that for both methods of obtaining a finite Chern-Simons propagator, the appearance of the second Friedman equation enforcing constant acceleration appeared.}

\subsection{Metric Representation Propagator}

\subsubsection{Lambda Calculation: Connection Space Path Integral}
Having shown that the schematic argument discussed at the beginning of this section holds for evaluating the $\mathcal{D}a^{2} \ dN$ path integrals to yield the propagator $G(b)$, we can assume that these same arguments will hold for evaluating the $\mathcal{D}b \ dN$ path integral to yield the propagator $G(a)$, expecting that it will correspond dot the Hartle-Hawking or Vilenkin wavefunction depending on the contour of $b$.
\begin{equation}
    G^{(0)}(a)= \int \mathcal{D}b \int dN \ e^{i \int dt (-(\dot{a^{2})}b-Na[-(b^{2}+k)-\frac{\Lambda}{3}a^{2}])}.
\end{equation}
By using approach 2 and reversing the order of integration, a Dirac delta term arises which fixes the value of $b$. The above path integral is reduced to,
\begin{equation}
    \int \mathcal{D}b \ \delta \Bigl(-(b^{2}+k)+\frac{\Lambda}{3}a^{2}\Bigl)e^{i \int dt (-2a\dot{a}b)}.
\end{equation}
This Dirac delta term, fixes the value of $b$ such that,
\begin{equation}
    b=\pm \sqrt{\frac{\Lambda}{3}a^{2}-k}\quad 
\end{equation}
Noting that the path integral is over $b$ and the Dirac delta term contains $b$. The functional equivalent of the Dirac delta identity
\begin{equation}
    \delta(f(b))=\frac{\sum_{i}\delta(b-b_{i})}{f'(b_{i})}
\end{equation}
where $b_{i}$ are the roots of the function $f(b)$. Using this identity, the propagator $G(a)$ is therefore
\begin{equation}
    G^{(0)}(a)=\frac{e^{\pm i\int^{1}_{0} dt (\dot{a^{2}})\sqrt{\frac{\Lambda}{3}a^{2}-k}}}{\sqrt{\frac{\Lambda}{3}a^{2}-k}}.
\end{equation}
From the choice of $b$ used to satisfy the Dirac Delta identity we see that $b$ is chosen to be real. Given that the $a^{2}$ which appears this equation is not fixed, it can be inferred that the contour for $b$ is the entire real line. A propagator $G(a)$ which is entirely real corresponds to the Hartle-Hawking wavefunction. The real part of the exponential can be taken with
\begin{align}
\begin{split}
    G^{(0)}(a)&= \frac{\cos\Biggl(\int^{1}_{0} dt \ (\dot{a^{2}})\sqrt{\frac{\Lambda}{3}a^{2}-k}\Biggl)}{\sqrt{\frac{\Lambda}{3}a^{2}-k}}\\
    &= \frac{\cos\Biggl(\frac{2}{\Lambda} \Bigl(\frac{\Lambda}{3}a_{f}^{2}-k\Bigl)^{3/2}\Biggl)}{\sqrt{\frac{\Lambda}{3}a^{2}-k}}.
\end{split}
\end{align}
At this point we run into an unavoidable problem.It was assumed that as in the case of $b$, there is some ratio of the path integral with the Hamiltonian constraint and the path integral for the classical action that ensures that the remaining path integral over $a^{2}$ which leads to infinities can be removed. However, if this is assumed, then we see that the scale factor in the denominator is still free and has not been fixed which is nonsensical. Despite these problems, for the sake of depiction, we will assume that that the $a^{2}$ in the denominator is fixed to be $a_{f}^{2}$. 

\begin{center}
    \textbf{Assumption 2: Delayed Gauge Fixing Mechanism}
\end{center}
The delayed gauge fixing and ratio methods both employ the Raychaudhuri equation in order to make the integral finite with the resulting answer for the two methods differing by a constant. While the ratio method was able to completely eliminate the final integral which led to divergences, the delayed gauge fixing method ensured that instead of being removed, the final integral becomes a regular integral in $a^{2}$. If we assume that a delayed gauge fixing approach can be taken such that the final integral becomes a normal integral, then the alternative form of the propagator is
\begin{align*}
\begin{split}
   G(a_{f})&=\cos\Biggl(\frac{2}{\Lambda} \Bigl(\frac{\Lambda}{3}a_{f}^{2}-k\Bigl)^{3/2}\Biggl) \int_{0}^{a_{f}^{2}}\frac{da^{2}}{\sqrt{\frac{\Lambda}{3}a^{2}-k}}\\
   &=\frac{3}{\Lambda}\Bigl(\frac{\Lambda}{3}a_{f}^{2}-k\Bigl)^{1/2}\cos\Biggl(\frac{2}{\Lambda} \Bigl(\frac{\Lambda}{3}a_{f}^{2}-k\Bigl)^{3/2}\Biggl)
\end{split}
\end{align*}
where the constant factor of resulting from the lower limit of the integral has been omitted. This is, of course, also not the Hartle-Hawking scenario. Even worse, the first term is to the power of a positive number. It is therefore expected that the amplitude of oscillations will increase roughly linearly with $a_{f}$. This is in complete opposition to the Hartle-Hawking universe in which the amplitude of oscillations decreases to the first term in the having a negative power. This behavior is seen by the graph below.\\

\textbf{Disproving the assumptions}\\
To see specifically how the ratio method fails, we start by assuming the existence of a propagator $G_{cl}(a_{f})$ corresponding to the path integral for the classical action. To get a connection representation propagator we used one of Hamilton's equations which in this case is the second friedmann equation to reduce the path integral for the classical action to just the path integral over $b$. In this case, we will use the fact that from the classical Poisson bracket we get the torsion free condition. Thus, the classical metric space propagator would take the form
\begin{equation}
    G_{cl}(a^{2}_{f})= \int \mathcal{D}b \int dN \int \mathcal{D}a^{2} \ e^{-i \int dt \ Nb^{2}}
\end{equation}
If the $N$ integral is taken first then one gets $\delta(b^{2})$ within the lapse integral which is undefined. If the lapse function integral is chosen first and the saddle point approximation is used then the classical propagator is
\begin{equation}
    G_{cl}(a^{2}_{f})= \int dN \int \mathcal{D}a^{2}= \infty
\end{equation}
This propagator is doubly divergent due to an $\infty$ coming from both integrals such that even if the path integral over $a^{2}$ where to cancel, there would still be a remaining divergence from the Lapse integral. There is also no way to reduce the full path integral to a normal integral by delayed gauge fixing. From this we see that, if instead of fixing $a^{2}$, we use the Dirac delta identity to fix $b$, then we have
\begin{equation*}
    b=\sqrt{\frac{\Lambda}{3}a^{2}-B-k}
\end{equation*}
In this case, the overall path integral contains a path integral over the auxiliary field
\begin{equation*}
    \int \mathcal{D}B \ e^{i \int dt\ ({\dot{a^{2}}})\sqrt{\frac{\Lambda}{3}a^{2}-B-k}}
\end{equation*}
which is not tractable analytically. As both methods of attack do not yield sensible answers, the most logical conclusion is that the logic of integrating over half of a conjugate pair is  not applicable in the case of $a^{2}$.

\subsection{Relation to Previous Rsults in The Metric Representation}

Given that trying to evaluate the path integral over half of phase space to find the metric space connection wave function is pathological, we will proceed by evaluating all of the integrals. Starting with the path integral over $b$, integrating by parts and using the fact that the action is then Gaussian in $b$, the saddle point approximation can be employed to find that
\begin{equation}
    \frac{\dot{(a^{2})}}{2N}=b \quad {\rm or} \frac{\dot{a}}{N}=b
\end{equation}
Substituting this back into the original action completes the use of the saddle point approximation. It is worth noting that action after the saddle point approximation the action becomes
\begin{equation}
S(a,\dot{a},N)=i\int dt \ \Biggl(-\frac{\dot{(a^{2})}^{2}}{4N}+N\Bigl( k-\frac{\Lambda}{3}a^{2} \Bigl)\Biggl)
\end{equation}
which is none other than the Einstein-Hilbert action. The use of the saddle point approximation in $b$ amounts to enforcing the Torsion free condition. In this case the Einstein-Cartan action for FRW reduces to the Einstein Hilbert action. The full path integral is given by
\begin{equation}
    \int \mathcal{D}a^{2} \int dN \ e^{i\int dt \ \Biggl(-\frac{\dot{(a^{2})}^{2}}{4N}+N\Bigl( k-\frac{\Lambda}{3}a^{2} \Bigl)\Biggl)}
\end{equation}
This is almost the path integral that Vilenkin considers in [Tunneling wavefunction of the universe] except that the measure in $a^{2}$ rather than $a$. However, the action is quadratic in the scale factor and its derivative so one should still arrive at the Tunneling wavefunction of the universe obtained by Vilenkin by the use of steepest-descent theory.  The first term may be integrated by parts to obtain
\begin{equation*}
    \int dt \ \dot{(a^{2})}^{2}=- \int dt \ a^{2}\ddot{(a^{2})}
\end{equation*}
Using the saddle point approximation in $a^{2}$ we arrive at 
\begin{equation*}
    \ddot{(a^{2})}=4N^{2}H^{2} \quad \text{{where}} \quad H=\sqrt{\frac{\Lambda}{3}}
\end{equation*}
This is the exact second order differential equation that Vilenkin starts with and uses tunneling boundary conditions to solve analytically so that the path integral is then of order 4 in $N$ and contains only a sclae factor dependence on $a(t_{1})=a_{1}$.
\begin{center}
    \textbf{{Where does the WKB approximation come in? }}
\end{center}
{The seemingly exact evaluation of the path integral yields the same answer as solving the WdW equation in the metric representation using the WKB approximation. This implies that at some point in the path integral calculation, the WKB approximation is employed. Using an unsavoury notion of canonical reasoning within the context of the a path integral we see that the Dirac delta is equivalent to,
\begin{equation}
    \delta \Bigl(-i\textbf{h}\frac{d}{da}-i\textbf{h}\frac{d^{2}}{da^{2}}+a^{2}\Bigl(\frac{\Lambda}{3}a^{2}-k\Bigl)\Bigl).
\end{equation}
Using the Dirac delta property,
\begin{equation*}
    \delta(x-z)\delta(z-y)=\delta(x-y),
\end{equation*}
Therefore, Dirac delta  can be rewritten as,
\begin{equation}
    \delta \Bigl(-i\textbf{h}\frac{d}{da}-i\textbf{h}\frac{d^{2}}{da^{2}}+a^{2}\Bigl(\frac{\Lambda}{3}a^{2}-k\Bigl)\Bigl)= \delta \Bigl(-i\textbf{h}\frac{d^{2}}{da^{2}}+a^{2}\Bigl(\frac{\Lambda}{3}a^{2}-k\Bigl)\Bigl) \cdot \delta  \Bigl(-i\textbf{h}\frac{d}{da} \Bigl)
\end{equation} Given that the overall path integral vanishes unless both of the Dirac delta terms are satisfied, both of terms are of significance. The second Dirac delta term corresponds to the WKB approximation. The first Dirac delta term is satisfied for solutions of the differential equation. Taking these conditions together ensures that the Green's function resulting from the Path integral is the same as the wave functional in the canonical quantisation scheme.
\begin{center}
    \textbf{The Chern-Simons functional and WKB Approximation}
\end{center} In the path integral formulation the WKB-but-exact nature of the Chern-Simons functional is hidden within the calculations. The Dirac delta function arising from the evaluation of the $a^{2}$ path integral can be considered as a shorthand for enforcing the WKB approximation. For a given path integral Z associated with a function $\phi(t)$
\begin{align*}
\begin{split}
    Z&=\int \mathcal{D}\phi \ e^{\frac{i}{\hbar} f(\phi)},\\
    &=\int \mathcal{D}\phi \ e^{\frac{i}{\hbar}(f(\phi^{*})+f'(\phi^{*})(\phi-\phi^{*})+\frac{1}{2}f''(\phi^{*})(\phi-\phi^{*})^{2})}
\end{split}
\end{align*}where $f(\phi)$ is some functional of the field $\phi(t)$ and the second line is a Taylor expansion to second order of this functional.  As these integrals are considered to be highly oscillatory in the $\hbar \rightarrow 0$ limit such that the exponential is considered to be zero except for its saddle points for which,
\begin{equation*}
    f'(\phi^{*})=0
\end{equation*}. The fluctuations about the saddle point can be defined by a field $\chi$ such that
\begin{align*}
\begin{split}
    (\phi-\phi^{*})&=\sqrt{\bar{h}}\chi,\\
    d\phi&=\sqrt{\bar{h}}d\chi.
\end{split}
\end{align*}Using the saddle point approximation and the fluctuation redefinition, the original path integral becomes a Gaussian integral in fluctuations
\begin{equation*}
    \sqrt{\hbar}e^{\frac{i}{\hbar}(f(\phi^{*}))}\int_{\infty}^{\infty}d\chi \ e^{\frac{i}{2}f''(\phi^{*})\chi^{2}}.
\end{equation*} This rationale can be applied to the connection representation propagator. The propagator can be rewritten as a Taylor expansion to second order.
\begin{equation}
    G^{(0)}(b_{f})=\int dN e^{i S\bigl(a^{*^{2}},b^{*}\bigl)+S'\bigl(a^{*^{2}},b^{*}\bigl)\bigl(a^{2}-a^{*^{2}}\bigl)+S''\bigl(a^{*^{2}}, B^{*}\bigl)\bigl(a^{2}-a^{*^{2}}\bigl)^{2}}
\end{equation}The saddle point approximation in the previous example can be adapted to the case where the variable of interest is $a^{2}$ as
\begin{equation}
    \frac{dS}{a(a^{2})}=0 \ \text{{or}} \ S'(a^{2})=0 \  \text{{when}} \ a^{2}=a^{*^{2}}.
\end{equation} The highest power of $a$ in the action is of degree which means that the second derivative of the action with respect to $a^{2}$ is zero. As a result, the only term present in the Taylor expansion of the action is the first term. We get the Chern-Simons function by noting that $\dot{b^{*}}=N\frac{\Lambda}{3}$}

\begin{center}
    \textbf{{Hartle-Hawking or Vilenkin?}}
\end{center}
{To find $G^{(0)}(a)$, a Dirac delta function originates from the evaluation of the lapse function path integral as shown previously. The Green's function given  corresponds to the Hartle-Hawking wavefunction. The Hartle-Hawking wavefunction is not the only possible outcome of the calculation but rather, results from the choice of b chosen to satisfy the Dirac delta condition. This choice corresponds to choosing a contour of integration with the two possible contours being
\begin{equation}
    b=\pm \sqrt{\frac{\Lambda}{3}a^{2}-k}\quad \text{or} \quad b=- i \sqrt{k-\frac{\Lambda}{3}a^{2}} \ \cup \ +\sqrt{\frac{\Lambda}{3}a^{2}-k}.
\end{equation}
Choosing the Former as was the case above leads to the Hartle-Hawking wavefunction whereas the choice of the second corresponds to the Vilenkin wavefunction.}

\section{Conclusion}
\begin{center}
    \textbf{General Outlook}
\end{center}
Currently, there has been no experimental detection of quantum mechanical phenomena exhibited by the universe as a whole. Due to this, there are no clues that could help us indirectly infer the relevant boundary conditions. On the other hand, we still lack a general theoretical framework which selects the boundary conditions and the evolution of the universe uniquely. Given this scenario, it appears as though the choice between the No Boundary and Tunneling wavefunction as the reasonable model for a closed FRW universe in minisuperspace is ultimately a matter of preference. Both wavefunctions can be derived in the context of Lorentzian path integrals using Picard-Lefshets theory with the distinguishing factor being which saddle points are chosen as the dominant contributions to the path integral in the semi-classical regime. The collection of saddle points taken is dependent on the contour in the complex $N$ plane taken.\\

The contour for $N$ taken is ultimately a matter of the interpretation of its physical significance. If one argues as FLT, that $N$ parameterises the proper time between the scale factor at different stages of the evolution of the universe, then the natural contour is along the positive real $N$ axis as this allows for causality. If $a(t_{i})<a(t_{f})$ then integration over $N$ in the proper-time gauge $\dot{N}=0$ amounts to an integration over all proper times between $a(t_{i})$ and $a(t_{f})$. This causality is made manifest by virtue of there being a Dirac delta $-i\delta(a(t_{f})-a(t_{i}))$ term in the inhomogeneous WdW equation of which the result of the path integral is a solution. If $N$ is taken as merely parameterising curves in minisuperspace and is not physical as Hartle \& Halliwell argue, then it is reasonable to take the contour as the entire real $N$ axis to obtain a solution to the homogeneous WdW equation.\\

In the final section of this thesis, we used the Einstein-Cartan action as the starting point in the Lorentzian path integral. From this we were able to derive the Chern-Simons state in agreement with Magueijo's results derived in the context of canonical quantisation. Using the fact the Einstein-Cartan action fixes $a^{2}$ and $b$ as conjugate variables a Fourier Transform of the Chern-Simmons state can be taken which yields the integral representation of Airy functions. The propagator between an initial fixed $b_{i}$ and a final fixed $b_{f}$ is simply the the product of an initial and final Chern Simons state. We used a Fourier Transform of one and both of these states to arrive at near identical results to those previously obtained by Halliwell. Computationally, this is much quicker than previous methods.\\

The Chern-Simons state being dual to both the No Boundary and Tunneling wavefunctions does not resolve the problem of choice. When taking the Fourier transform, one must specify the contour for $b$ for which the integral representation of the Airy function is defined over leading to the result of the Fourier Transform being one of thee two metric representation wavefunctions. Thus what was initially a debate the physical significance of $N$ shifts to a debate concerning which contour in $b$ is correct to take.

\begin{center}
    \textbf{Future Work}
\end{center}
The little explored option of starting with the Einstein-Cartan is an intriguing one. The calculations shown in this thesis are for the case in which there is no torsion. Adding torsion and a momentum conjugate to torsion that is specified to be zero on shell requires adding a Lagrange multiplier to the Einstein-Cartan action. From the poisson brackets of these conjugate variables we see second class constraints. Second class constraints are notoriously difficult to work with in the context of path integrals. Future work could investigate the method by which one obtains the results derived by Magueijo for torsionful No Boundary and Tunneling wavefunctions by starting with a path integral. In addition to this, the calculations in this case wee carried out for the case of no matter fields. Fundamentally new results are not expected since the degrees of freedom are ultimately the same as the case of the Einstein-Hilbert action with a minimally coupled scalar field. However, as in this case, using this form of the action may yield the same results but a more concise way revealing alternative insights.

\newpage
\bibliographystyle{unsrt}
\bibliography{cit}
\end{document}